\documentclass[orivec,runningheads]{llncs}
\usepackage{ecir25-evaluating-chs-answers-frame}
\graphicspath{{../ecir25-evaluating-chs-answers-figures/}}

\begin{document}
\title{Ranking Generated Answers}
\subtitle{On the Agreement of Retrieval Models with Humans on Consumer Health Questions}

\author{
Sebastian Heineking\inst{1}\orcidlink{0000-0002-7701-3294} \and
Jonas Probst\inst{1} \and
Daniel Steinbach\inst{4}\orcidlink{0000-0003-2364-598X} \and
Martin~Potthast\inst{2,3}\orcidlink{0000-0003-2451-0665} \and
Harrisen Scells\inst{1,2}\orcidlink{0000-0001-9578-7157}}
\authorrunning{Heineking et al.}

\institute{
Leipzig University \and
University of Kassel and hessian.AI \and
ScaDS.AI \and
University of Leipzig Medical Center}

\maketitle

\setcounter{footnote}{0}

\begin{abstract}
Evaluating the output of generative large language models~(LLMs) is challenging and difficult to scale. Many evaluations of LLMs focus on tasks such as single-choice question-answering or text classification. These tasks are not suitable for assessing open-ended question-answering capabilities, which are critical in domains where expertise is required. One such domain is health, where misleading or incorrect answers can have a negative impact on a user's well-being. Using human experts to evaluate the quality of LLM~answers is generally considered the gold standard, but expert annotation is costly and slow. We present a method for evaluating LLM~answers that uses ranking models trained on annotated document collections as a substitute for explicit relevance judgements and apply it to the CLEF 2021 eHealth dataset. In a user study, our method correlates with the preferences of a human expert (Kendall's $\tau=0.64$). It is also consistent with previous findings in that the quality of generated answers improves with the size of the model and more sophisticated prompting strategies.%
\footnote{Code and data:\,\url{https://github.com/webis-de/ECIR-25}.} 

\end{abstract}

\section{Introduction and Related Work}

Search engines are used to ask questions in domains where a wrong answer can mean a high risk for the user~\cite{roegiest:2024}. Examples include health, medicine, finance, and law. Due to the many factors that influence answers in these domains, open-ended questions---as opposed to single-choice or factual questions---are more common here. The use of LLMs for retrieval-augmented generation~(RAG) and as chatbots in conversational search engines, as well as the tendency of some models to confirm user bias~\cite{berberette:2024,koopman:2023,wester:2023} call for a careful, large-scale evaluation of question-answering systems that require domain expertise. In fact, this requirement may go beyond expert domains: \citet{ouyang:2022} found that~57\% of GPT-3 API calls are open-ended questions.

While benchmarks with single-choice or factual questions allow a comparably simple automatic evaluation of LLMs~\cite{fan:2019,xu:2023}, their quality measures, such as text overlap or text similarity, are not suited for evaluating open-ended questions and their complex, nuanced answers. Here, automatic evaluation cannot compete with manual evaluation by humans~\cite{xu:2023}. As a remedy, crowdworkers are often employed for evaluation, but they require extensive training and knowledge of the domain in question. In this context, \citet{krishna:2021} note that the evaluation of answers to open-ended questions is much more challenging than that of answers to single-choice questions due to their length. Longer answers increase the time needed to process an example, leading to low annotator agreement when choosing between two answers to the same question. Another factor contributing to disagreement among annotators is that the quality of answers often cannot be reduced to a single dimension but depends on several factors~\cite{krishna:2021}, requiring a multifaceted evaluation. \citet{sakai:2023} and \citet{gienapp:2024a} propose to evaluate the quality of conversational systems based on criteria such as fluency, soundness, and explainability. These criteria, however, currently require a human-in-the-loop evaluation, while automated, scalable solutions remain an open research question. \citet{farzi:2024} propose to evaluate LLM answers using a grading rubric of questions that answers must address. The method, however, is focused only on relevance, while other criteria are not accounted for.

We propose a new evaluation method for generated answers to bridge the gap between high-quality human assessments and more efficient automatic approaches. Our method uses ranking signals from annotated corpora to measure the effectiveness of generated answers. For this method, we take inspiration from a ranking-based evaluation method for machine translation, where the translations generated by different models are ranked together with reference translations to compare the models' effectiveness~\cite{duh:2008,guzman:2015,li:2013}. To our knowledge, ranking-based evaluation has yet to be used to evaluate other tasks. Since manual annotations are not required for each new answer, this method allows for scalable evaluation of LLMs, including comparing different prompting strategies and model sizes. Furthermore, it facilitates a consistent evaluation of generated answers by limiting the room for subjective judgements to the initial annotations. We conduct experiments on the CLEF 2021 eHealth dataset~\cite{goeuriot:2021} and analyze how model size and prompting strategy influence the effectiveness of LLMs as measured by our rank-based evaluation. We further validate our ranking method in a user study by correlating its results with the ranking of a healthcare professional, as suggested by~\citet{arabzadeh:2024}.

\section{\mbox{\fontsize{11.7pt}{10.7pt}\selectfont NRP: Normalized Rank Position for Generated Answers}}

We propose normalized rank position~(NRP), an automatic method to evaluate answers generated by~LLMs. It requires a set of queries~$Q$ and a set of documents~$D$ with expert judgements for the queries. These judgements can express the document's relevance to the query or other quality dimensions. First, a set of retrieval models is evaluated on ranking documents in accordance with the expert judgements. Second, a set of LLMs is prompted to generate an answer~$a_q$ for each query~$q$ in the dataset. 
Third, the most effective retrieval model is used to rank~$a_q$ alongside all documents~$D_q\subseteq D$ with expert judgements for~$q$. The effectiveness of an LLM for~$q$ is evaluated based on the rank position of the answer it generated. The normalized rank position~(NRP) is calculated based on the absolute rank position~$r$ of~$a_q$ and the number of documents in~$D_q$: 
$$
\text{NRP}=1-\frac{r}{|D_q|},\ \ \mbox{where} \quad D_q = \{d \in{D} |~d~\text{has expert judgements for}~q\}.
$$
For NRP scores to be in the range~$[0,1]$, the rank positions need to be based on zero-based numbering, so that the highest rank position is~0 instead of~1.

If the expert judgements encompass multiple quality dimensions, different retrieval models might be the most effective at ranking documents in accordance with a given dimension. In these cases, a dimension-specific NRP can be calculated using the most effective retrieval model for each dimension. The individual scores are then combined in a weighted sum that corresponds to the importance of each quality dimension for the given use case. For our experiments, we report a single NRP scores as the same retrieval model was the most effective on all quality dimensions.
\section{Experimental Setup}
\label{sec:experimental-setup}

\paragraph{Data Collection and Preparation.}\label{sec:dataset}
We used the CLEF eHealth\,2021 dataset~\cite{goeuriot:2021} for evaluation. It comprises 55~health-related queries, obtained from medical experts and social media discussions, with judgements for web documents as well as Reddit and Twitter posts. The judgements reflect the quality dimensions relevance, readability, and credibility. With API changes at Reddit and Twitter, the content from these platforms has become unavailable. We therefore obtained only the original web~documents from CommonCrawl, discarded those containing fewer than 50~characters in the HTML~body, and extracted plain text using the Resiliparse library.%
\footnote{\url{https://resiliparse.chatnoir.eu/en/stable/}}
We were able to restore 6,692~web documents with judgements, omitting 5~queries exclusively paired with Twitter and Reddit posts.

\paragraph{Retrieval Pipeline.}
For ranking, we used two lexical retrieval models, TF-IDF and DPH, and four transformer-based models: \mbox{\texttt{ColBERTv1}} \cite{khattab:2020}, \mbox{\texttt{ColBERTv2}} \cite{santhanam:2022}, \mbox{\texttt{monoT5}} \cite{nogueira:2020}, and \mbox{\texttt{duoT5}} \cite{pradeep:2021}. Except for \mbox{\texttt{ColBERTv2}}, the versions available via PyTerrier~\cite{macdonald:2020} were used. For \mbox{\texttt{ColBERTv2}}, the implementation is the one provided by the authors of the model.%
\footnote{\url{https://github.com/stanford-futuredata/ColBERT}}
All transfor\-mer-based models are fine-tuned on MSMARCO. The most effective model in terms of nDCG@10 across the three quality dimensions of relevance, readability, and credibility is \mbox{\texttt{monoT5}} with scores of~$0.645$, $0.813$, and~$0.722$, respectively. Hence, \mbox{\texttt{monoT5}} is used in all further experiments and we calculate a single NRP based on its rankings.

\paragraph{Generating Answers.}
To generate answers, we selected LLMs that differ in 
\Ni 
number of parameters, 
\Nii 
amount and type of training data, and 
\Niii 
pre-training and fine-tuning.
We used the base, medium, large, and XL~variants of \mbox{\texttt{GPT-2}}, the instruction-tuned \mbox{\texttt{Falcon}}~7B, \mbox{\texttt{LLaMA-2}}~7B, and~13B, as well as \mbox{\texttt{ChatGPT}} based on \texttt{GPT-3.5-\allowbreak{}turbo-0613}. Across all models, we fixed the maximum number of new tokens to~512, the temperature to~0.75, top-$k$ to~50, top-$p$ to~0.95, and the repetition penalty to~1.2. For \mbox{\texttt{ChatGPT}}, we could only set the temperature and maximum number of new tokens. 
Prior research has shown the prompt formulation to strongly affect answer quality~\cite{reynolds:2021}. Therefore, after a pilot study, we included multiple prompts in our experiments for comparison:
\begin{description}
\item[\textbf{No Prompt:}]
\textit{query}
\item[\textbf{Short QA Prompt:}]
Q: \textit{query} A:
\item[\textbf{Long QA Prompt:}]
Question: \textit{query} Answer:
\item[\textbf{MultiMedQA Prompt~\cite{nori:2023}:}]
You are a helpful medical knowledge assistant. Provide useful, complete, and scientifically grounded answers to common consumer search queries about health. Question: \textit{query} Complete Answer:
\end{description}
We generated ten answers for each query and LLM to reduce the effect of random variations in generated answers.

\paragraph{User Study.}
Finally, we validated NRP by correlating its effectiveness estimation of different LLMs with annotations by an expert (a medical doctor and co-author of this paper). To keep the annotation task feasible, we sampled 20~of the 50~queries and provided five options per query. For each query, the expert received the most relevant web document according to \mbox{\texttt{monoT5}} and the top-ranked answer by \mbox{\texttt{ChatGPT}}, \mbox{\texttt{Llama-2 13B}}, \mbox{\texttt{GPT-2 XL}}, and \mbox{\texttt{GPT-2}}. They then created a joint ranking for all three quality dimensions from most relevant, readable, and credible to least. Answers were generated using the MultiMedQA prompt and agreement between the expert and \mbox{\texttt{monoT5}} was measured using Kendall's~$\tau$.

\section{Results}

\begin{figure}[t]
    \centering
        \begin{minipage}[t]{.66\textwidth}
            \vspace{0pt}
            \centering
            \includegraphics[width=\textwidth]{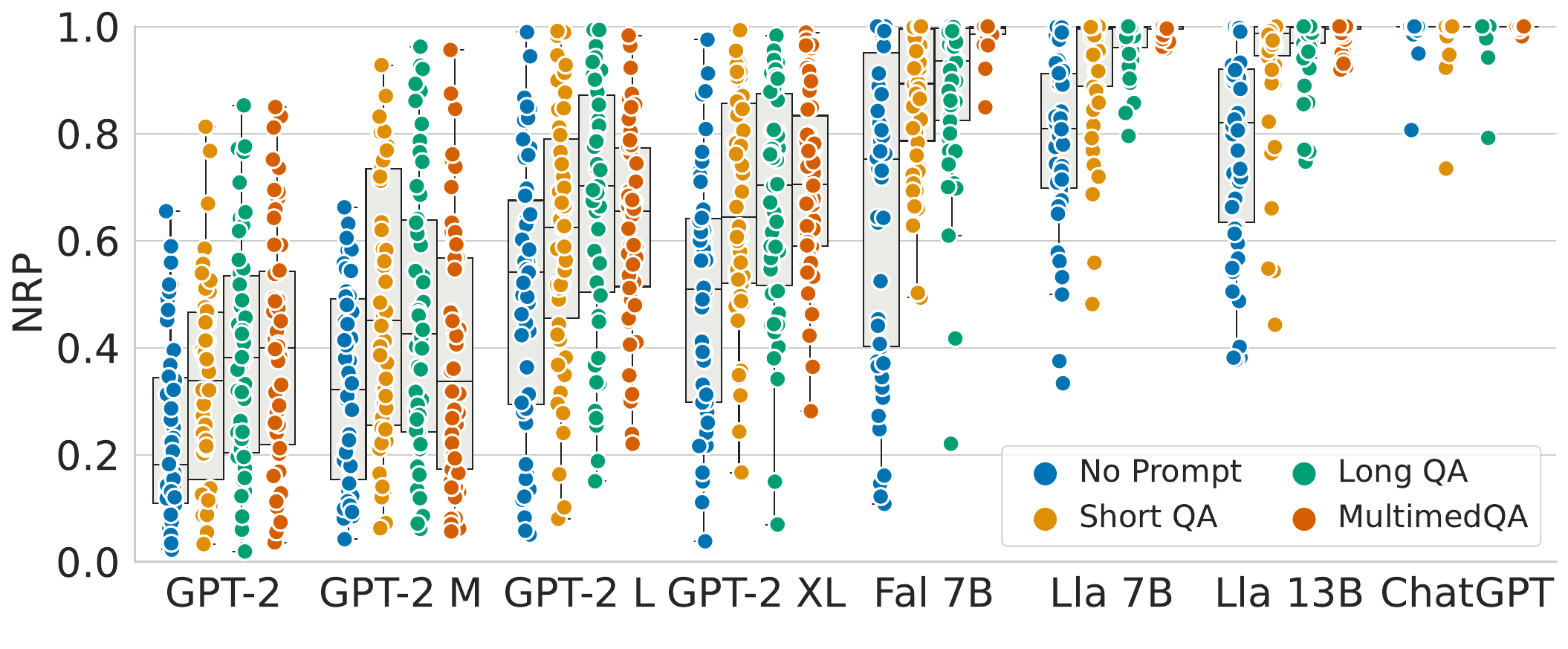}
            \caption{NRP of LLM answers, averaged over the ten generated answers per question, grouped by prompt. Points are single answers. \mbox{\texttt{GPT-2 L}} and \mbox{\texttt{Llama-2 7B}} not shown.}
            \label{fig:weighted_position_boxplot}
        \end{minipage}
        \hfill
        \begin{minipage}[t]{.3\textwidth}
            \vspace{0pt}
            \centering
            \includegraphics[width=\textwidth]{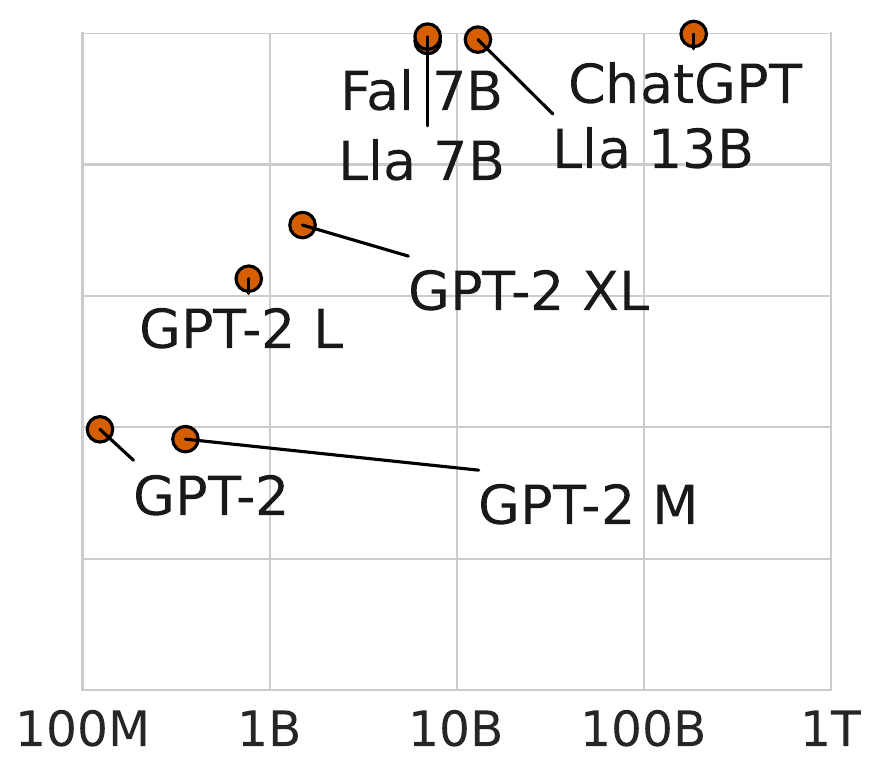}
            \vspace{-2ex}
            \caption{Number of model parameters vs. NRP using the MultiMedQA prompt.}
            \label{fig:weighted_position_vs_model_size}
        \end{minipage}%
    \end{figure}

\paragraph{Factors Influencing LLM Answer Quality.}
We investigated the effect of prompting strategy and model size on~NRP. In total, we ranked 16,000~generated answers: Each of the 8~models generated 10~answers for 4~different prompting strategies and 50~queries. Figure~\ref{fig:weighted_position_boxplot} shows the NRP scores for each~LLM, grouped by prompt. As illustrated in the figure, the choice of prompting strategy influences the position of generated answers, especially for models with more parameters. When no context is given to the model (``No Prompt''), the effectiveness decreases for many models, with the exception of \mbox{\texttt{ChatGPT}}. 
Many small models are able to achieve a higher effectiveness with the MultiMedQA prompt than the next larger model with no prompt. This gain in effectiveness underlines the importance of prompting and is particularly visible for the instruction-tuned models \mbox{\texttt{Falcon}} and \mbox{\texttt{Llama-2}}.
Of all models, \mbox{\texttt{ChatGPT}} is the most effective with a mean NRP of~0.998. \mbox{\texttt{Falcon~7B}} is the least effective of the fine-tuned models with the \mbox{\texttt{Llama}} models being slightly more effective. The \mbox{\texttt{GPT-2}} variants show the lowest effectiveness overall, but improves as the number of parameters increases. Figure~\ref{fig:weighted_position_vs_model_size} visualises this trend by comparing the number of parameters to NRP scores when using the MultiMedQA prompt. The largest improvements in NRP are between \mbox{\texttt{GPT-2 Medium}} and \mbox{\texttt{GPT-2 Large}}, and between \mbox{\texttt{GPT-2 XL}} and \mbox{\texttt{Falcon~7B}}/\mbox{\texttt{Llama-2~7B}}. For the small to medium-sized fine-tuned models, specifically the 7B~variant of \mbox{\texttt{Falcon}} and the~7B and 13B~variants of \mbox{\texttt{Llama-2}}, the trend is less clear. \mbox{\texttt{Llama-2~7B}} is more effective than the 7B~variant of \mbox{\texttt{Falcon}}, achieving an NRP of~0.995 compared to~0.988 by \mbox{\texttt{Falcon~7B}}. While they have the same number of parameters, the observed differences could be explained by differences in training data and the more sophisticated fine-tuning of \mbox{\texttt{Llama-2}}. \mbox{\texttt{Llama-2~7B}} ranks slightly better than the larger variant, which has nearly twice as many parameters. The similar rankings could indicate a saturation of the \mbox{\texttt{Llama-2}} model's effectiveness around that parameter count for our specific task.
To summarize, prompting and model size have a large influence on the effectiveness of generated answers and these effects are measurable with~NRP.

\paragraph{Agreement with Human Expert Preferences.}
\newcommand{\provideranks}[1]{\raisebox{-0.5ex}[0em][0em]{\includegraphics[height=1em,width=5em, trim=0pt 125pt 0pt 125pt, clip=True]{rank_visualization/#1}}}
\newcommand{\provideexample}[1]{\raisebox{-0.5ex}[0em][0em]{\includegraphics[height=1em,width=1em, trim=10pt 125pt 365pt 125pt, clip=True]{rank_visualization/example-#1}}}

\begin{table*}[t!]
\centering%
\begin{tabular}{@{}rccccccr@{}}
\toprule
\bfseries Query & \multicolumn{5}{c}{\bfseries Rank Positions of an Answer / Document} & \bfseries Correlation \\
\cmidrule(l@{\tabcolsep}r@{\tabcolsep}){2-6}
qid & ChatGPT & Lla 13B & GPT-2 XL & GPT-2 & Document & Kendall's $\tau$ \\
\midrule
  1 &   \provideranks{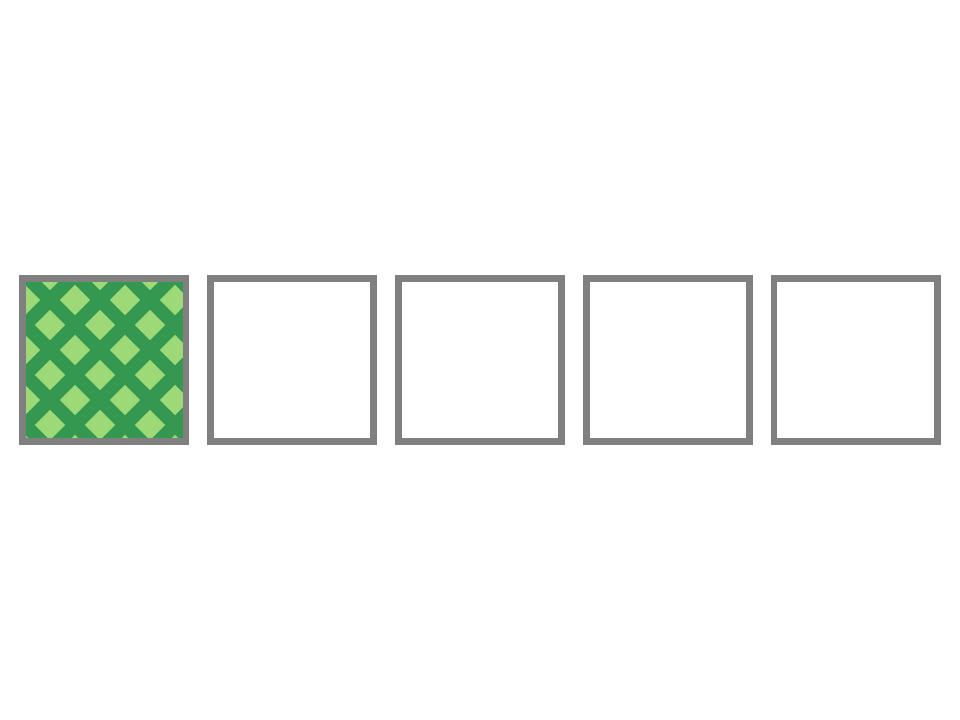} &   \provideranks{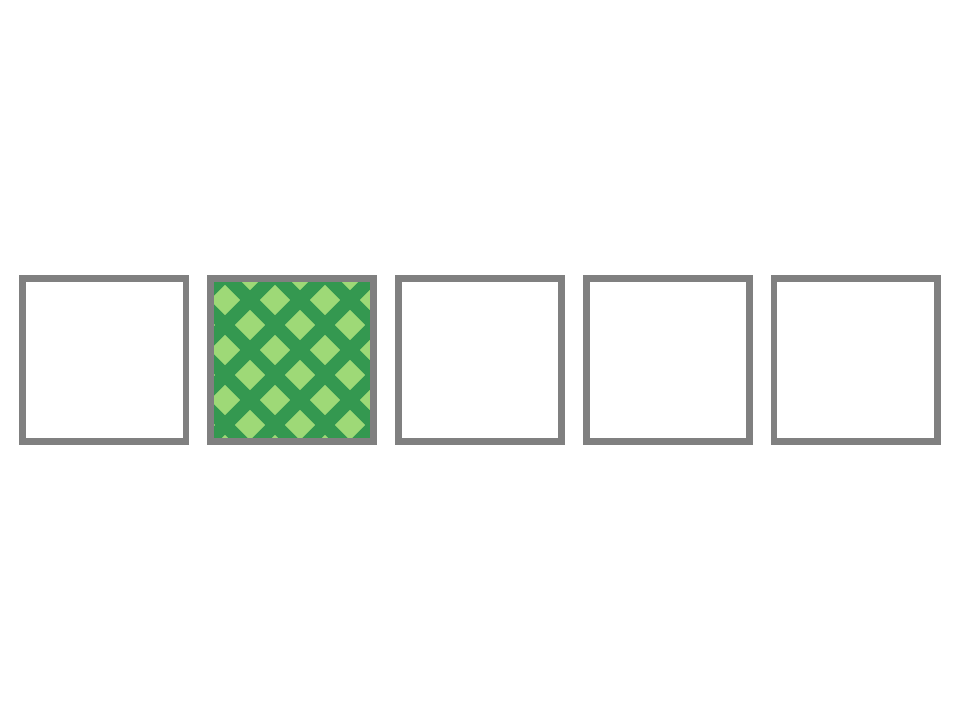} &   \provideranks{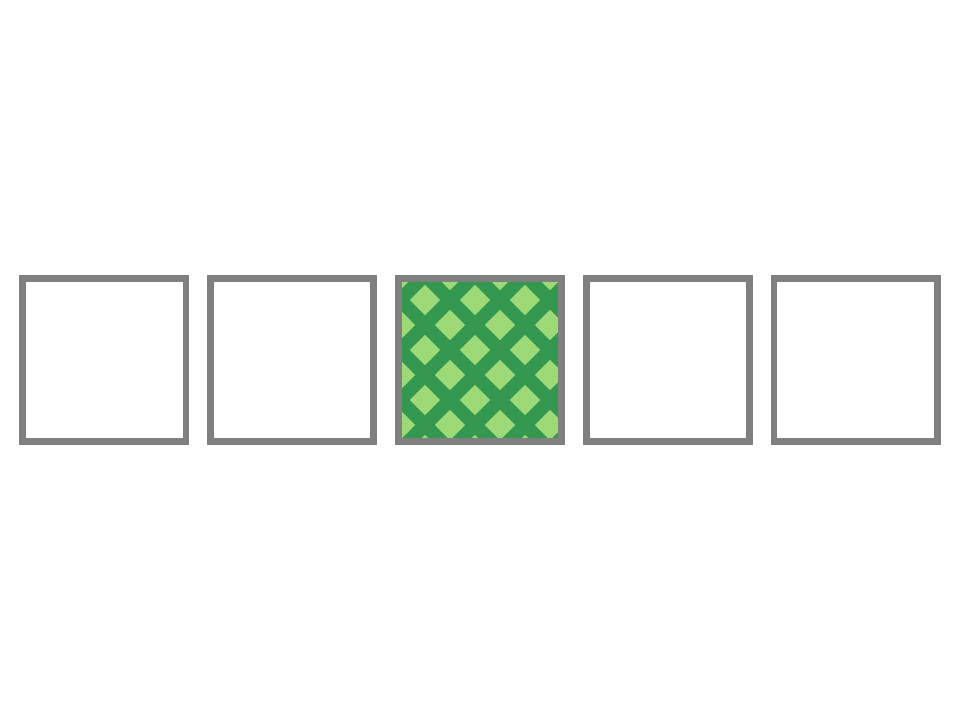} &   \provideranks{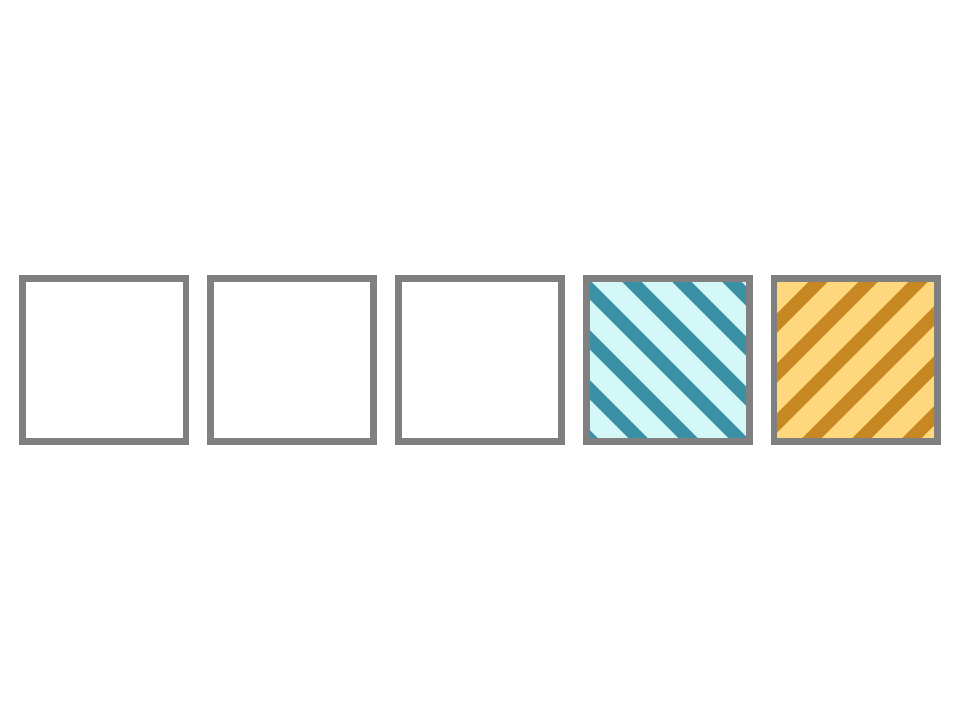} &   \provideranks{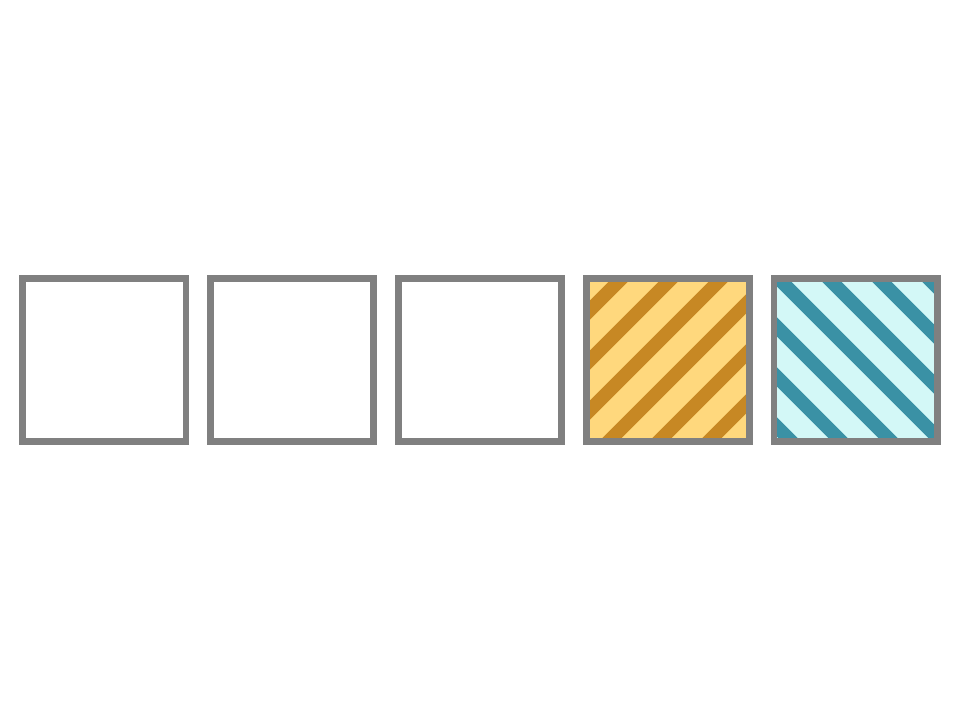} & \phantom{-}0.80 \\
 22 &  \provideranks{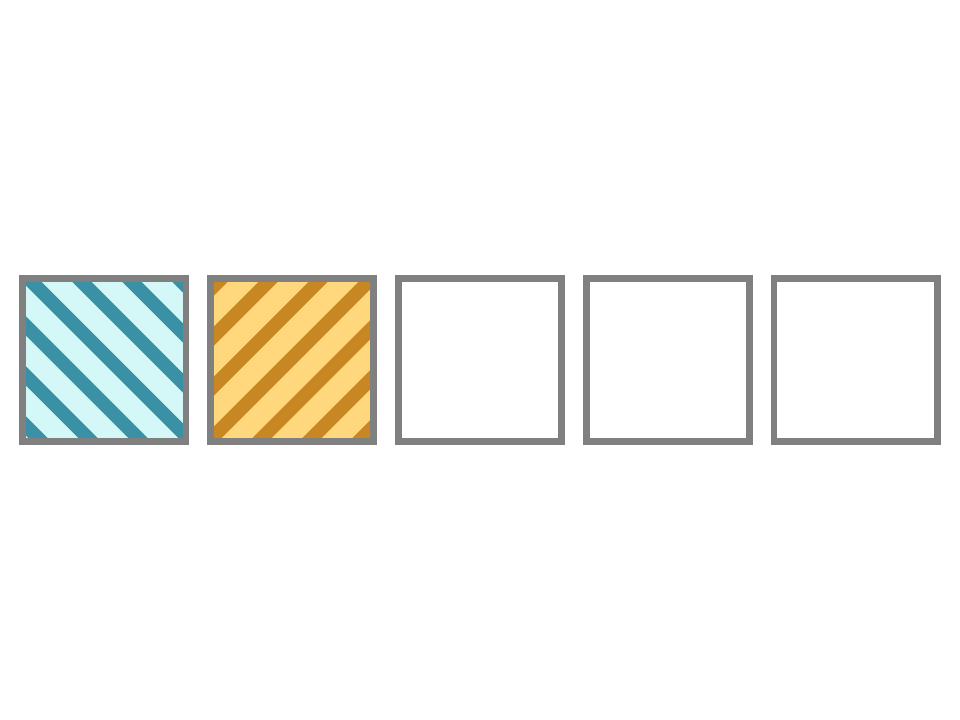} &  \provideranks{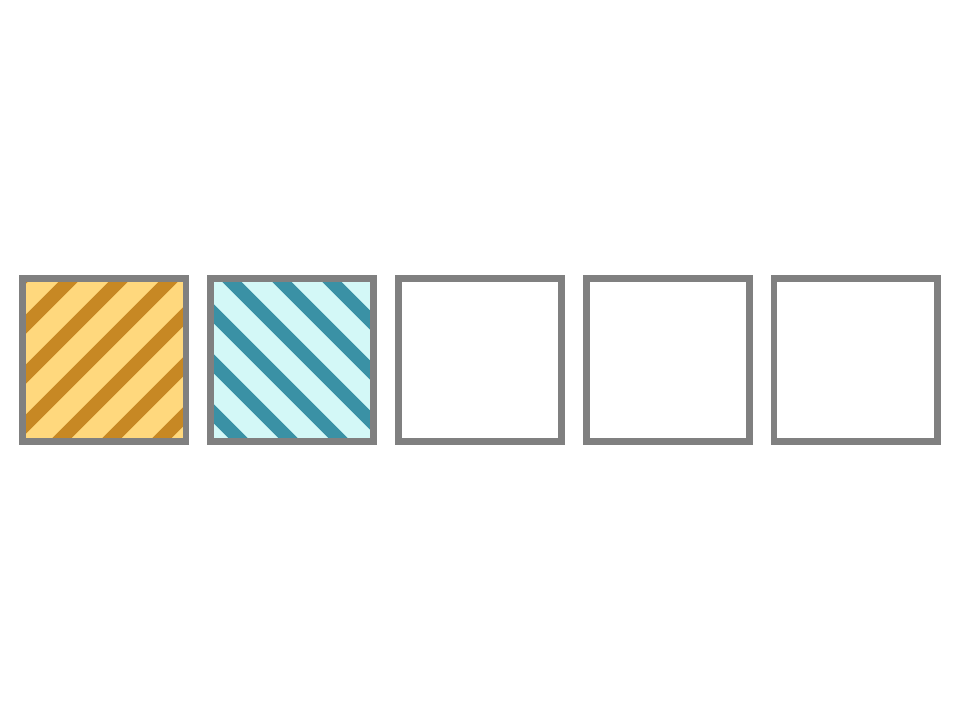} &  \provideranks{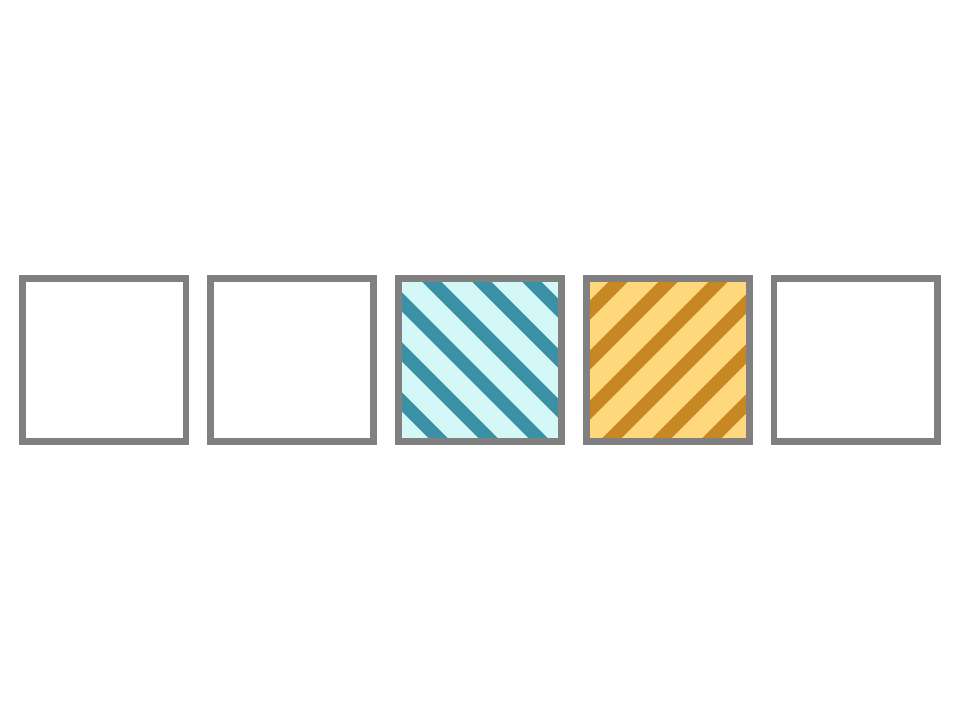} &  \provideranks{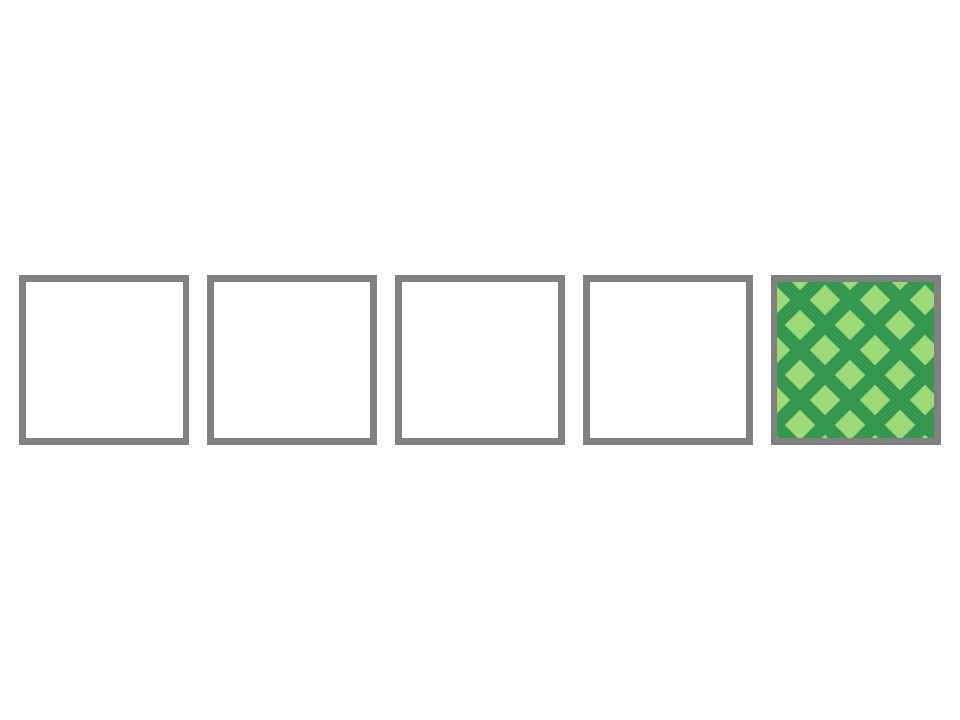} &  \provideranks{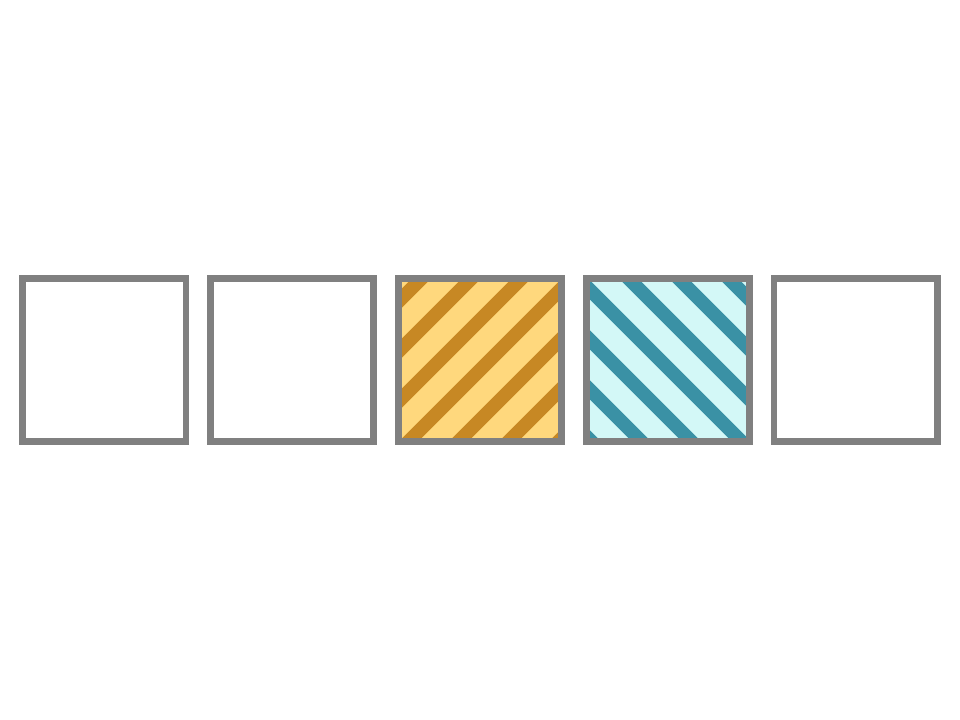} & \phantom{-}0.60 \\
 35 &  \provideranks{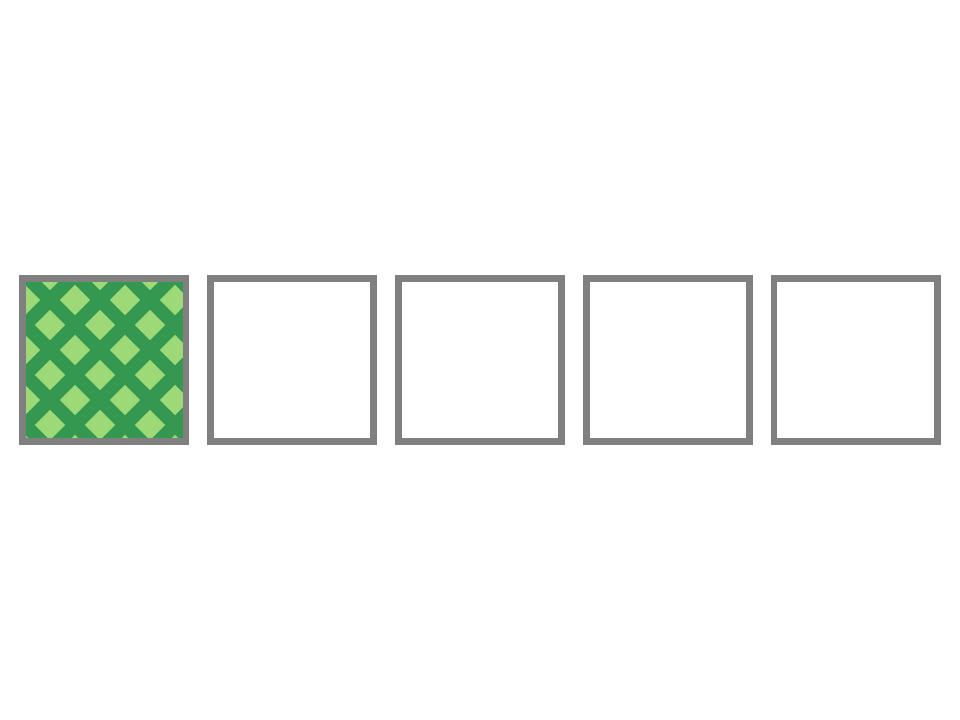} &  \provideranks{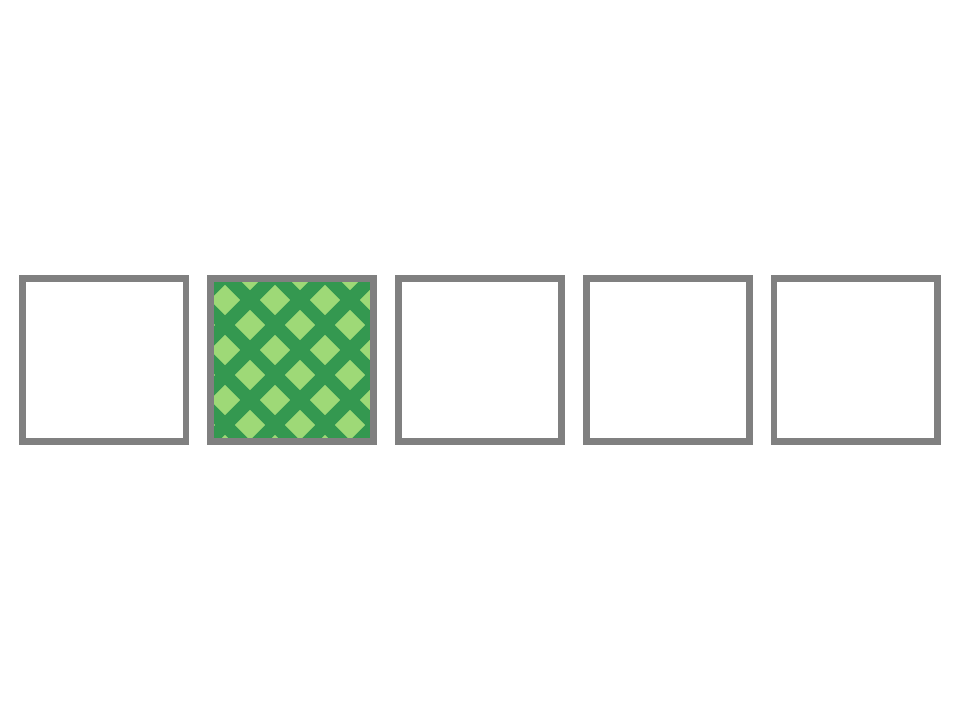} &  \provideranks{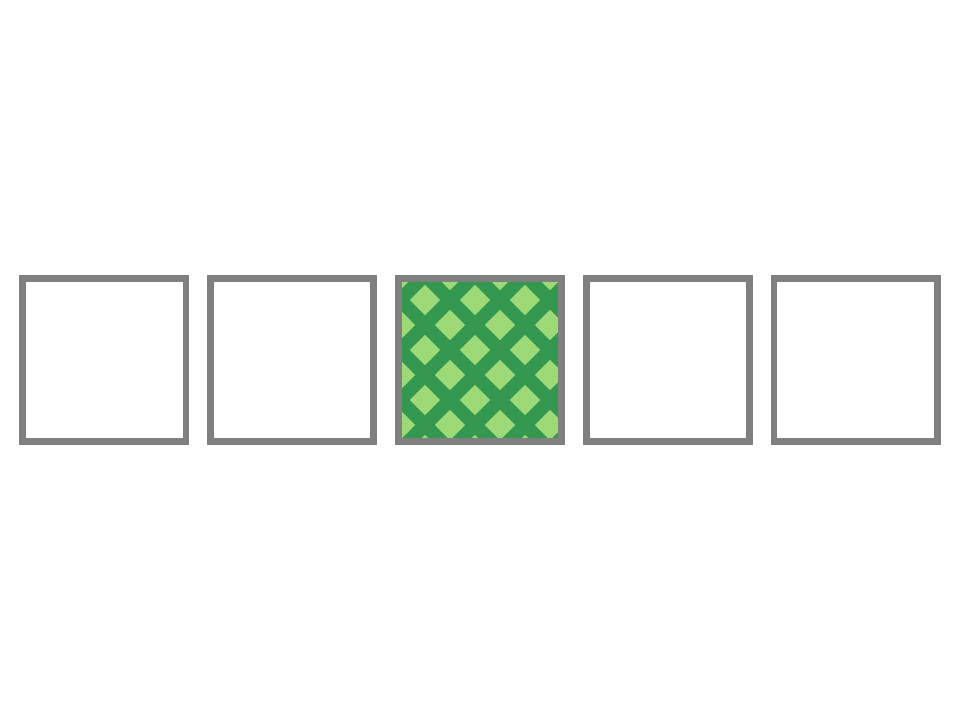} &  \provideranks{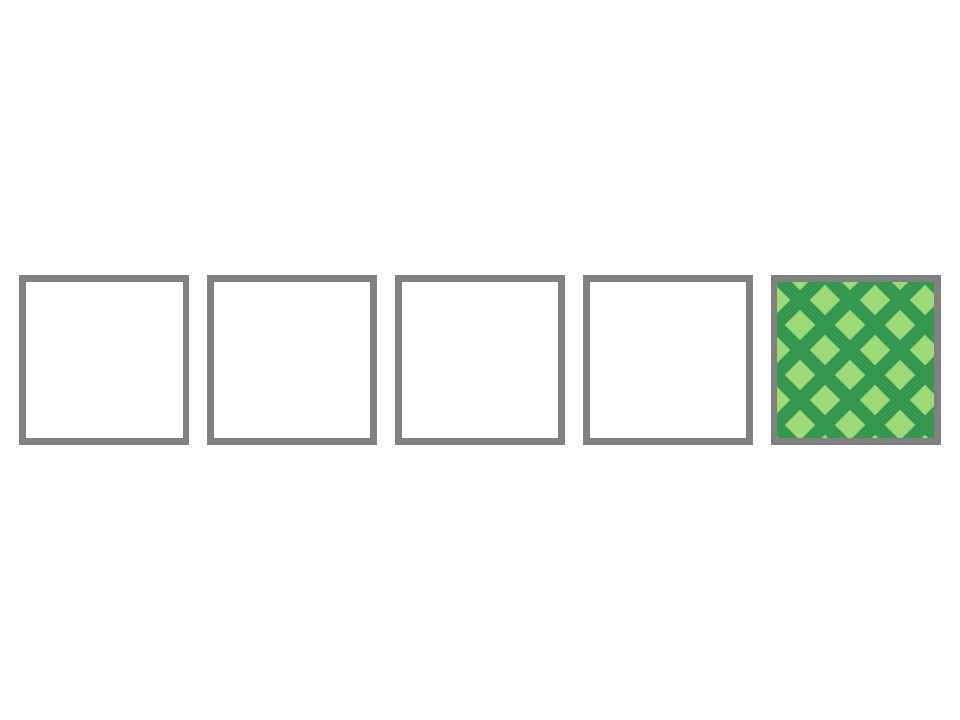} &  \provideranks{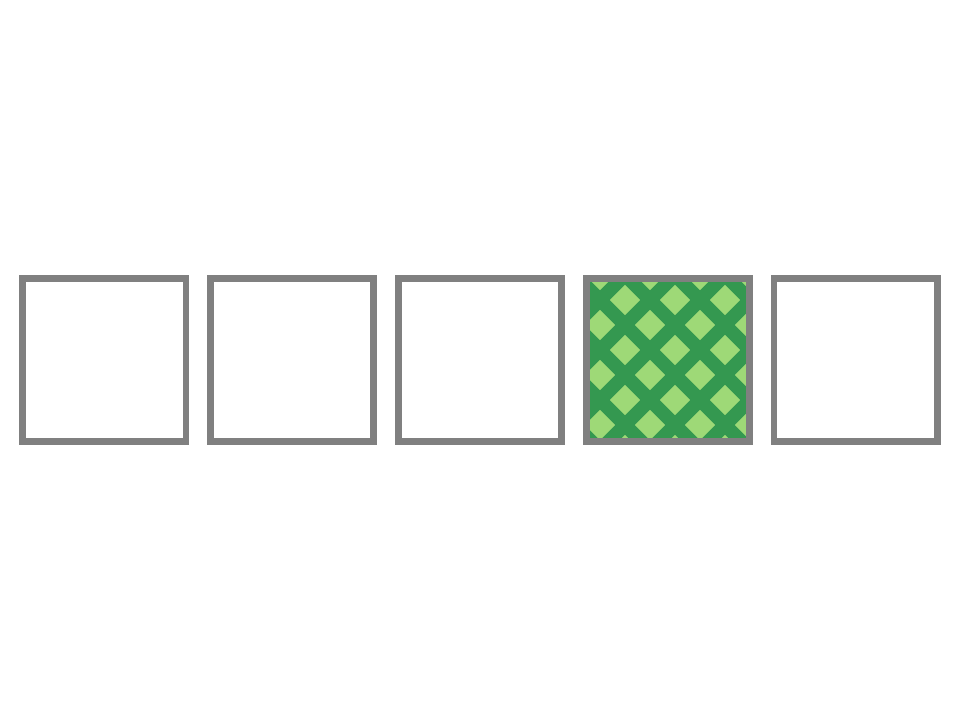} & \phantom{-}1.00 \\
 52 &  \provideranks{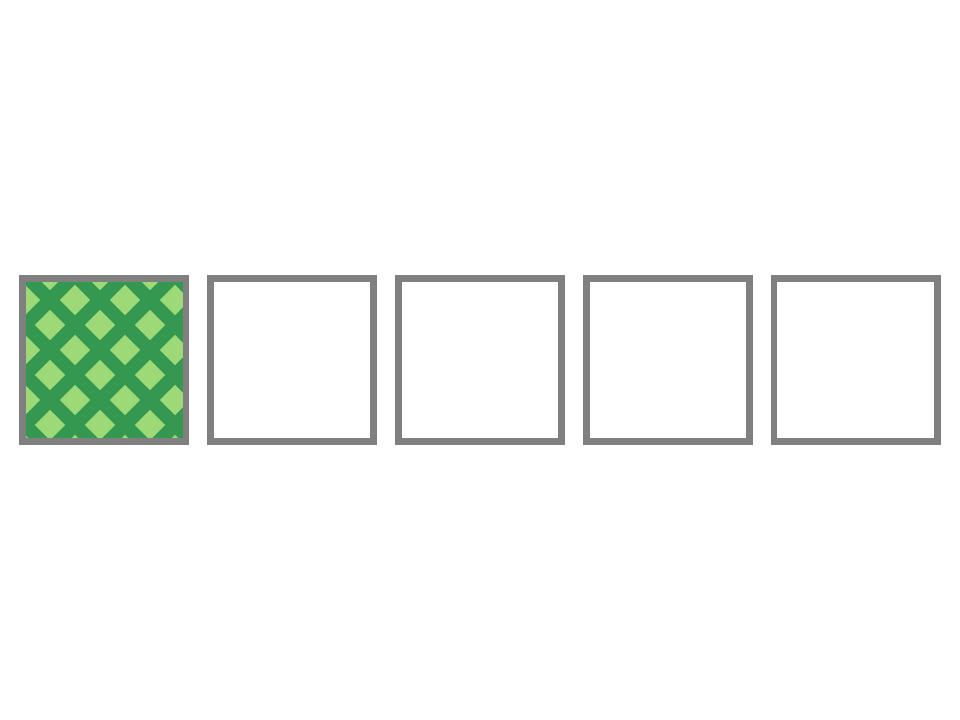} &  \provideranks{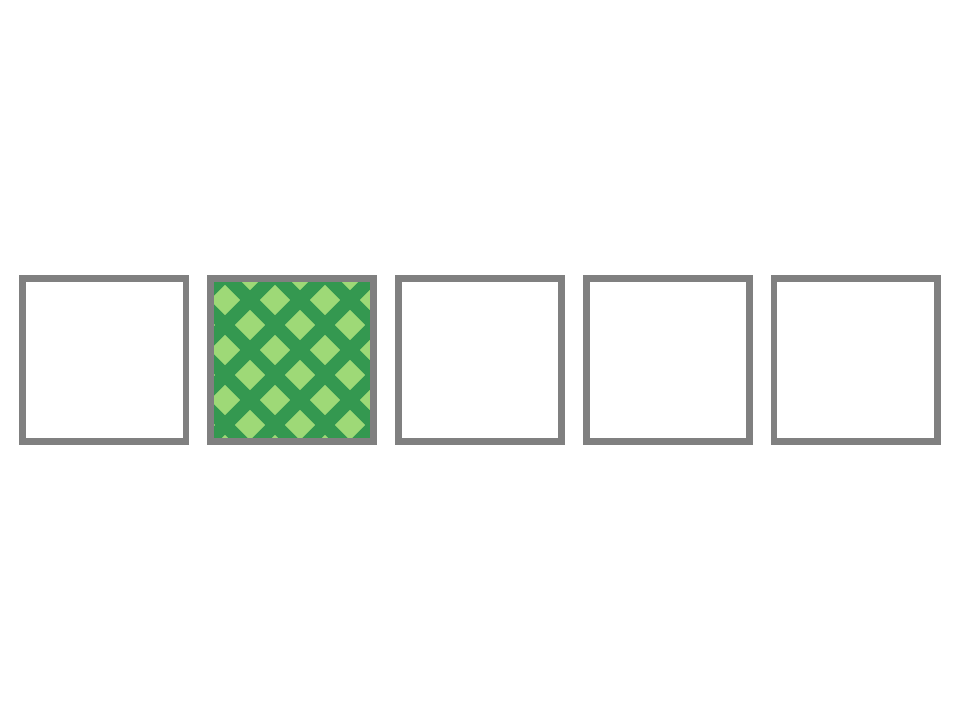} &  \provideranks{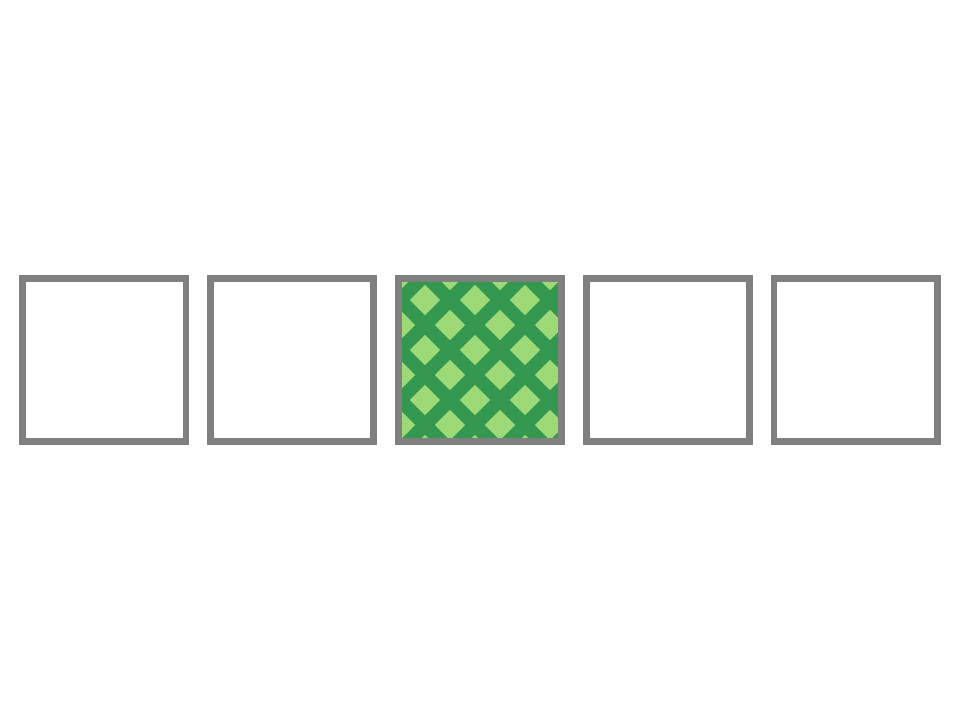} &  \provideranks{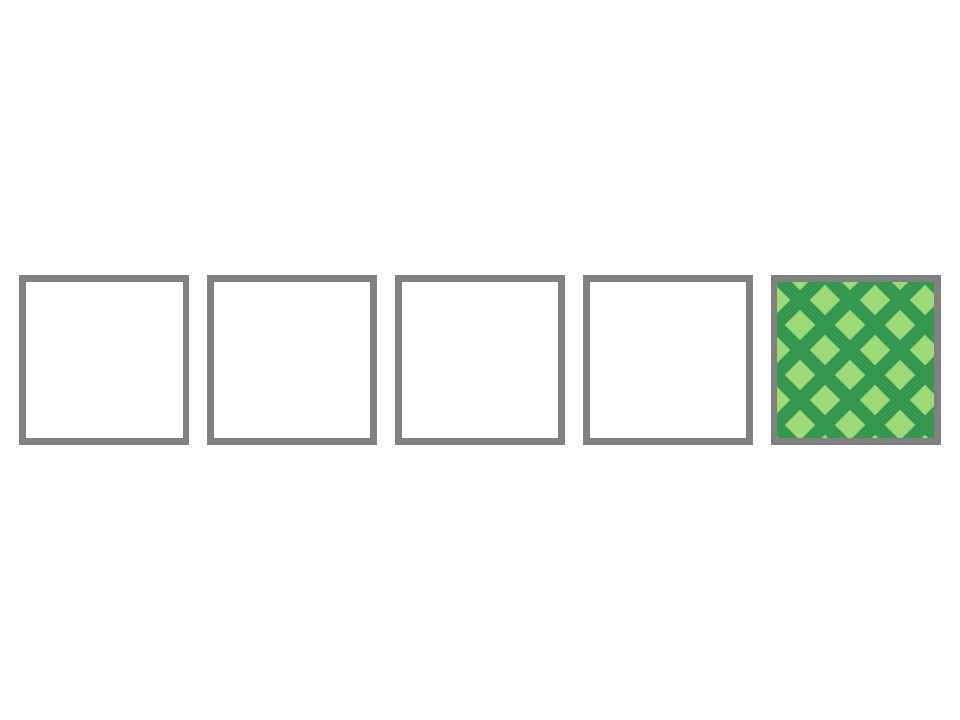} &  \provideranks{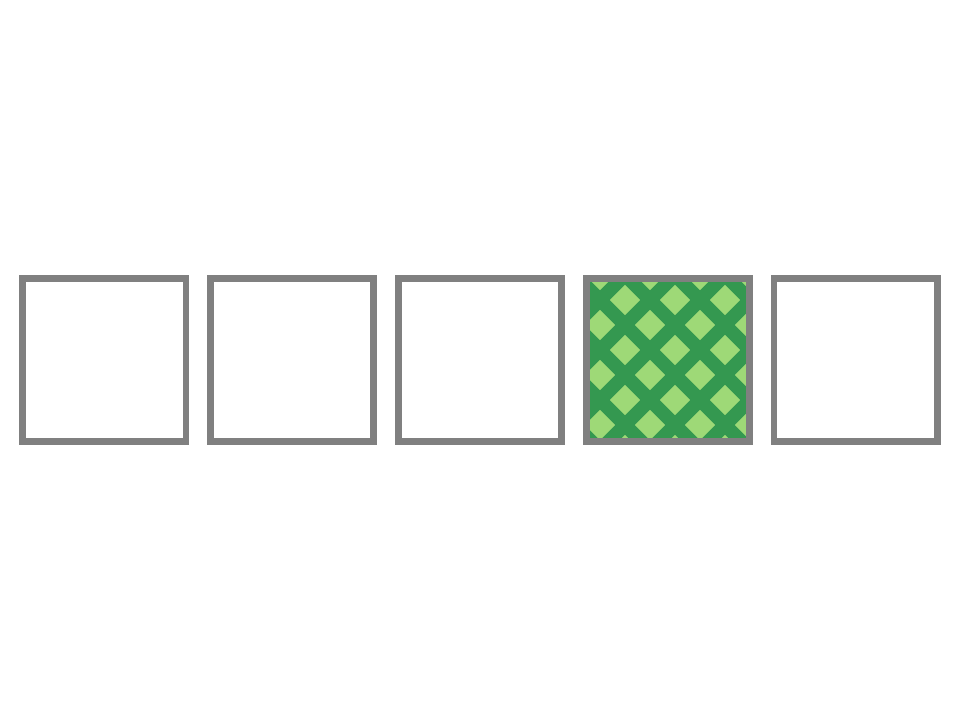} & \phantom{-}1.00 \\
 54 &  \provideranks{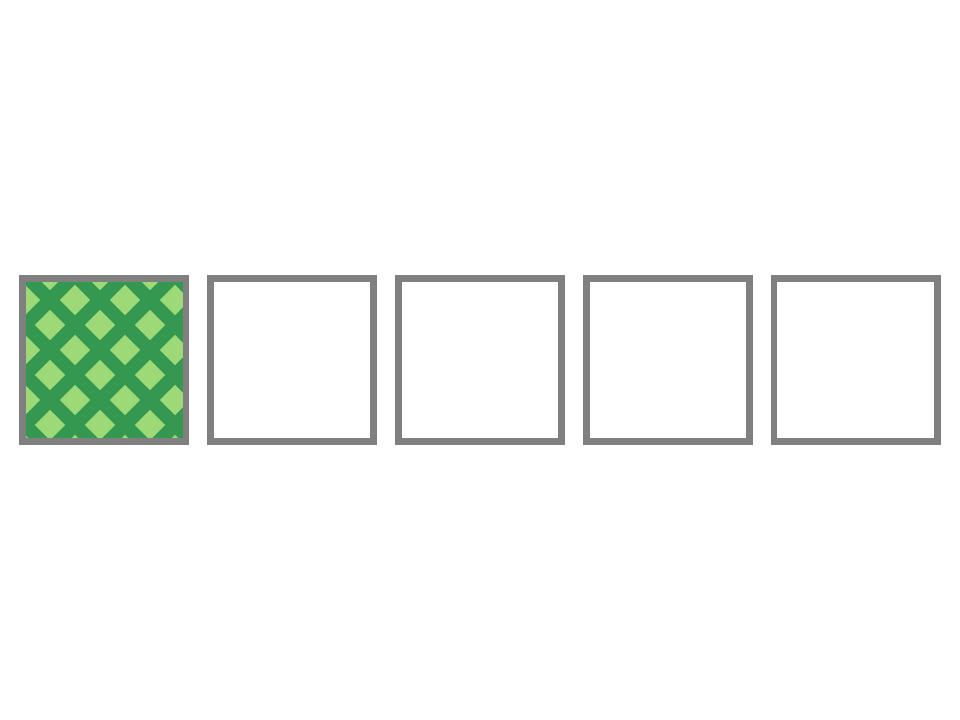} &  \provideranks{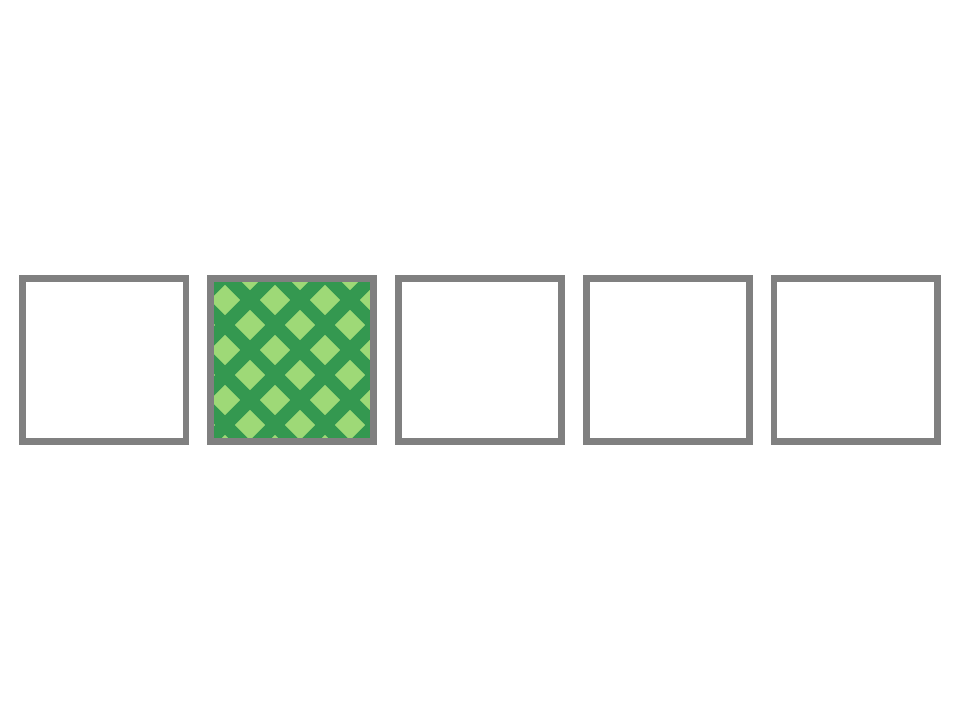} &  \provideranks{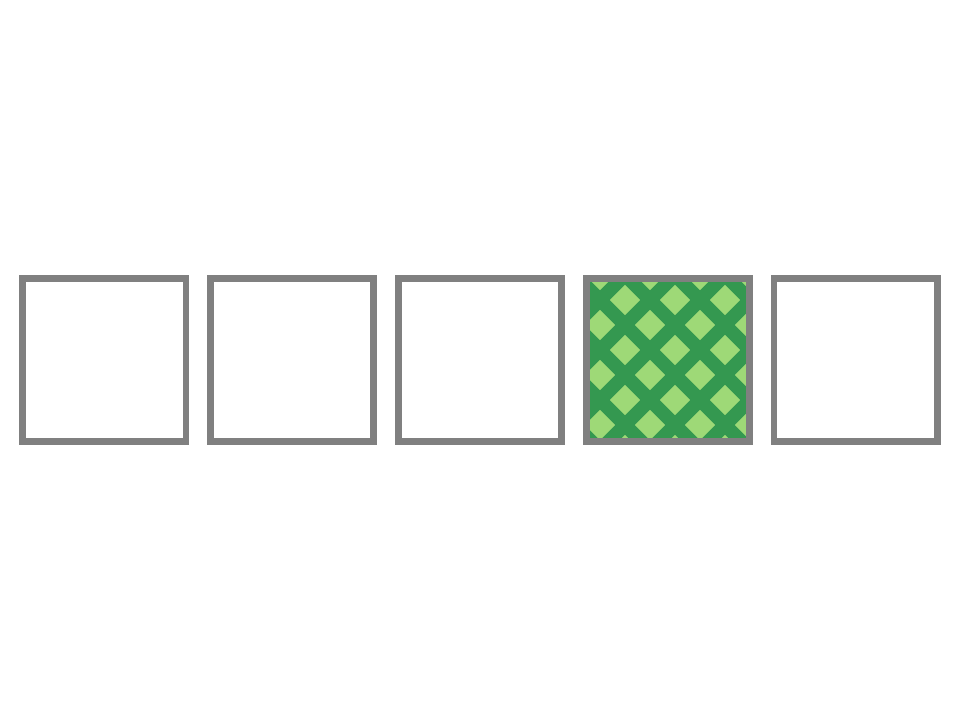} &  \provideranks{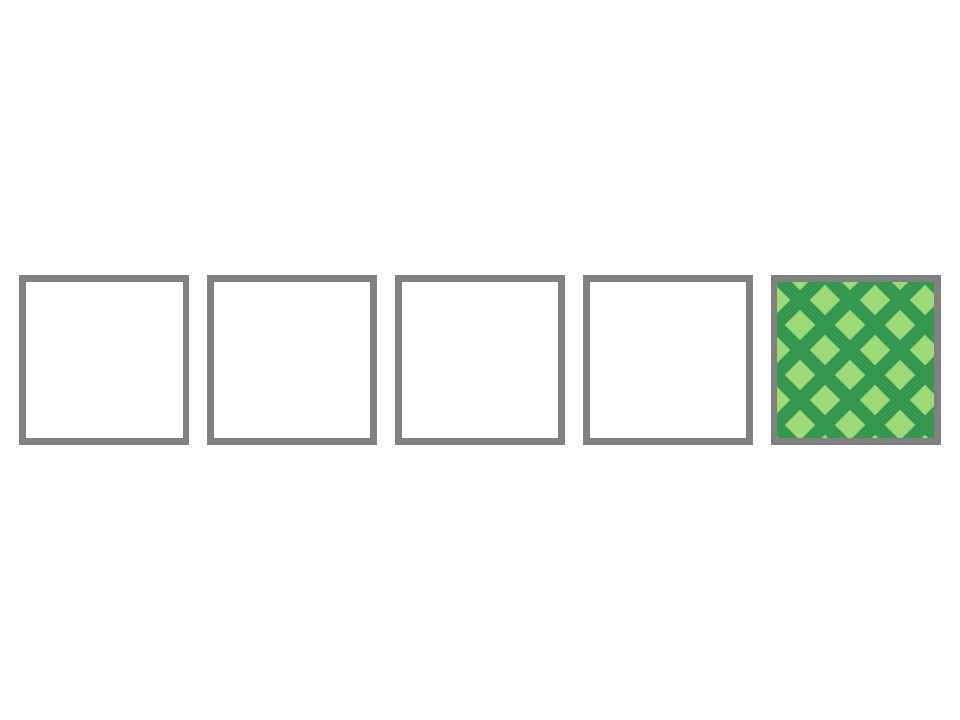} &  \provideranks{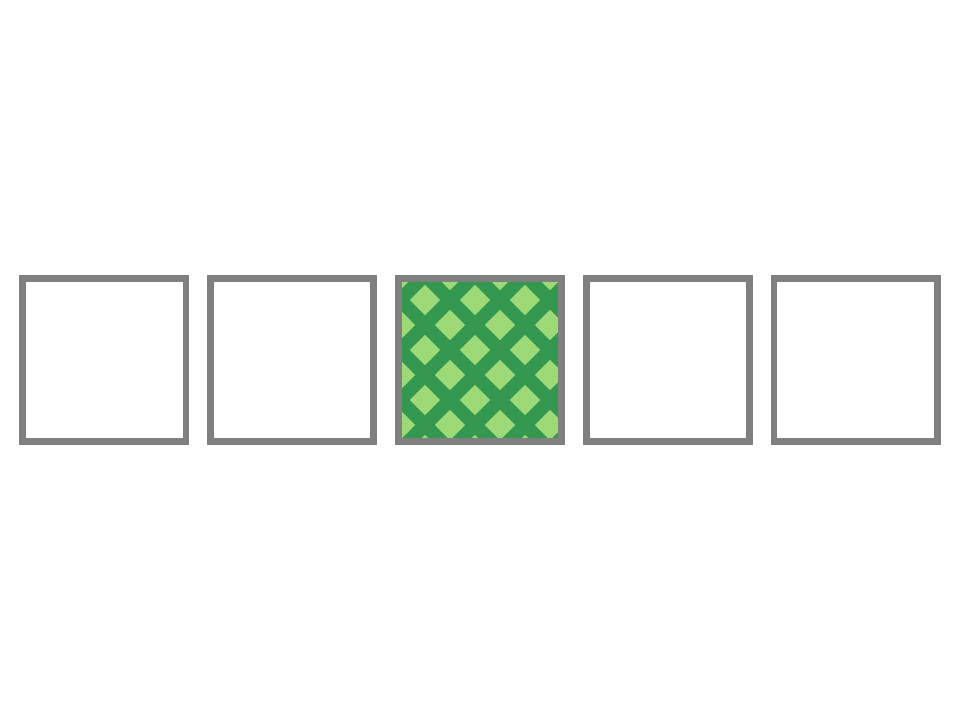} & \phantom{-}1.00 \\
 55 &  \provideranks{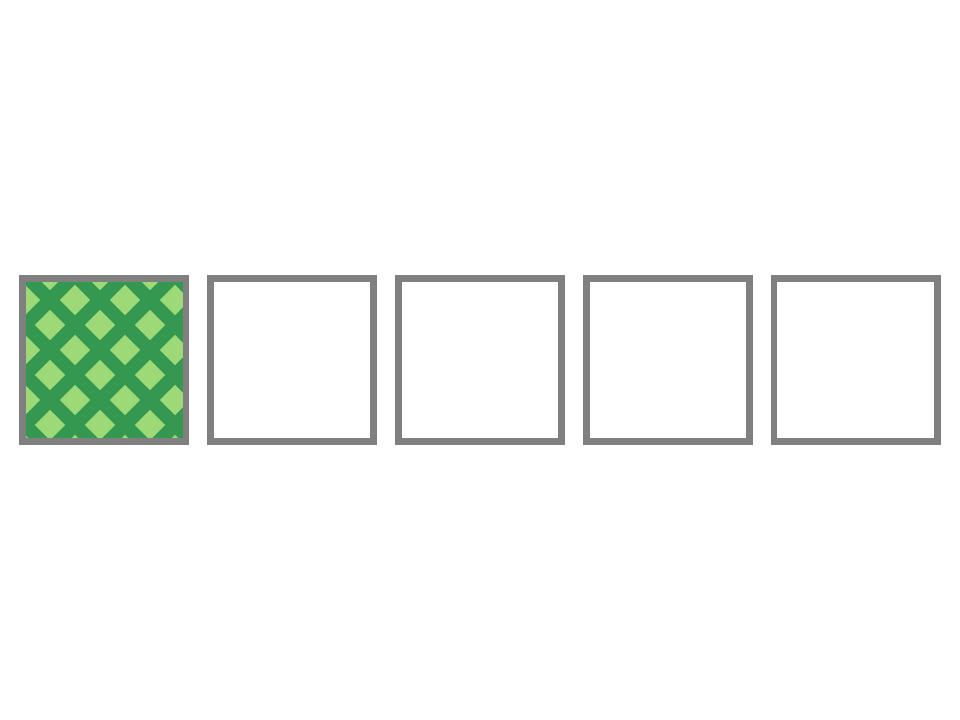} &  \provideranks{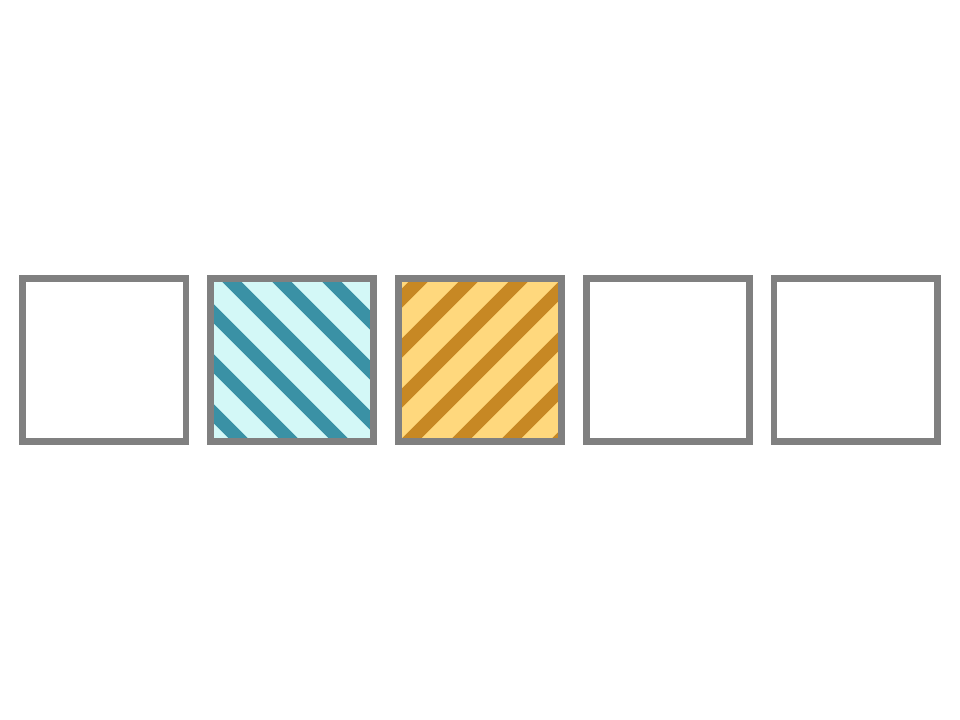} &  \provideranks{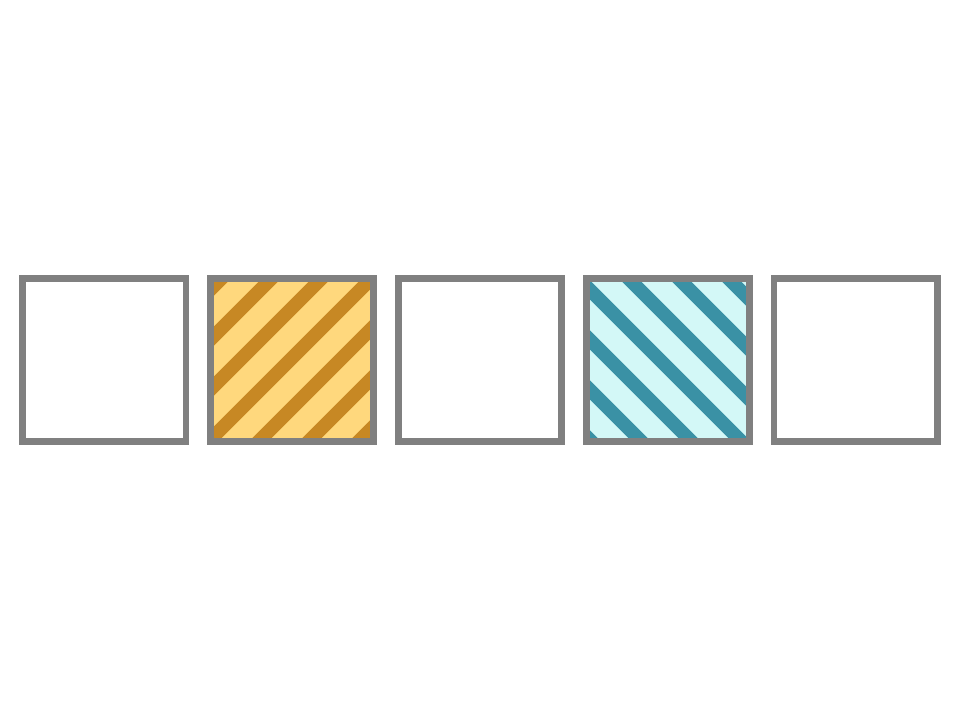} &  \provideranks{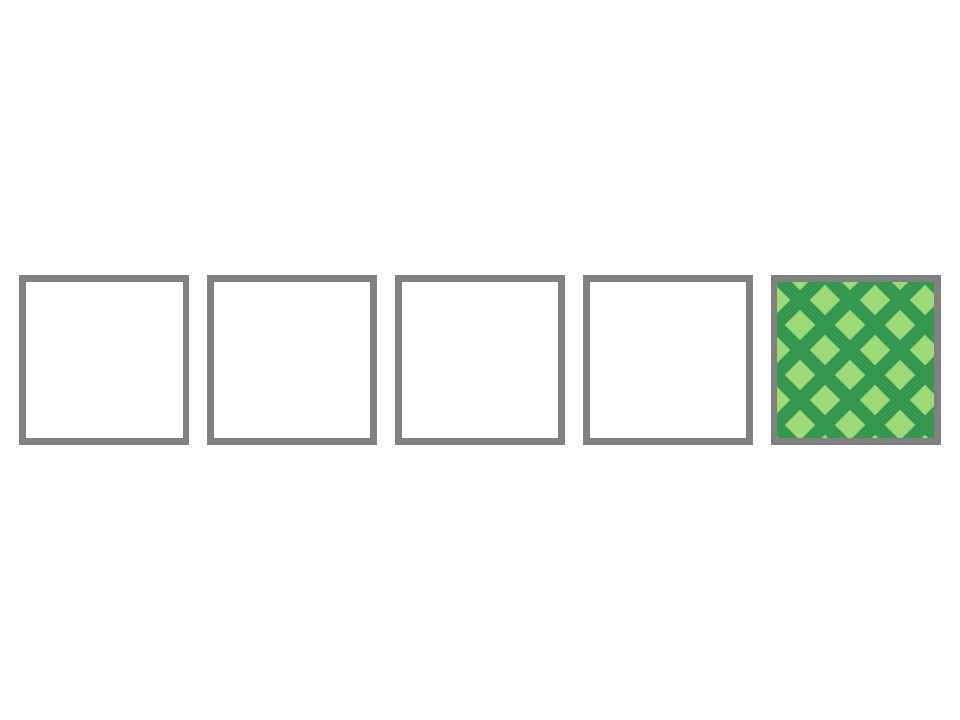} &  \provideranks{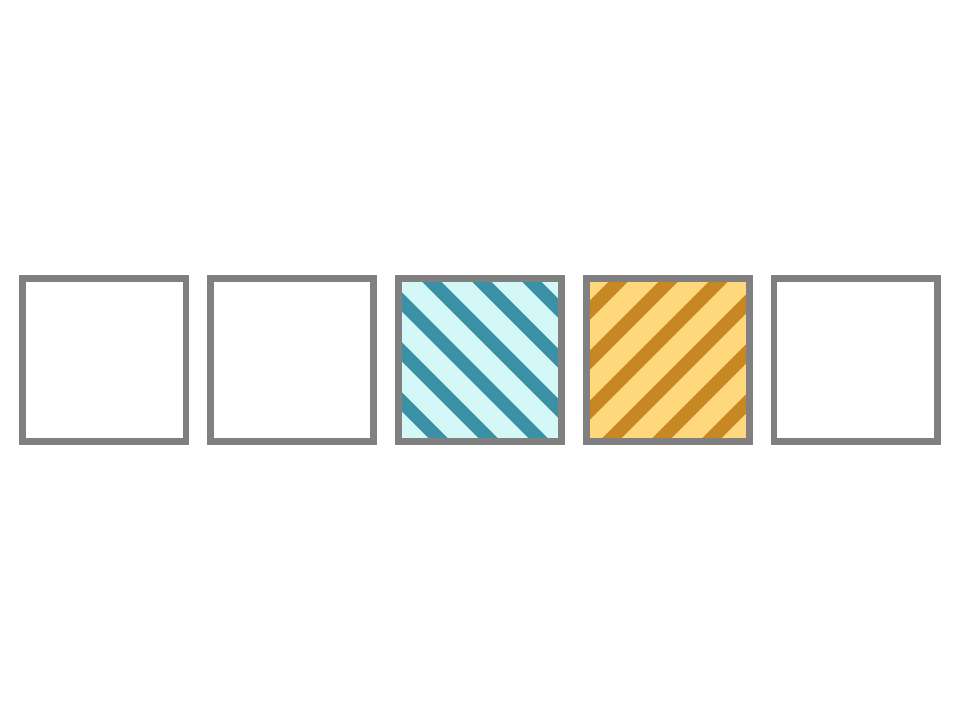} & \phantom{-}0.60 \\
 57 &  \provideranks{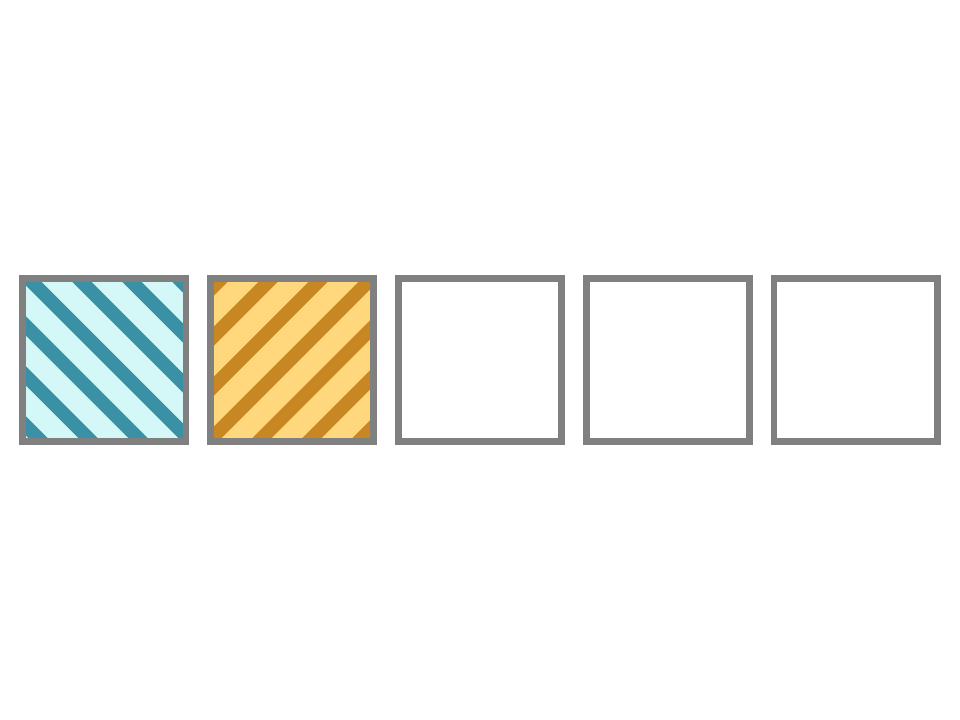} &  \provideranks{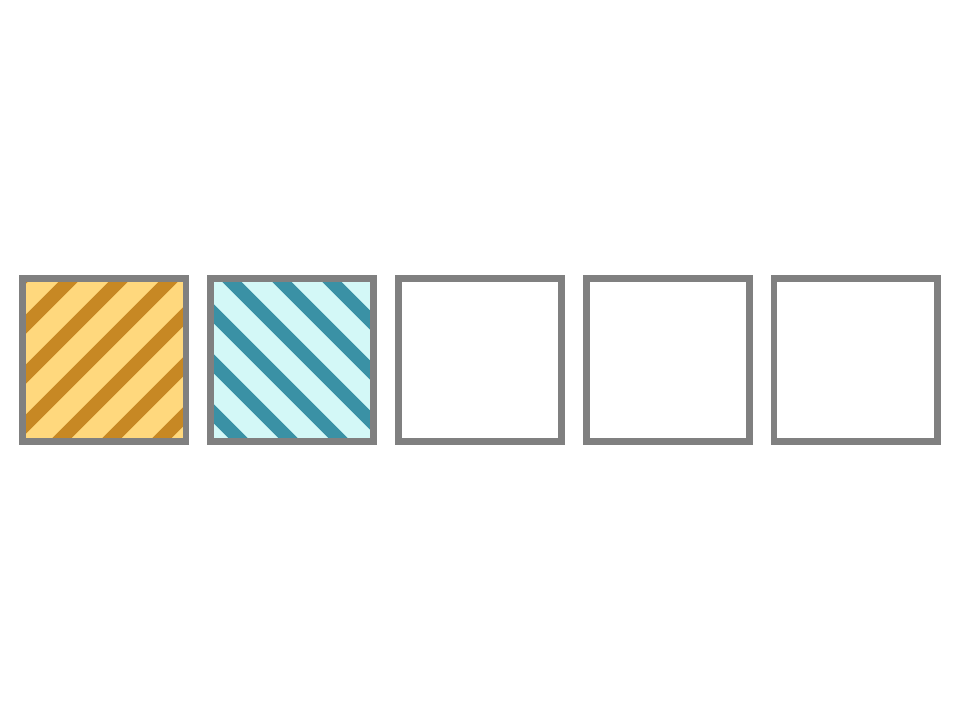} &  \provideranks{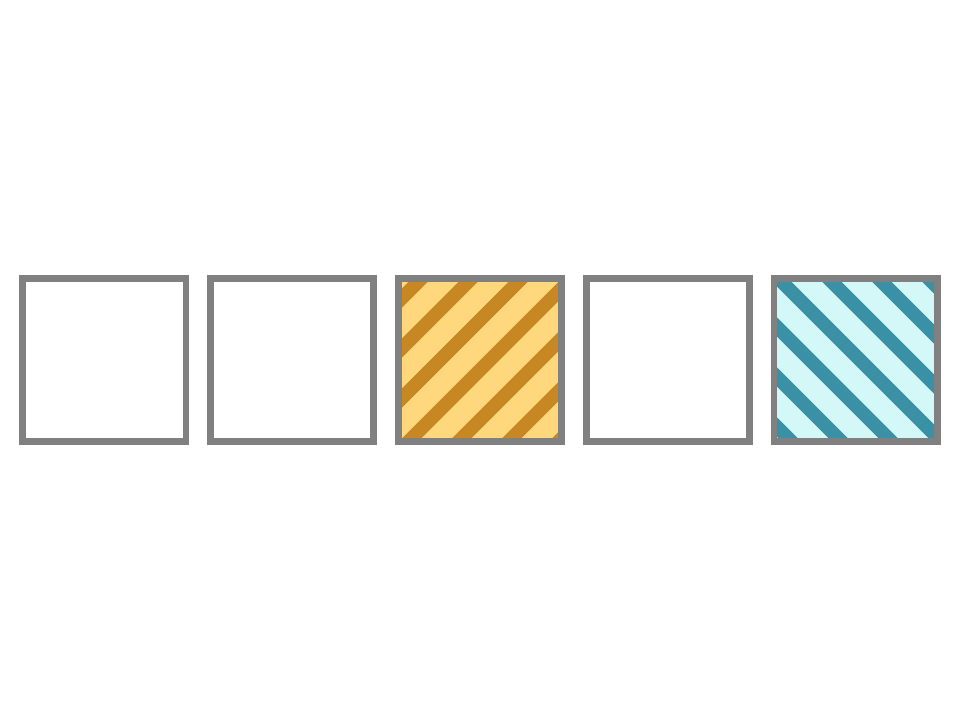} &  \provideranks{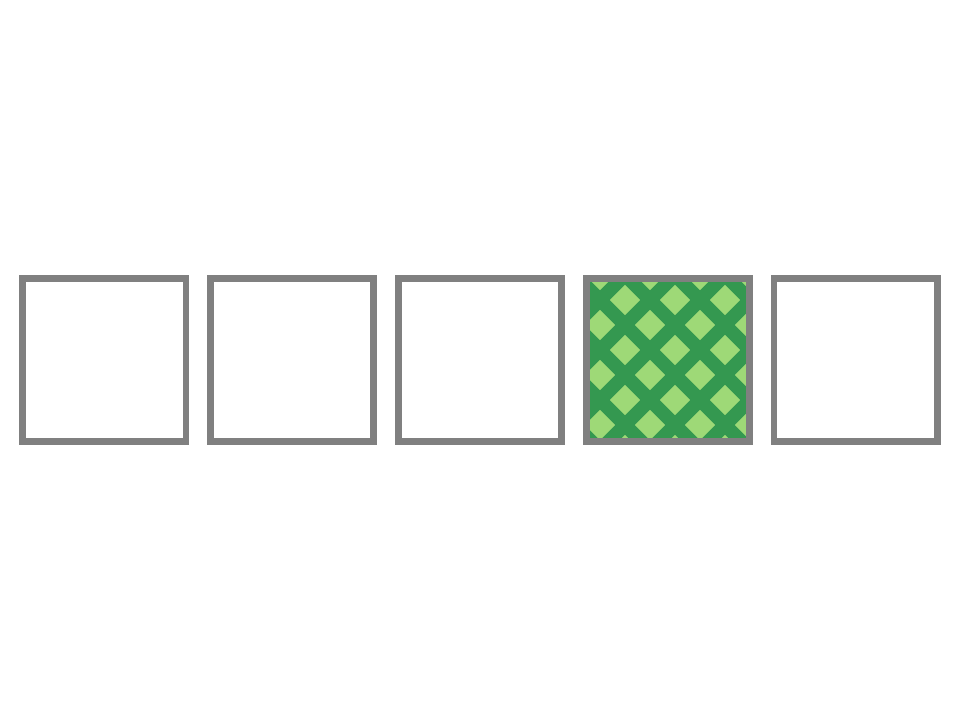} &  \provideranks{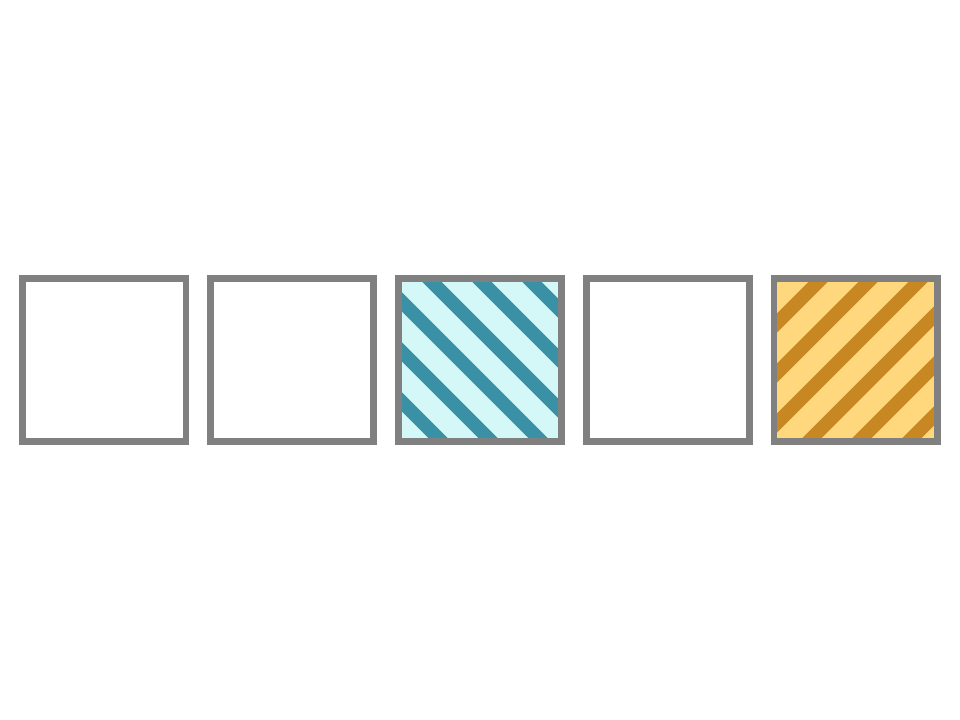} & \phantom{-}0.20 \\
 68 &  \provideranks{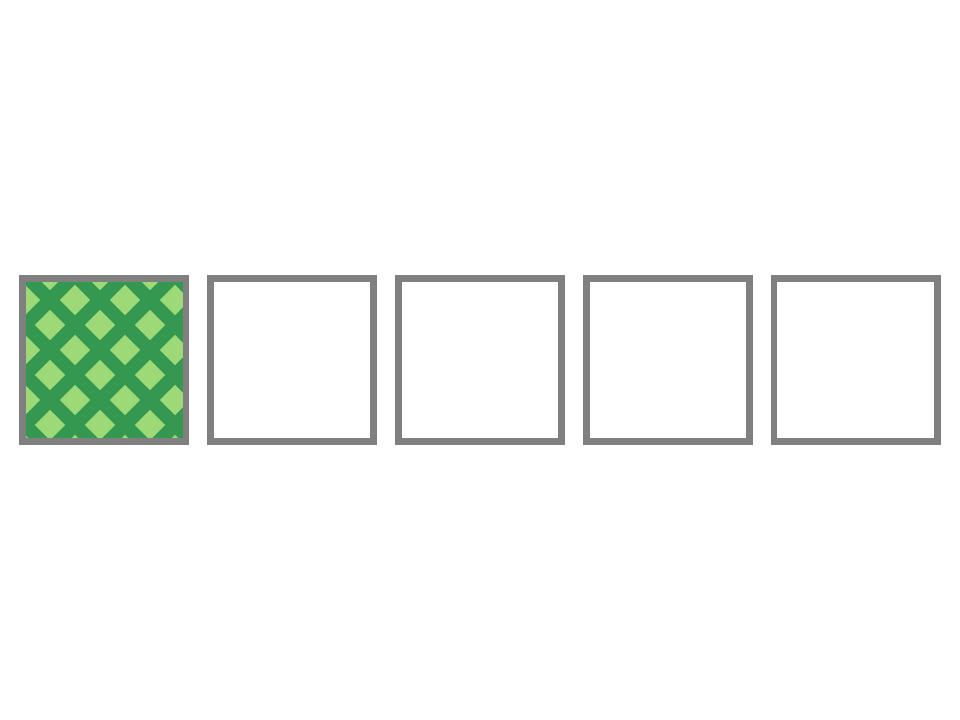} &  \provideranks{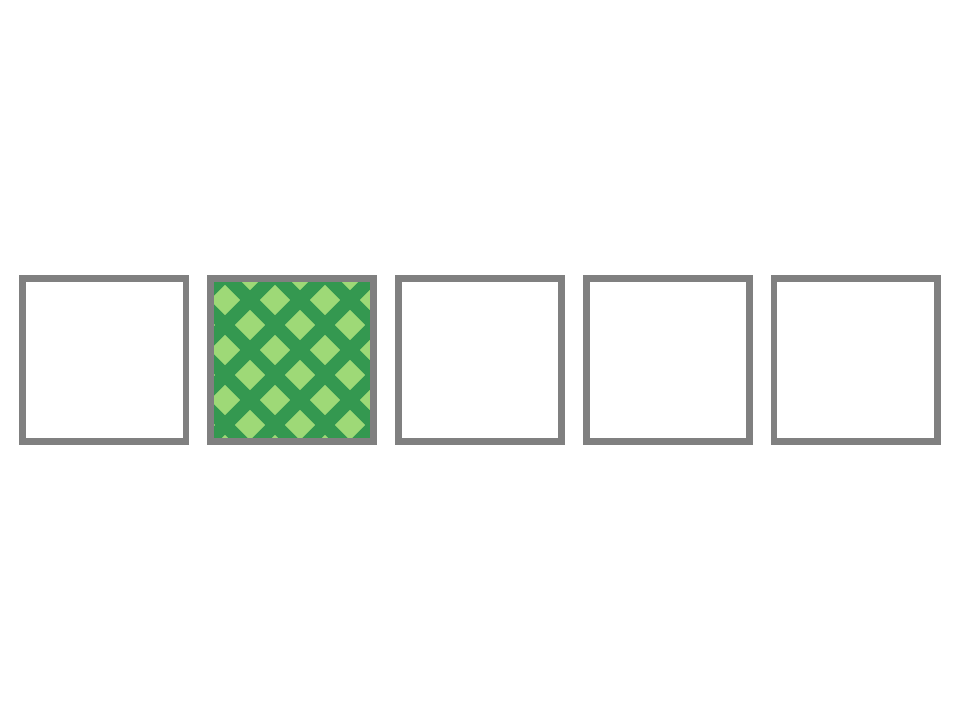} &  \provideranks{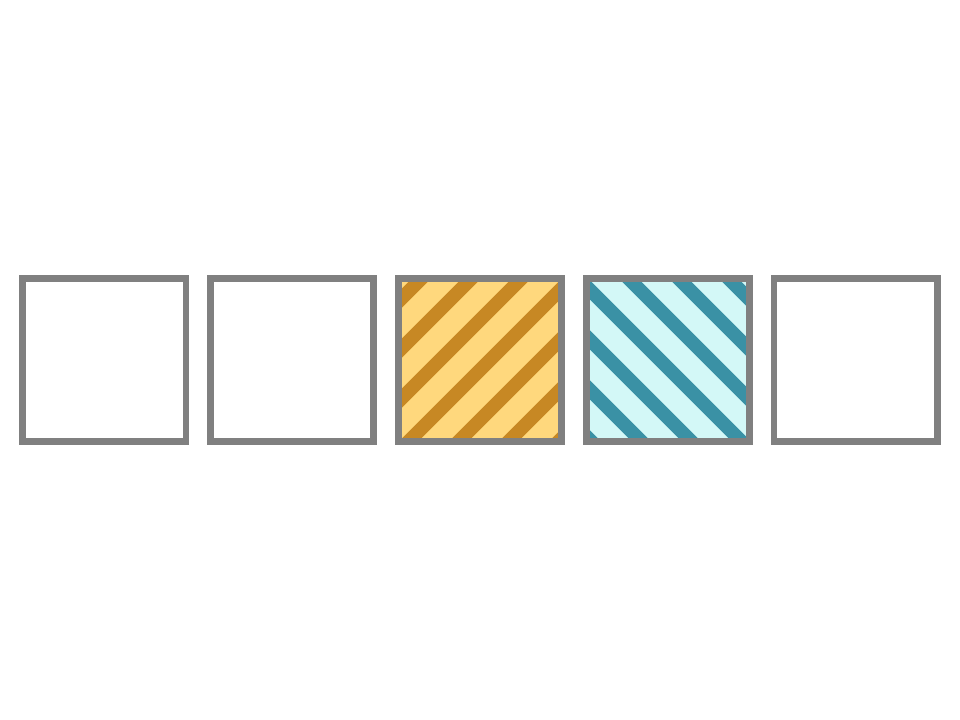} &  \provideranks{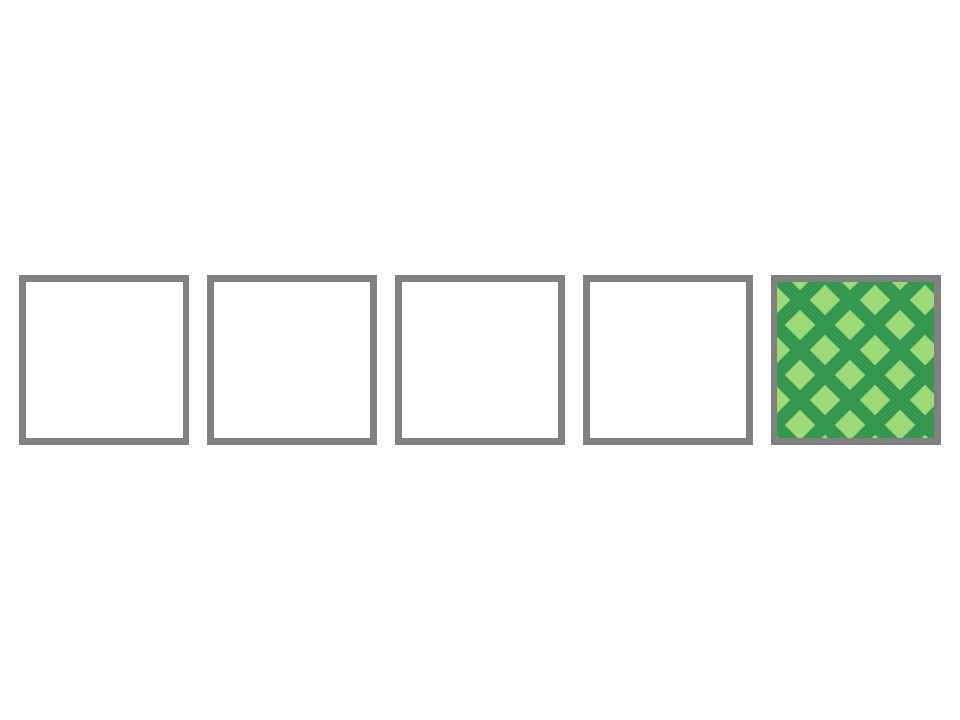} &  \provideranks{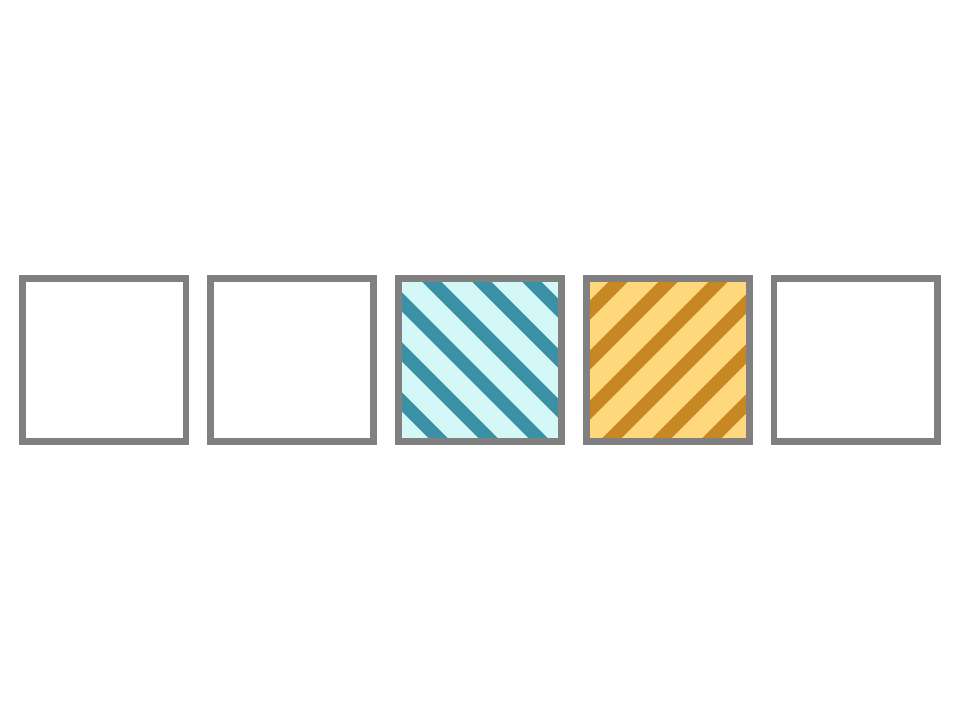} & \phantom{-}0.80 \\
 81 &  \provideranks{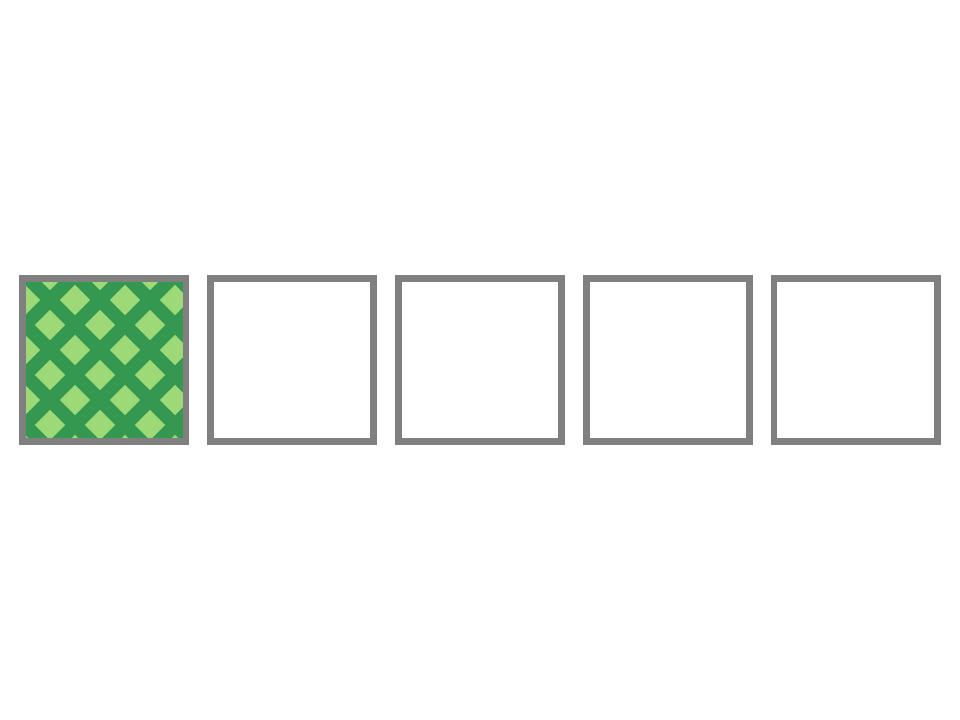} &  \provideranks{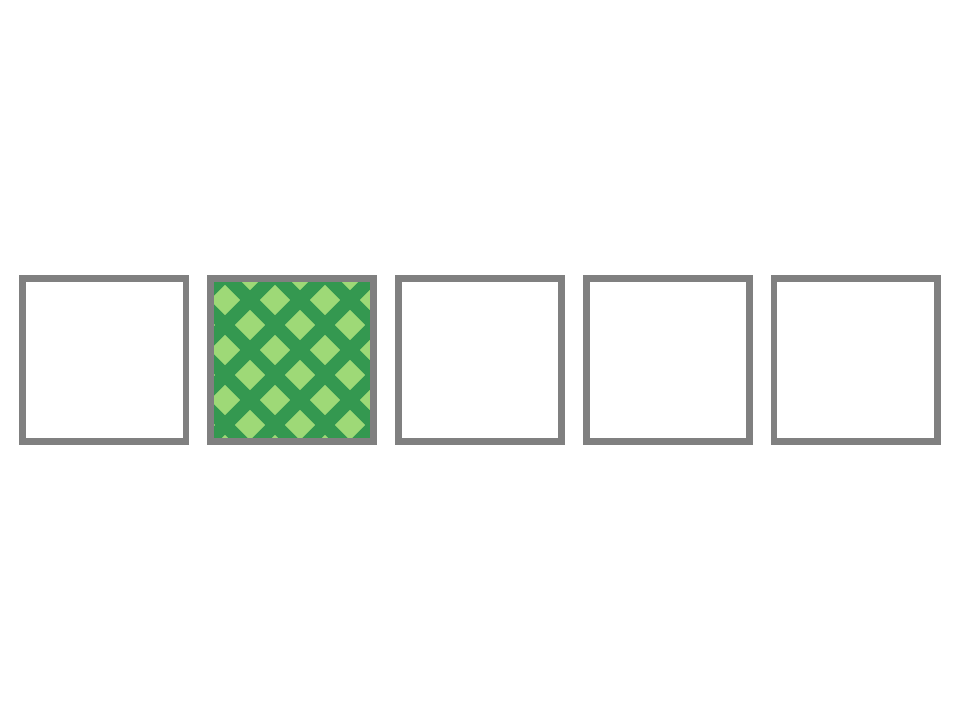} &  \provideranks{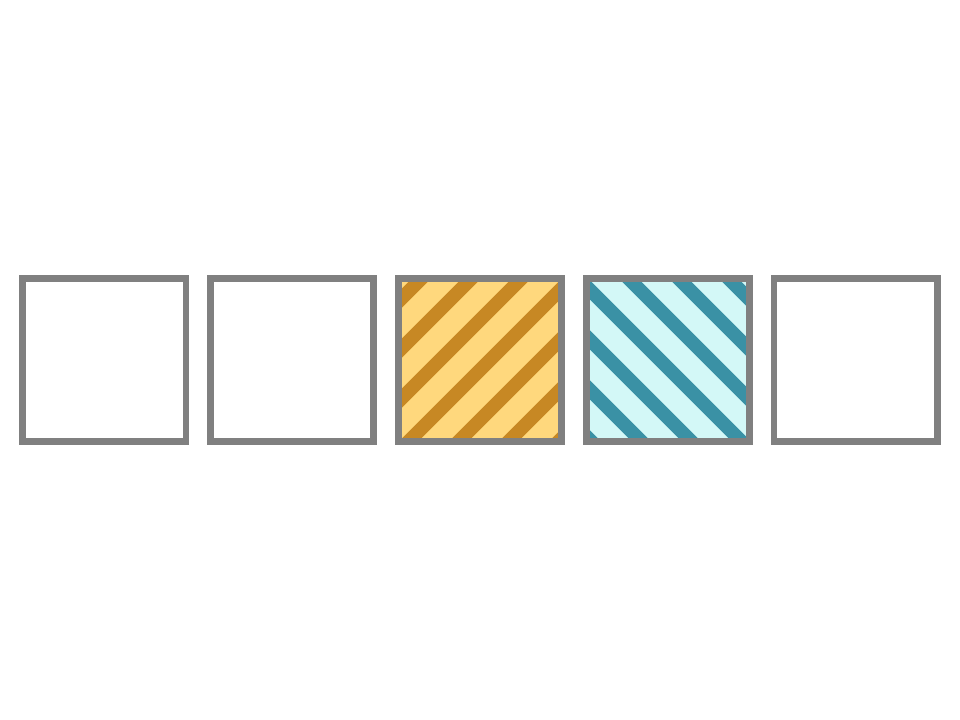} &  \provideranks{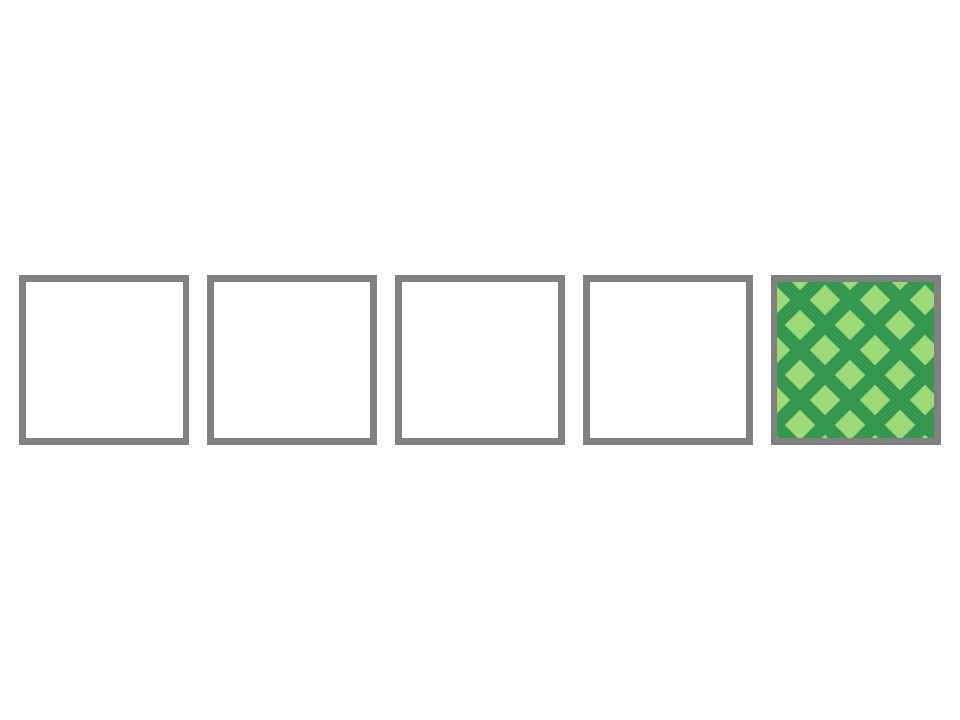} &  \provideranks{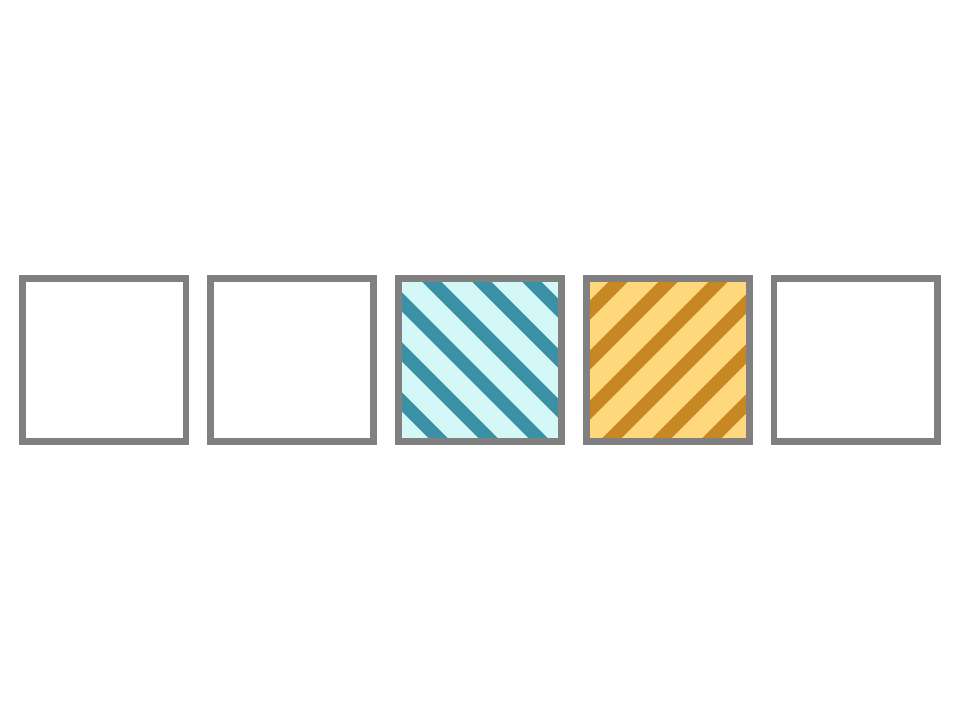} & \phantom{-}0.80 \\
 83 &  \provideranks{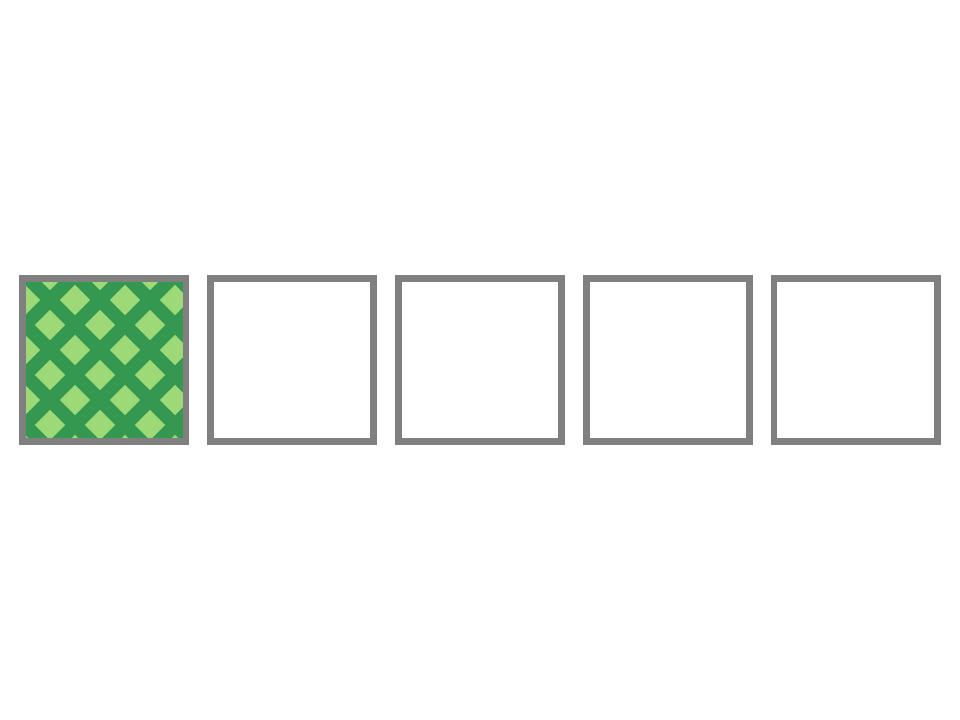} &  \provideranks{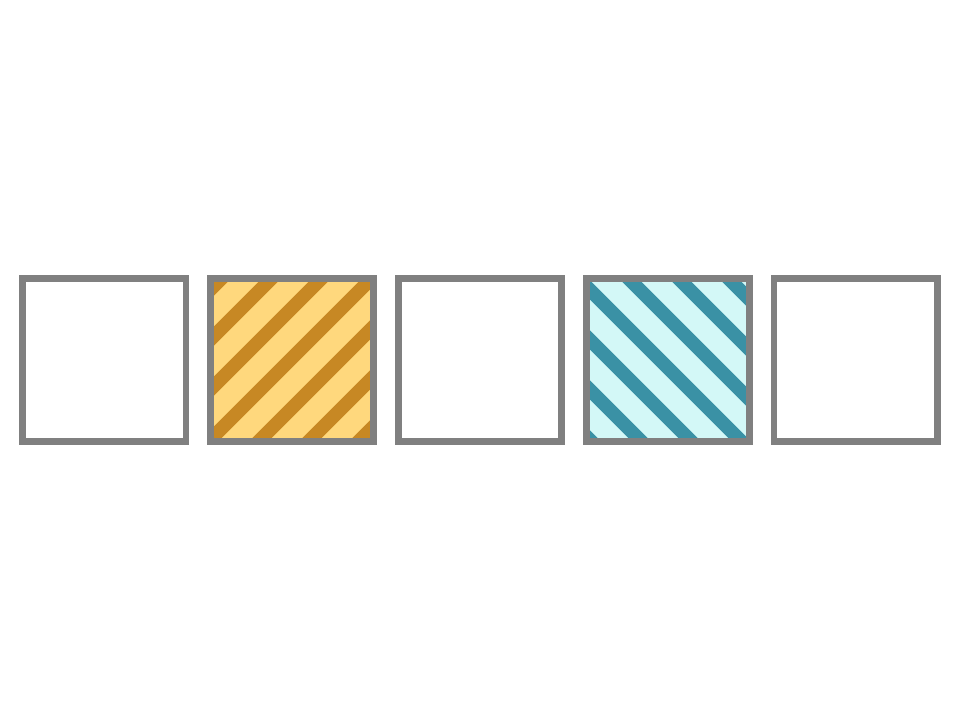} &  \provideranks{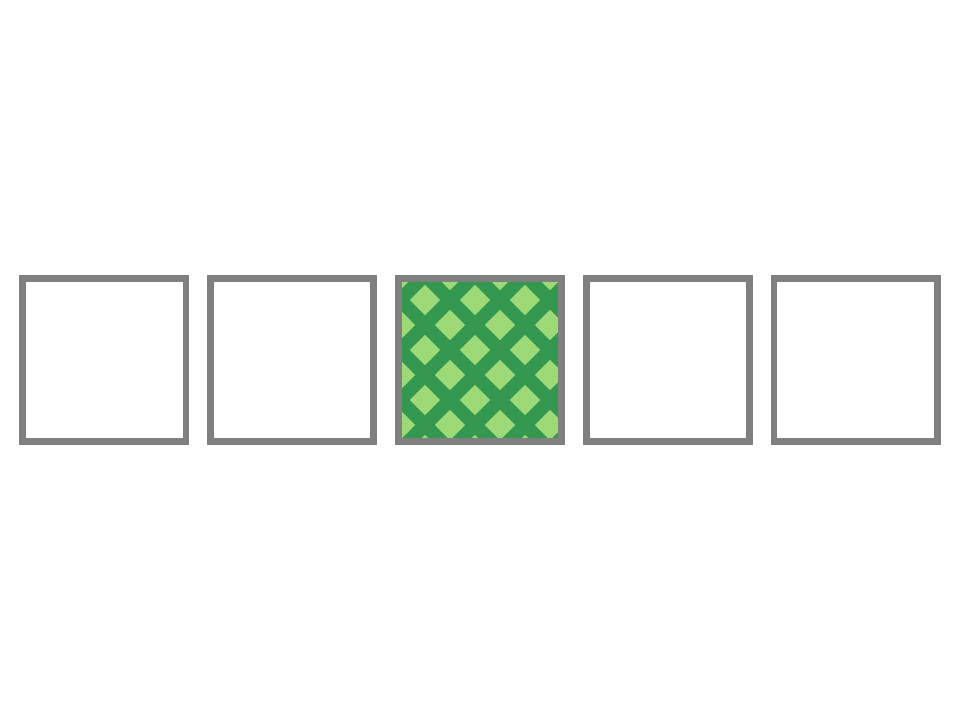} &  \provideranks{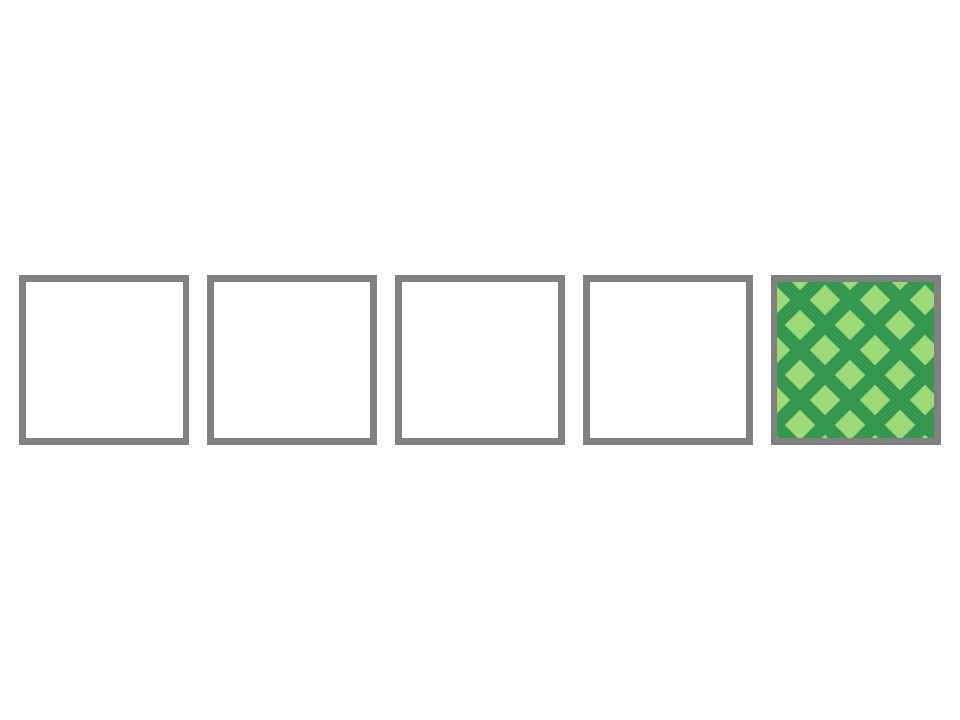} &  \provideranks{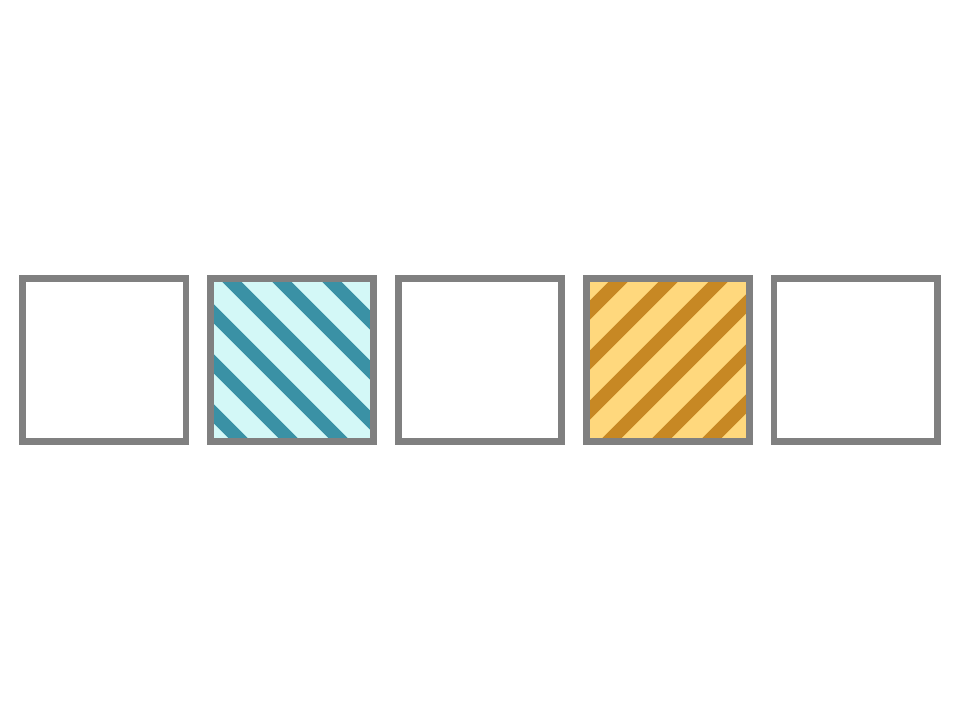} & \phantom{-}0.40 \\
 85 &  \provideranks{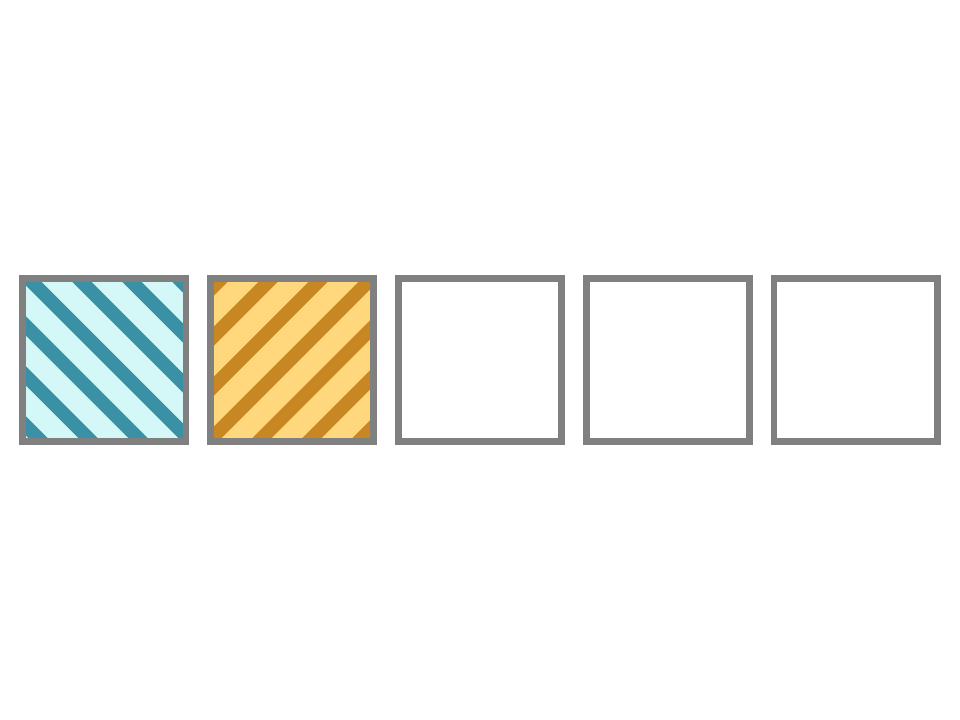} &  \provideranks{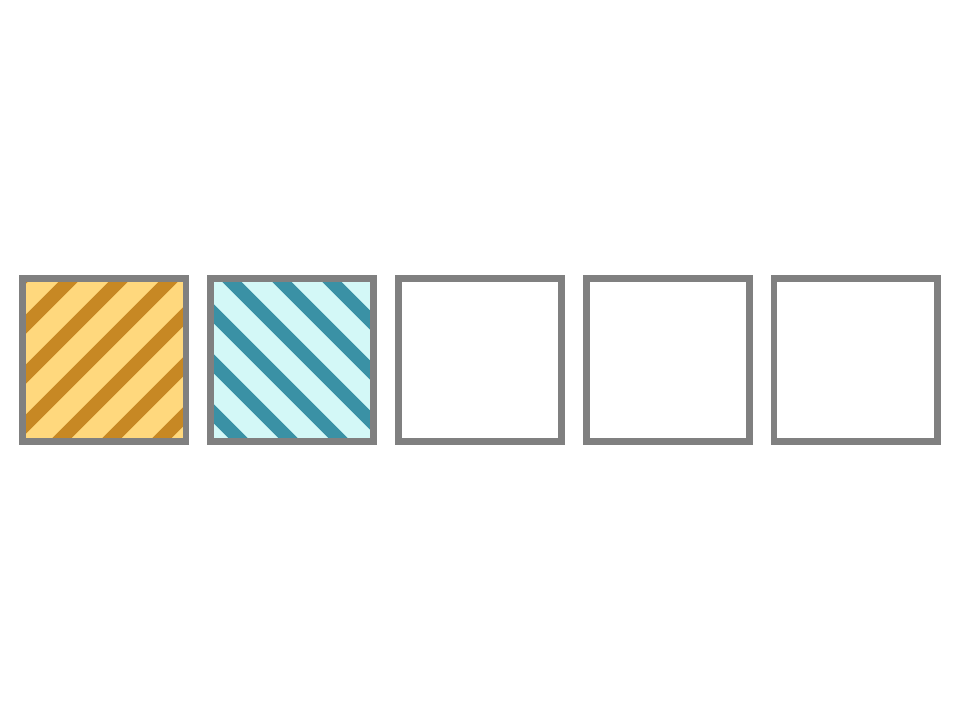} &  \provideranks{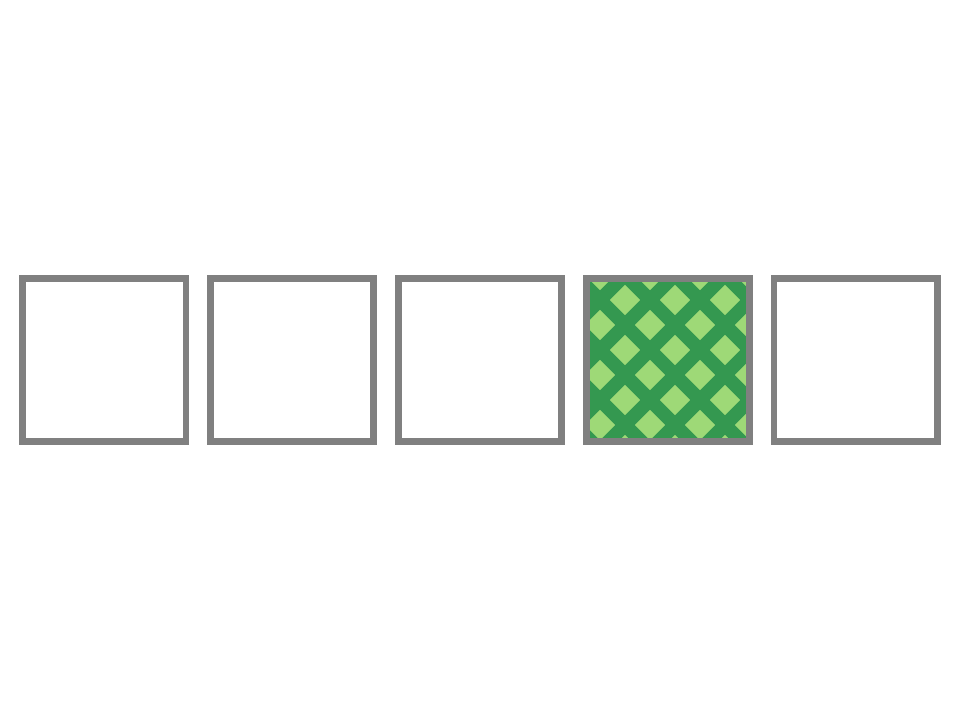} &  \provideranks{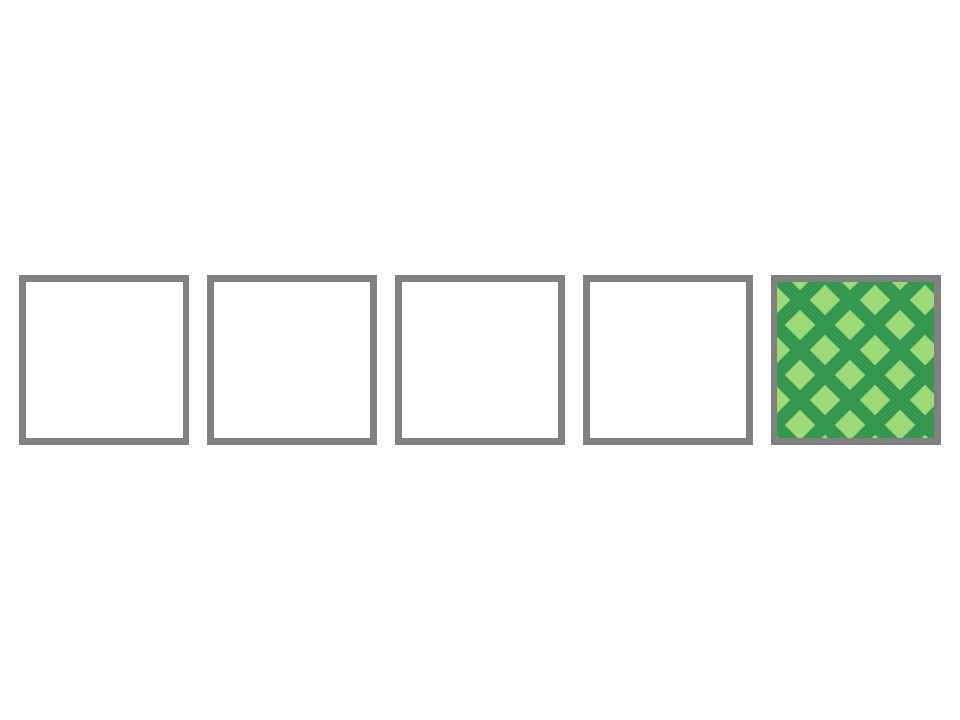} &  \provideranks{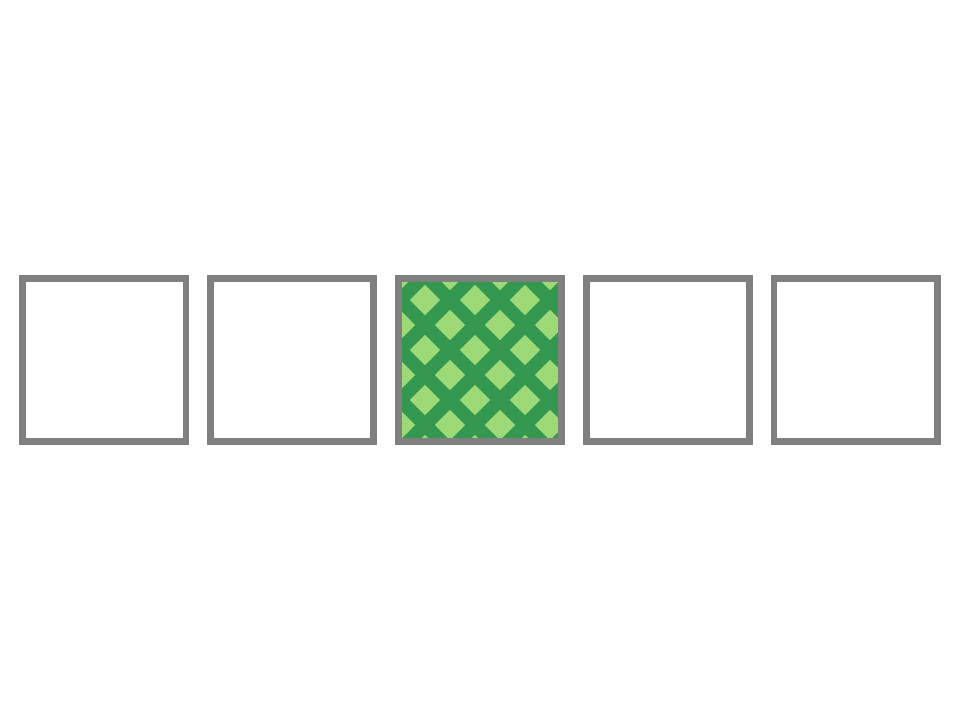} & \phantom{-}0.80 \\
 94 &  \provideranks{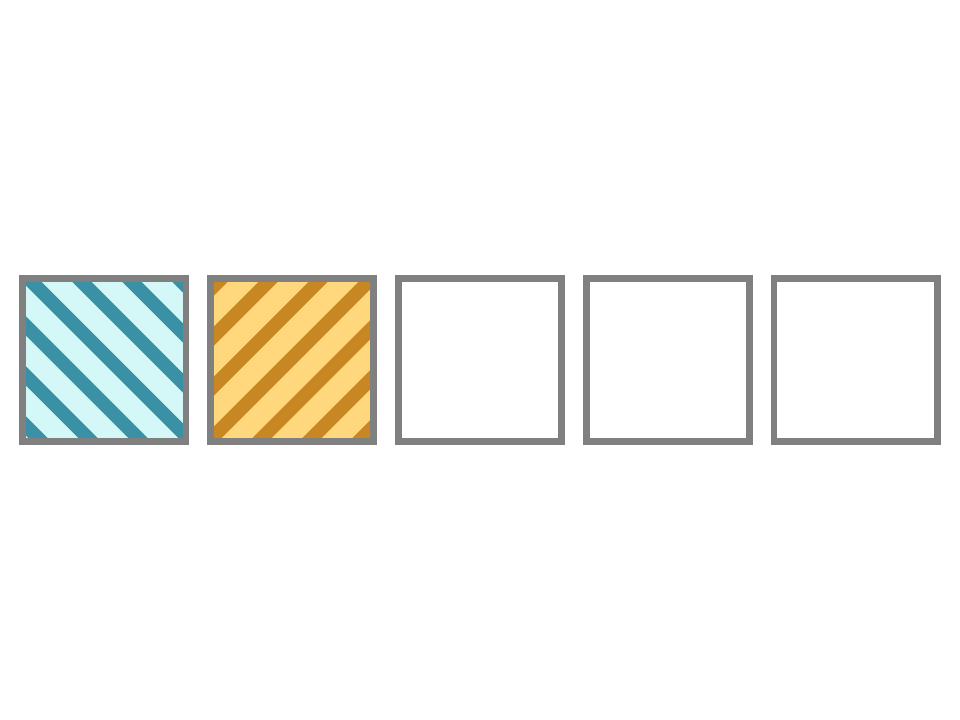} &  \provideranks{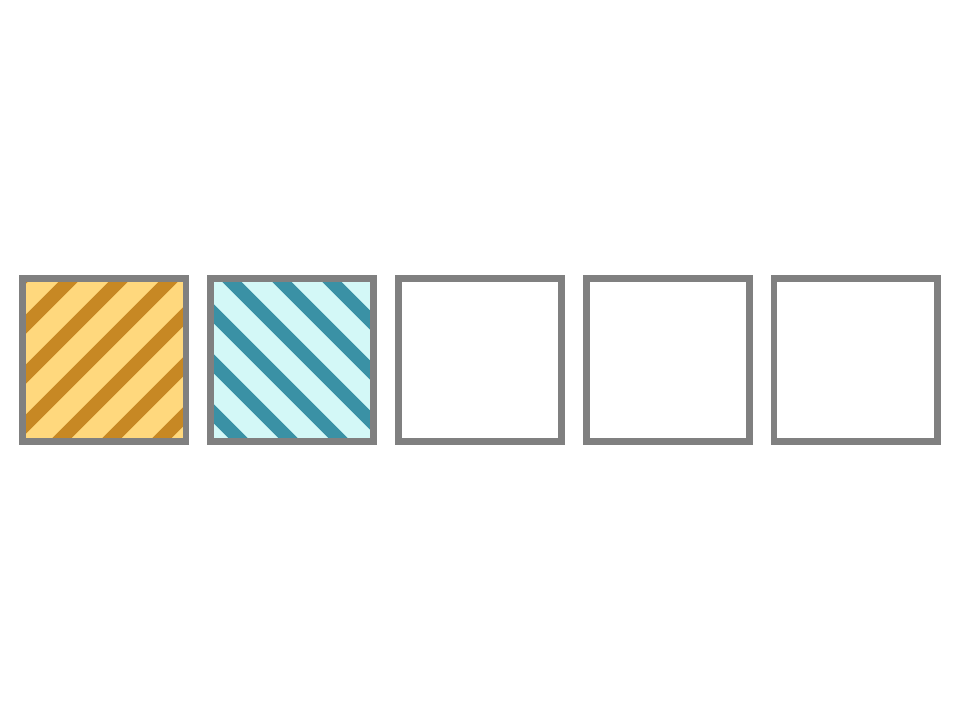} &  \provideranks{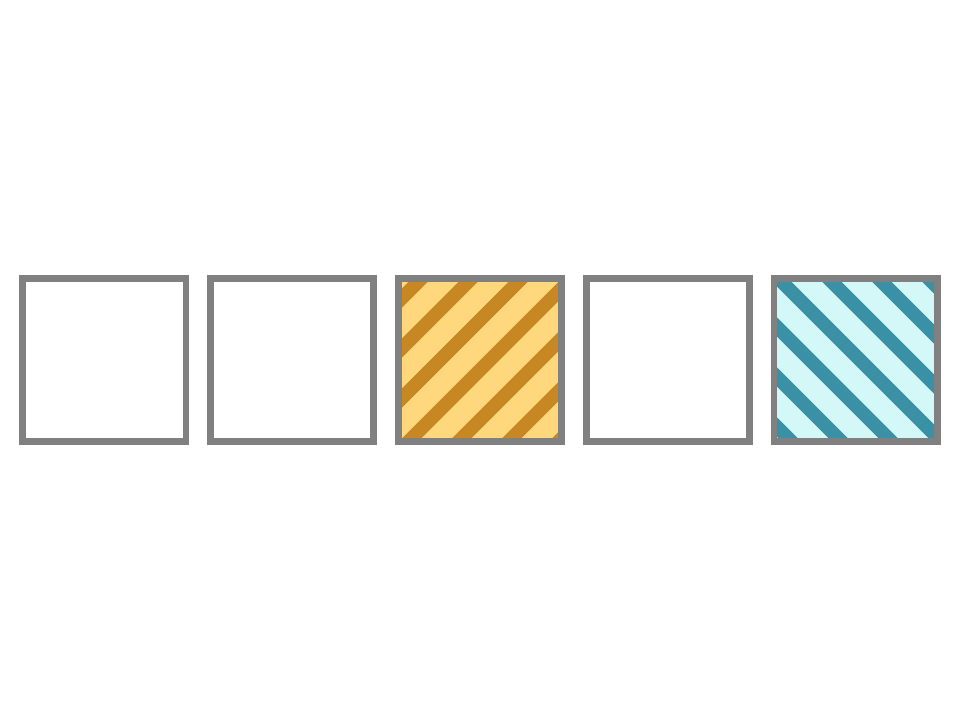} &  \provideranks{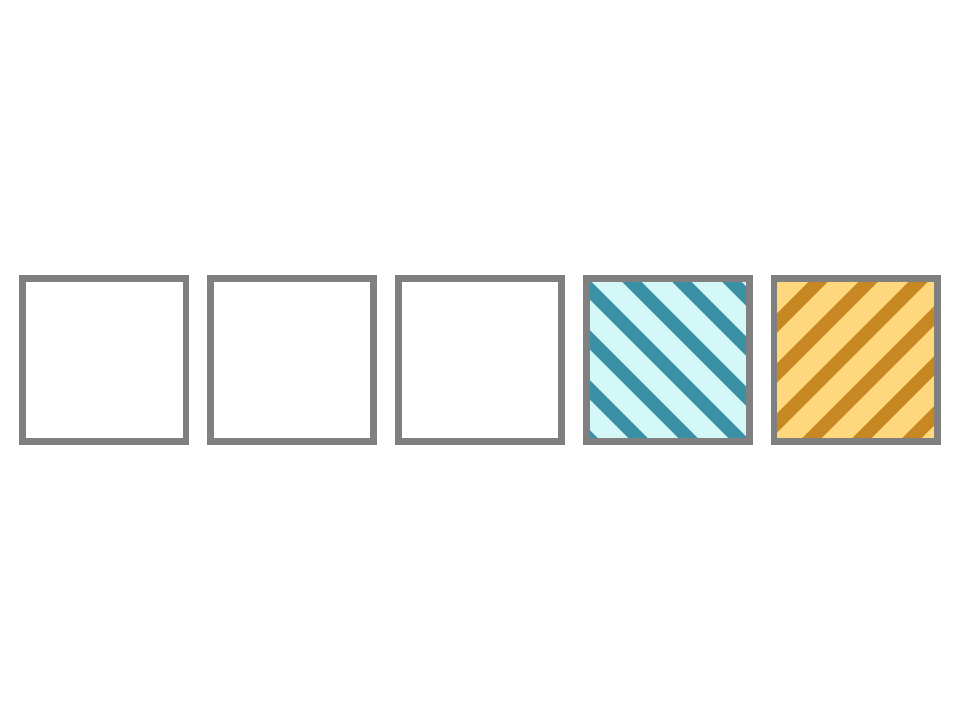} &  \provideranks{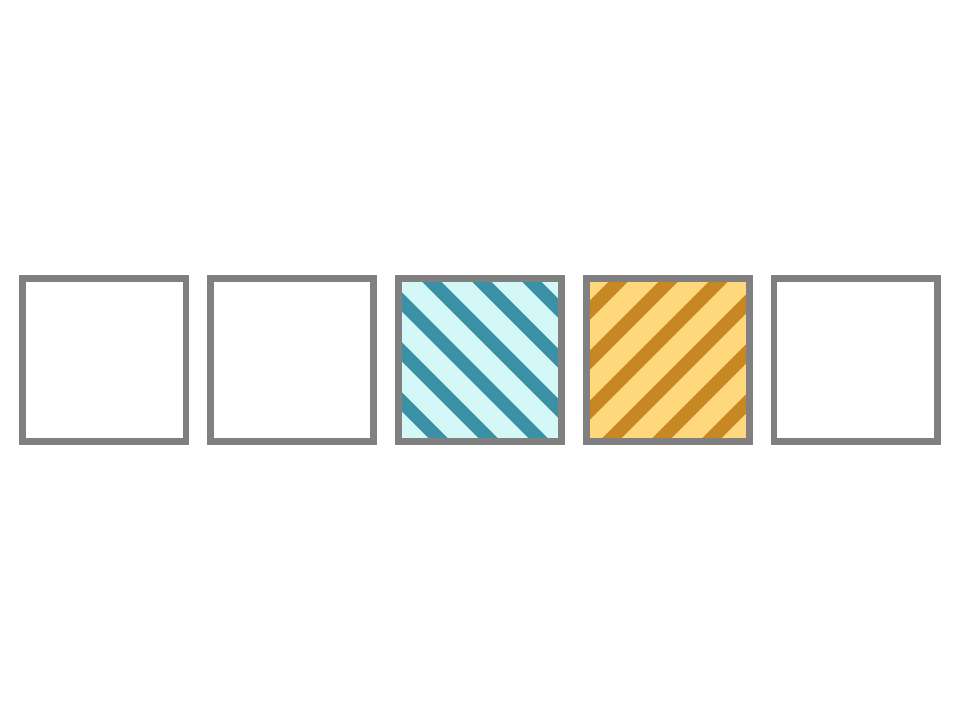} & \phantom{-}0.40 \\
 95 &  \provideranks{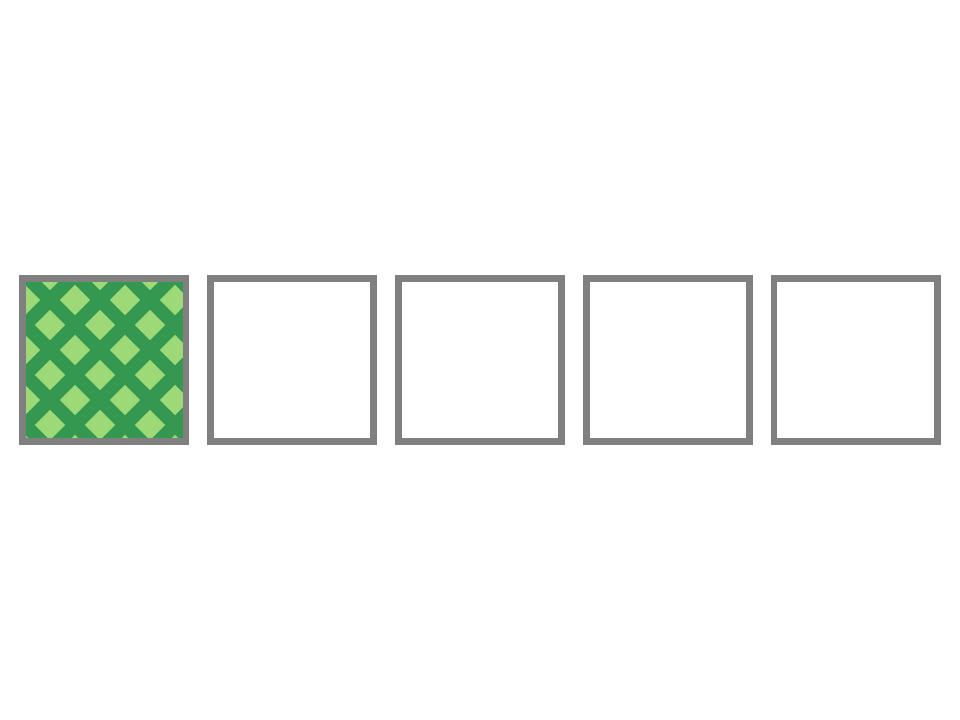} &  \provideranks{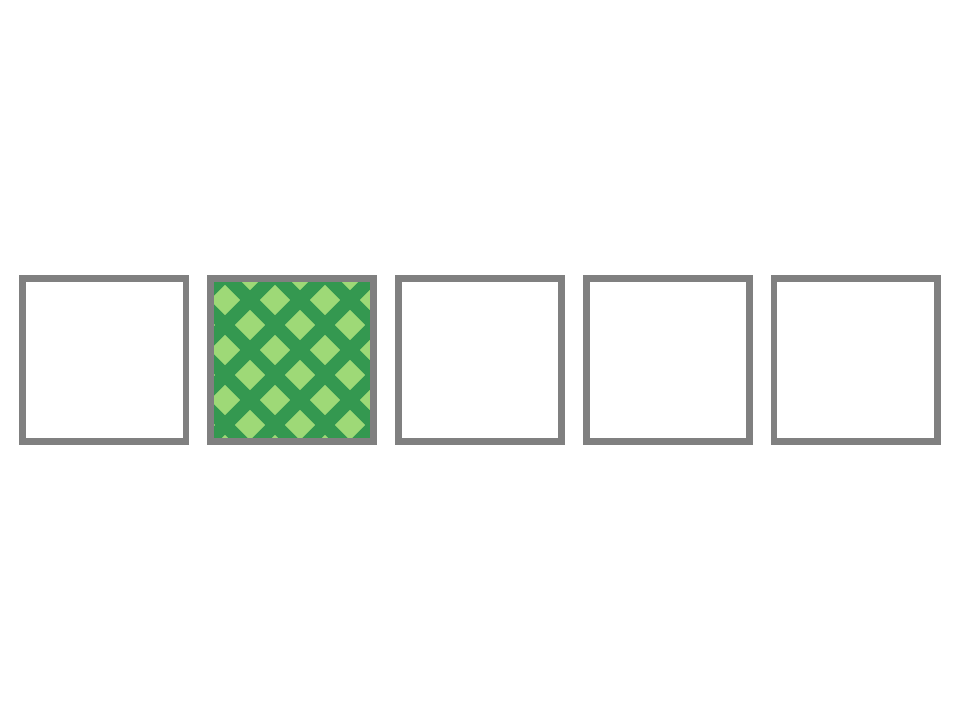} &  \provideranks{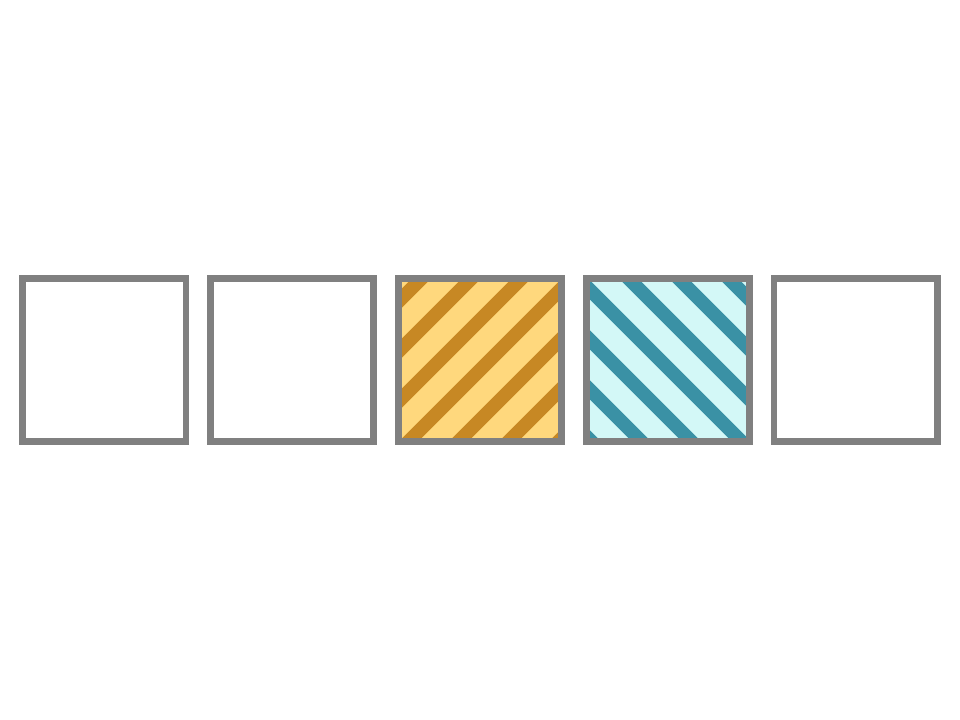} &  \provideranks{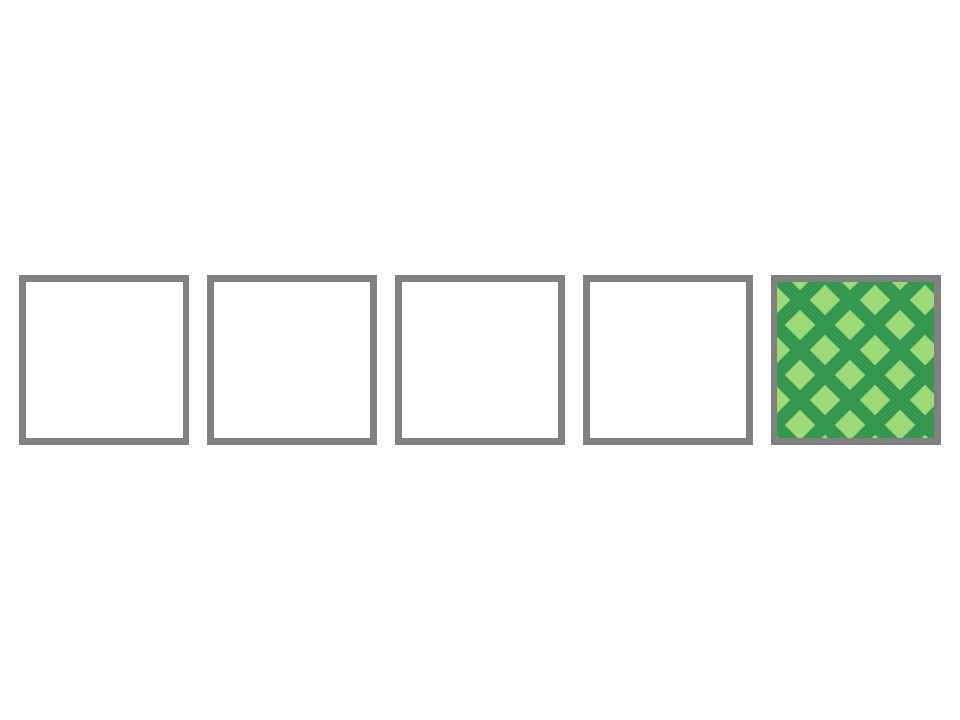} &  \provideranks{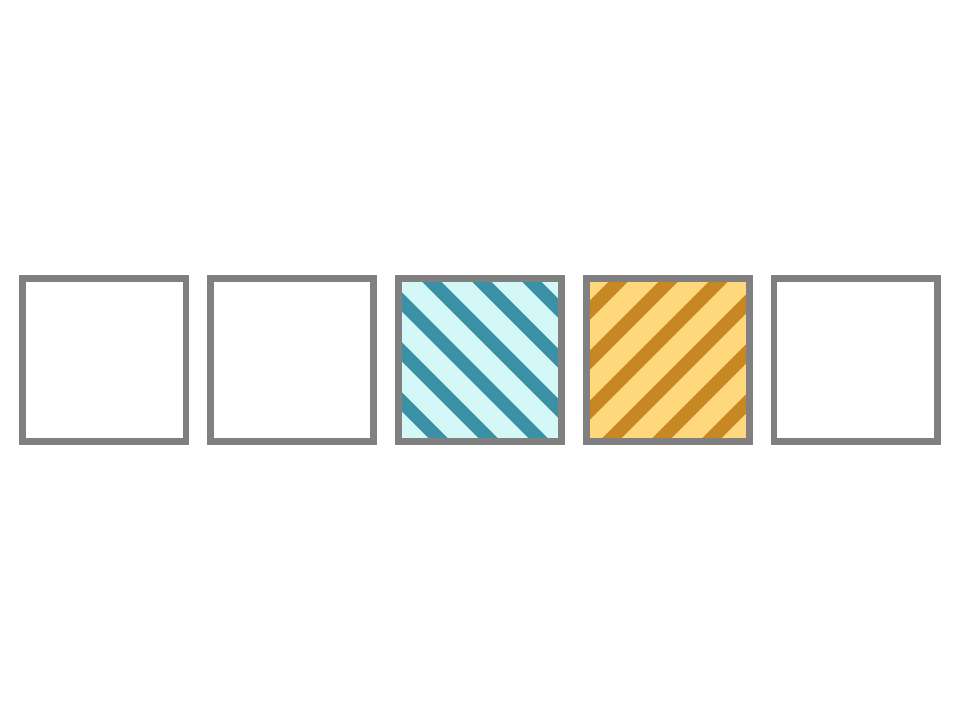} & \phantom{-}0.80 \\
 96 &  \provideranks{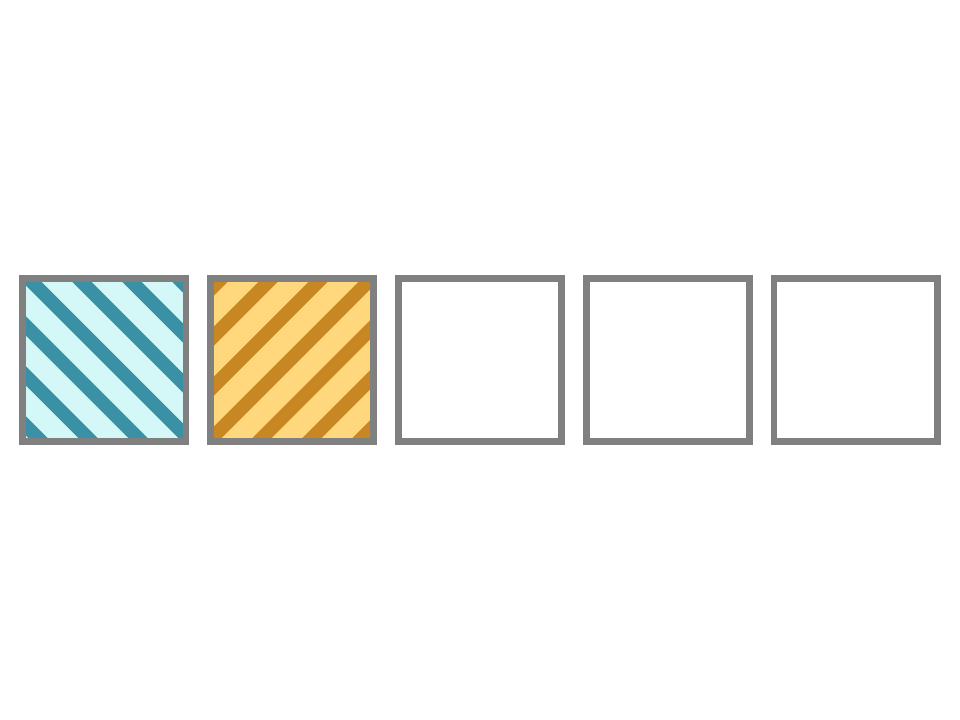} &  \provideranks{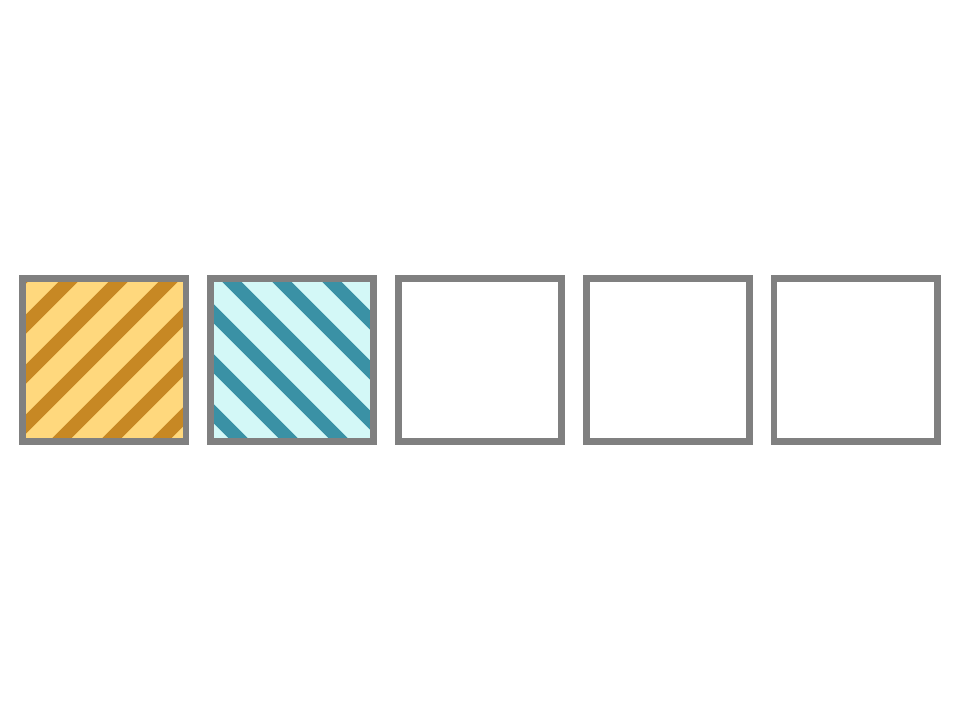} &  \provideranks{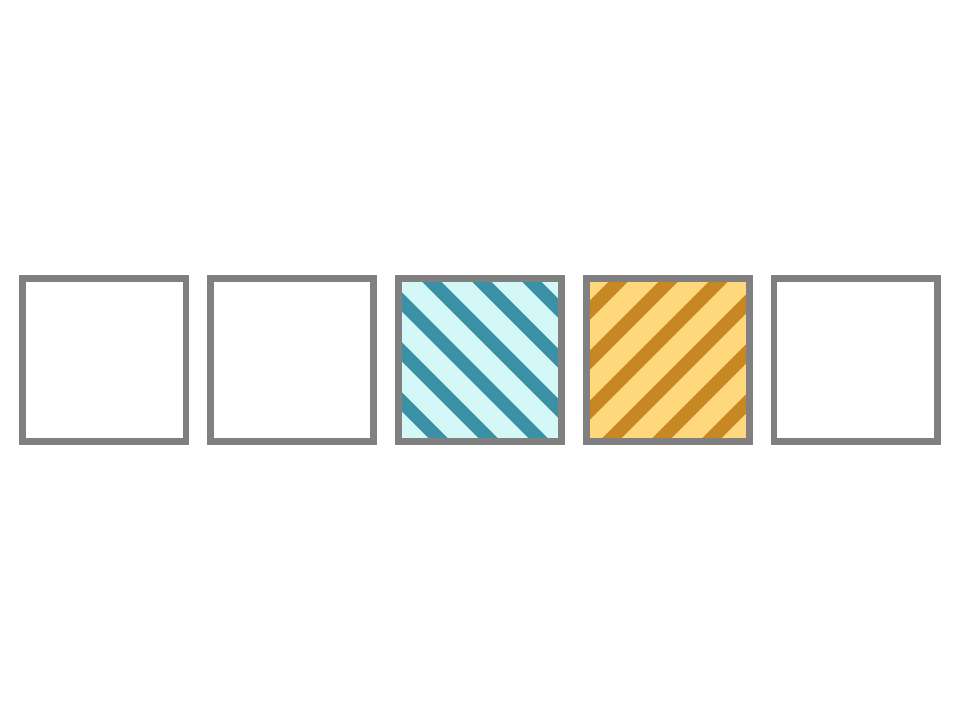} &  \provideranks{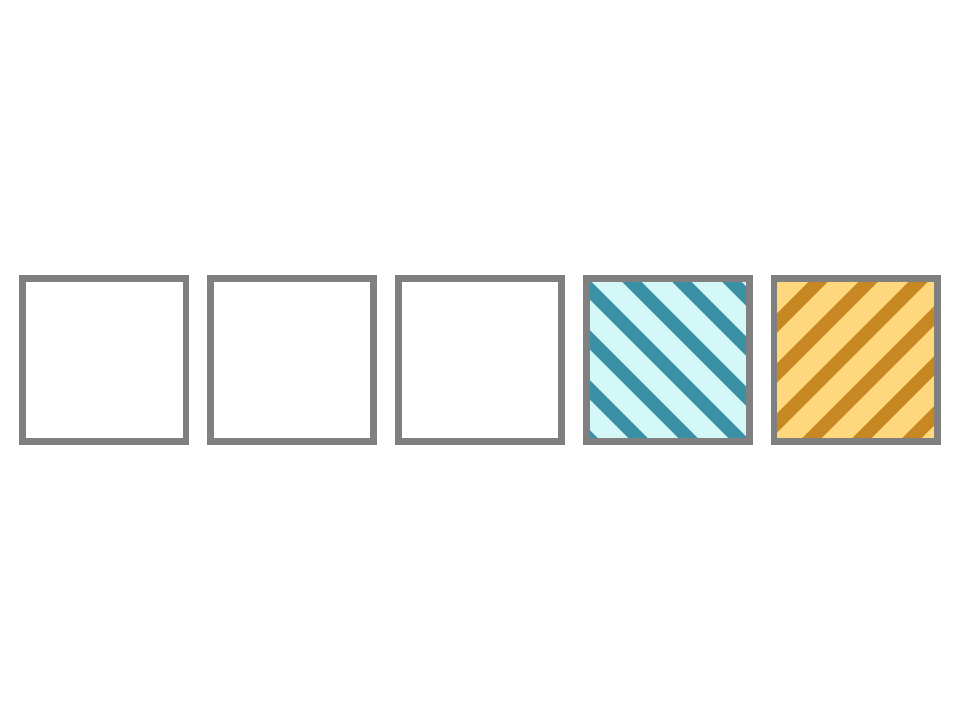} &  \provideranks{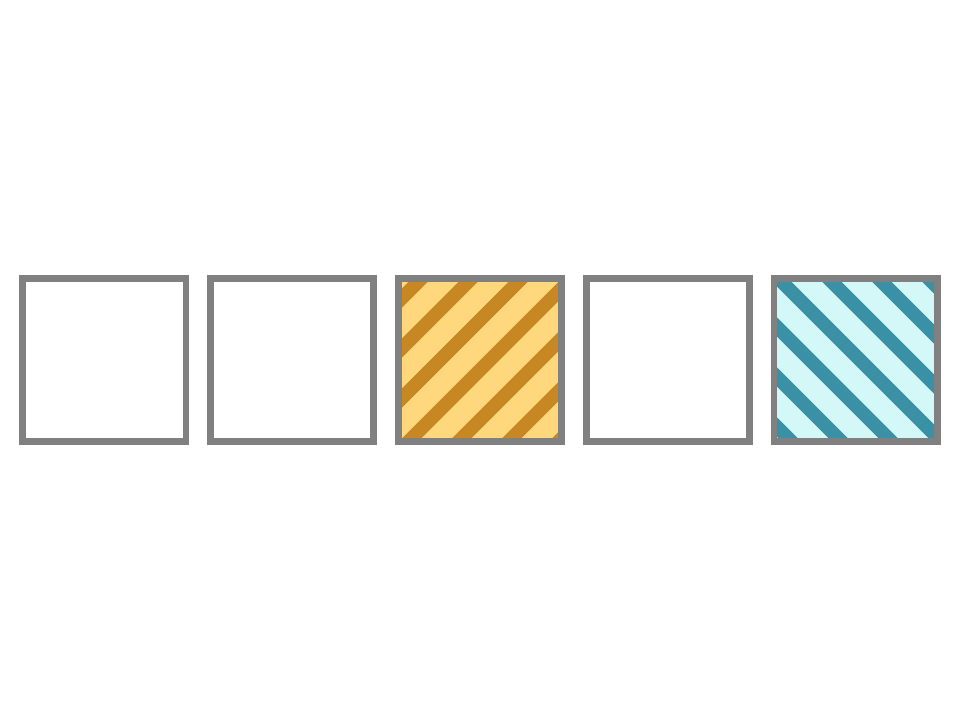} & \phantom{-}0.40 \\
 97 &  \provideranks{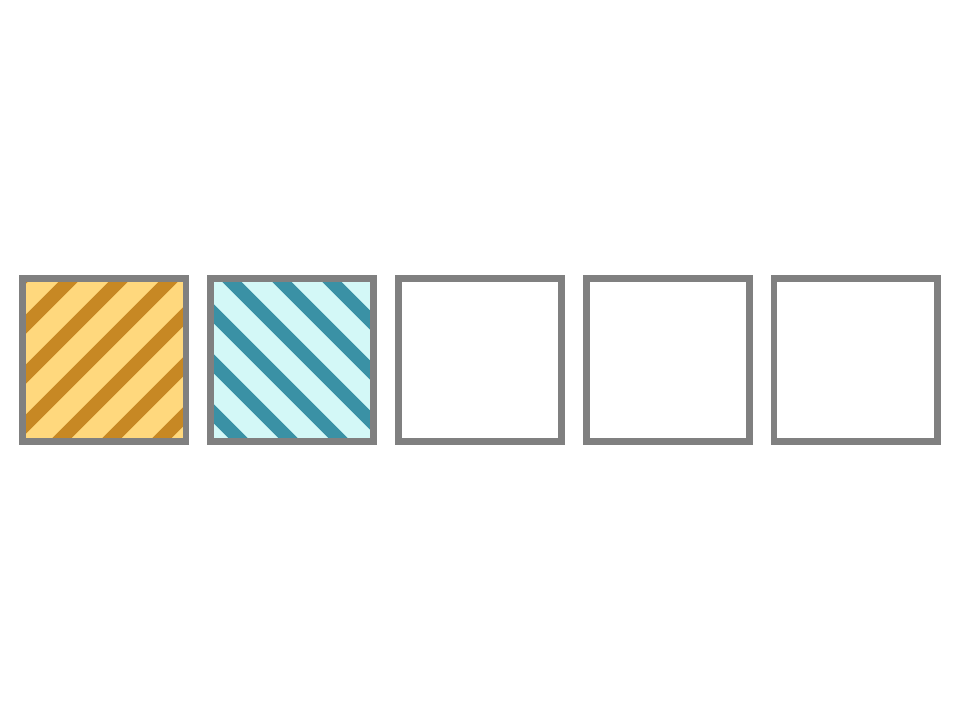} &  \provideranks{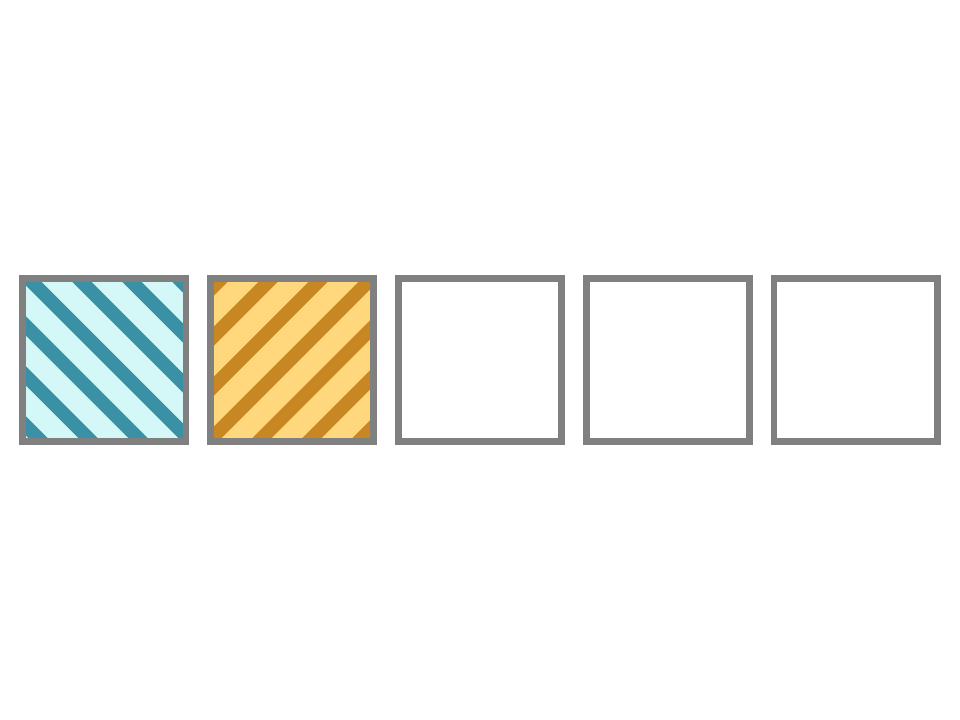} &  \provideranks{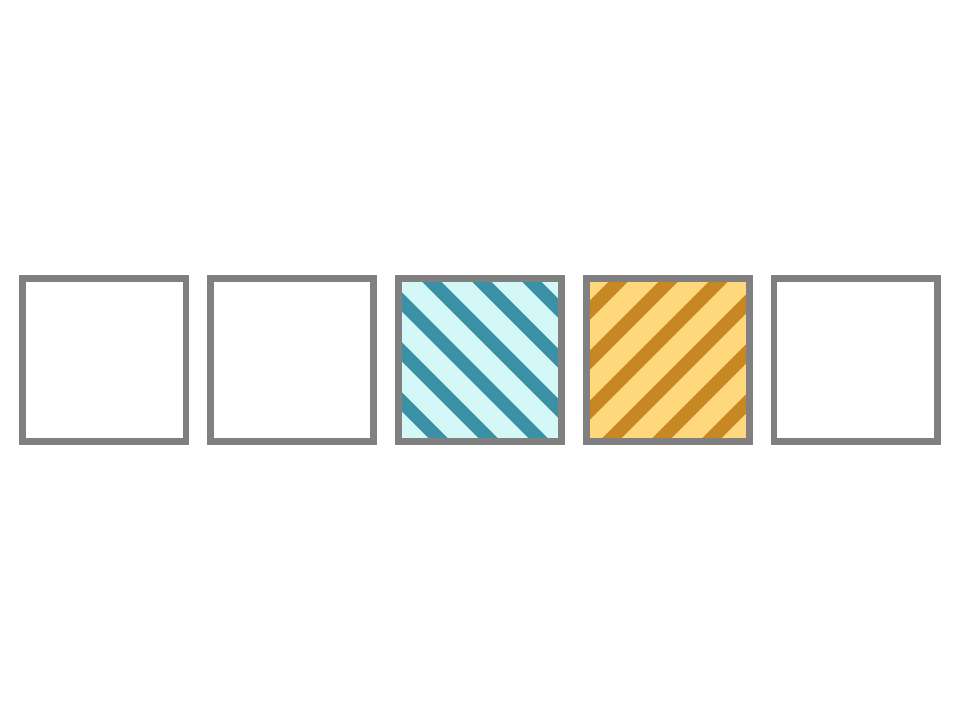} &  \provideranks{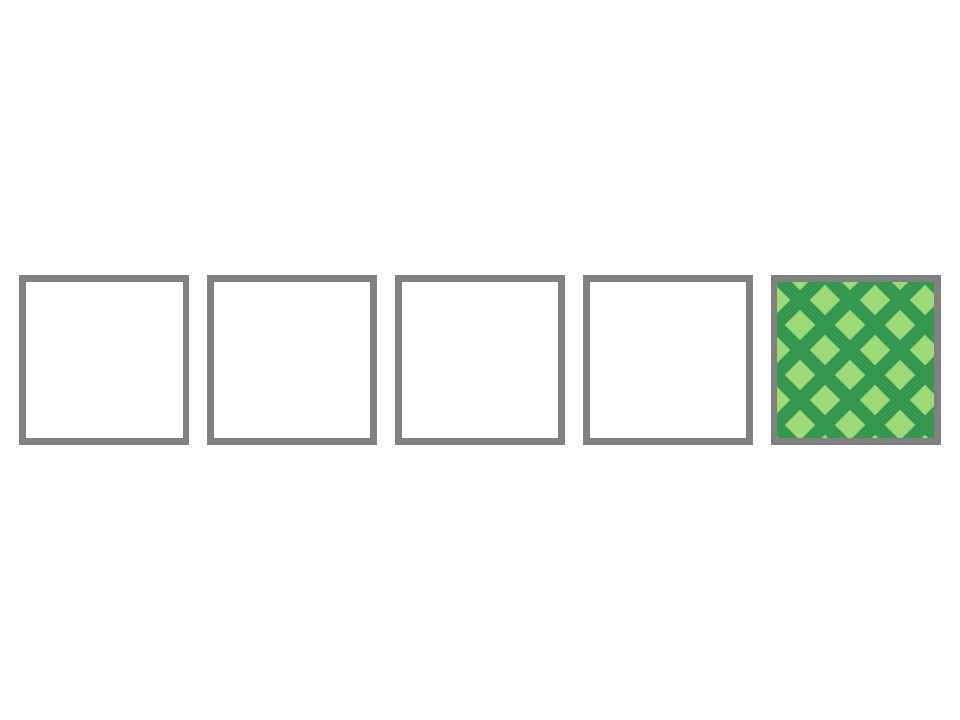} &  \provideranks{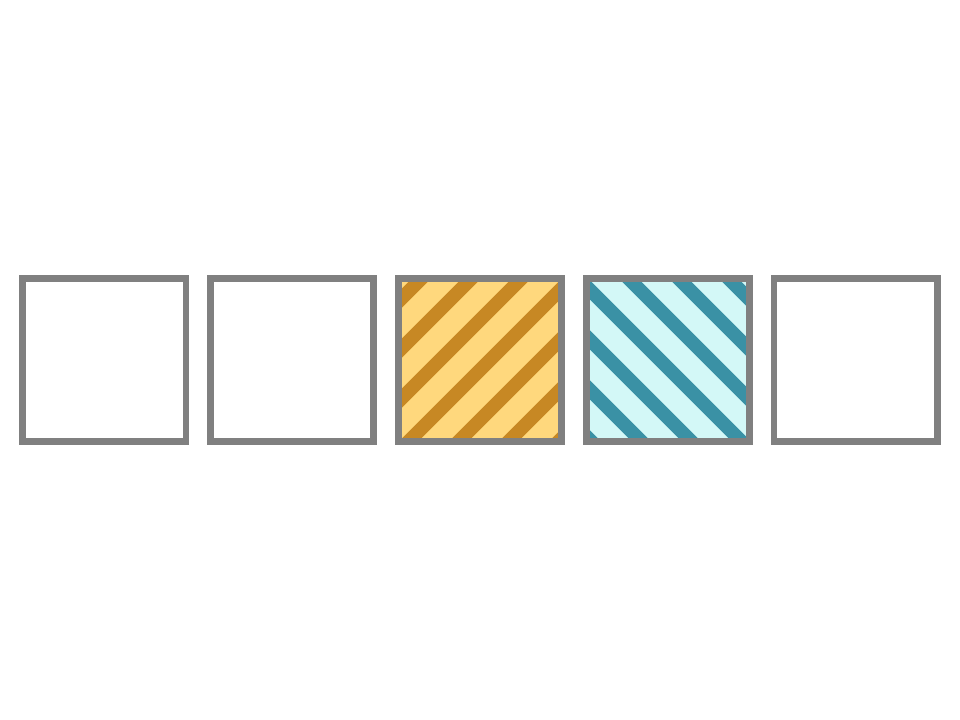} & \phantom{-}0.60 \\
101 & \provideranks{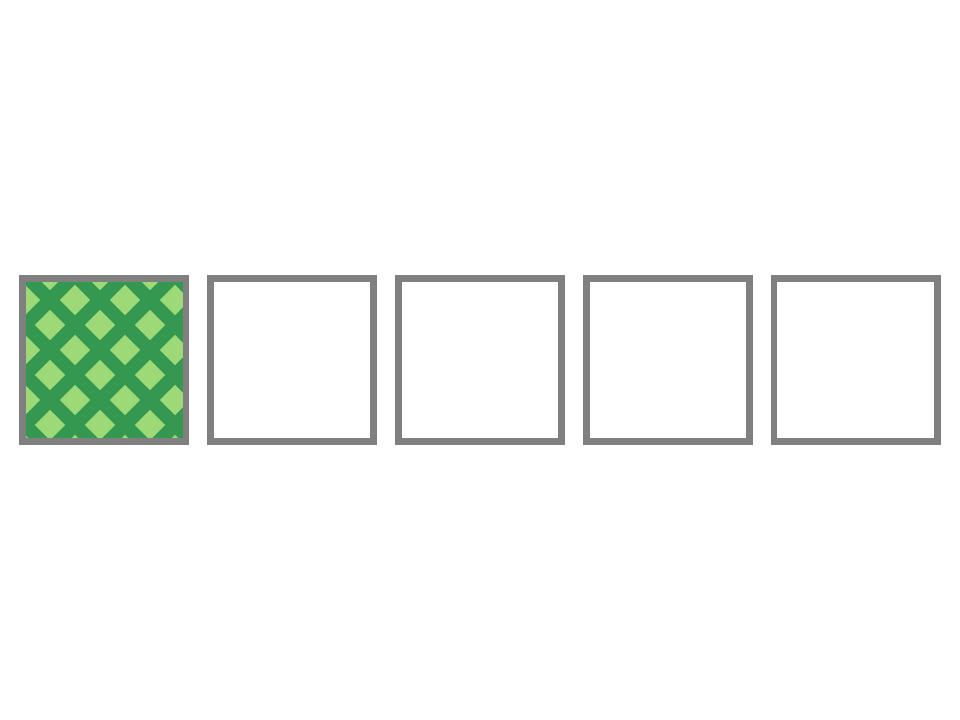} & \provideranks{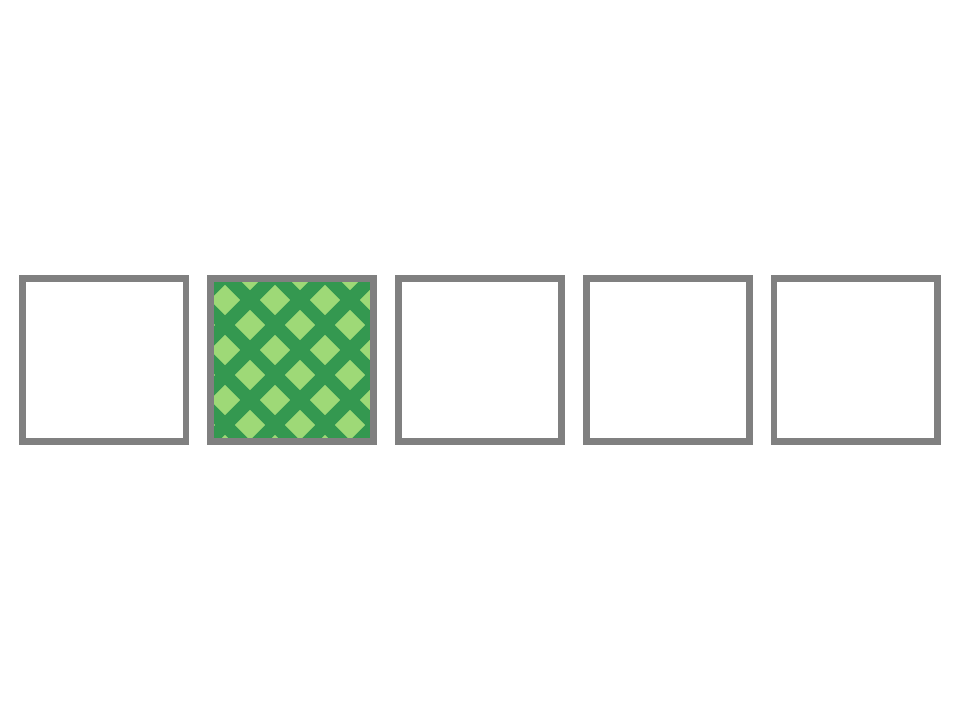} & \provideranks{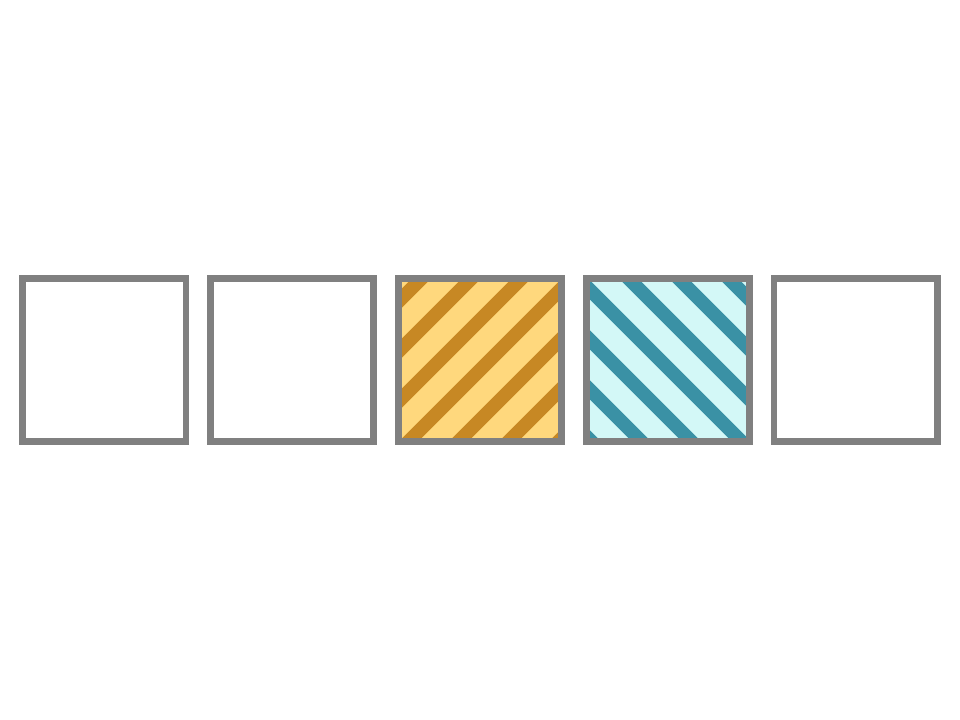} & \provideranks{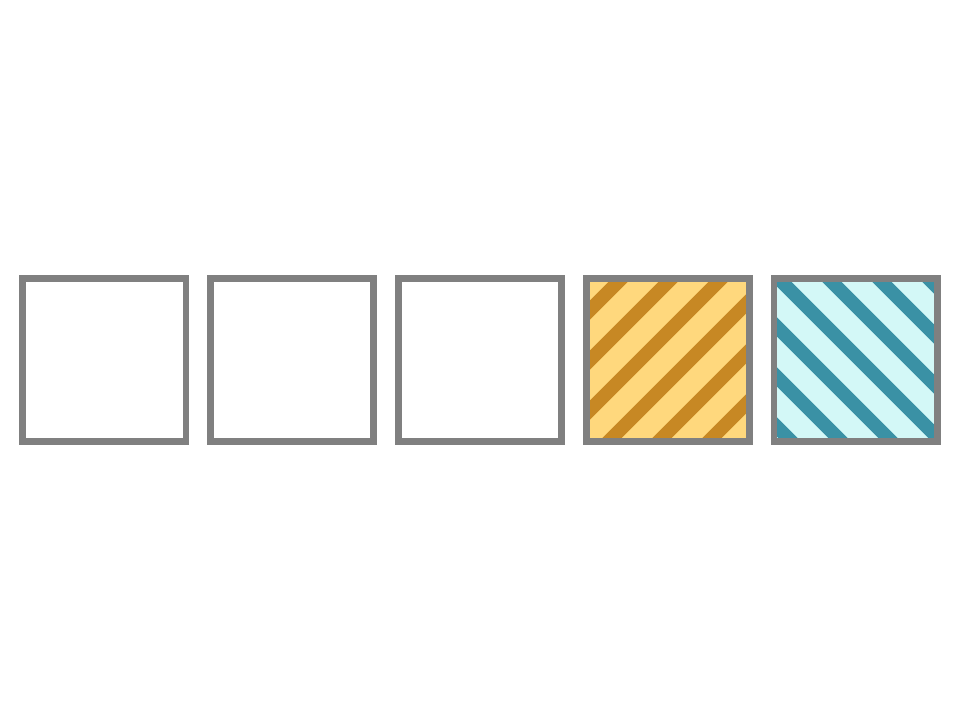} & \provideranks{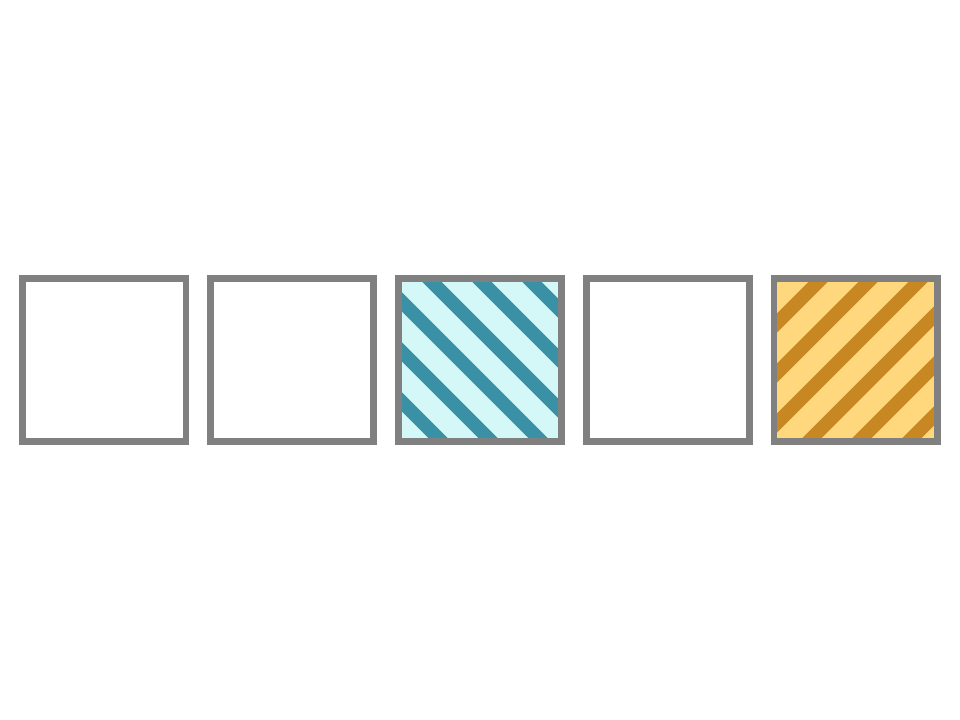} & \phantom{-}0.60 \\
102 & \provideranks{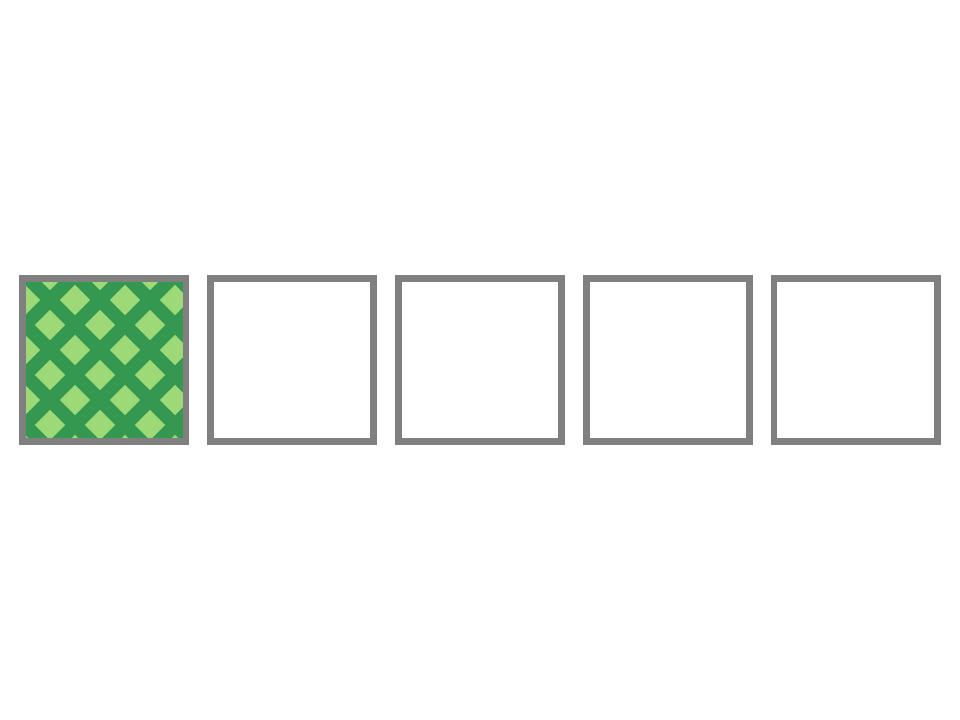} & \provideranks{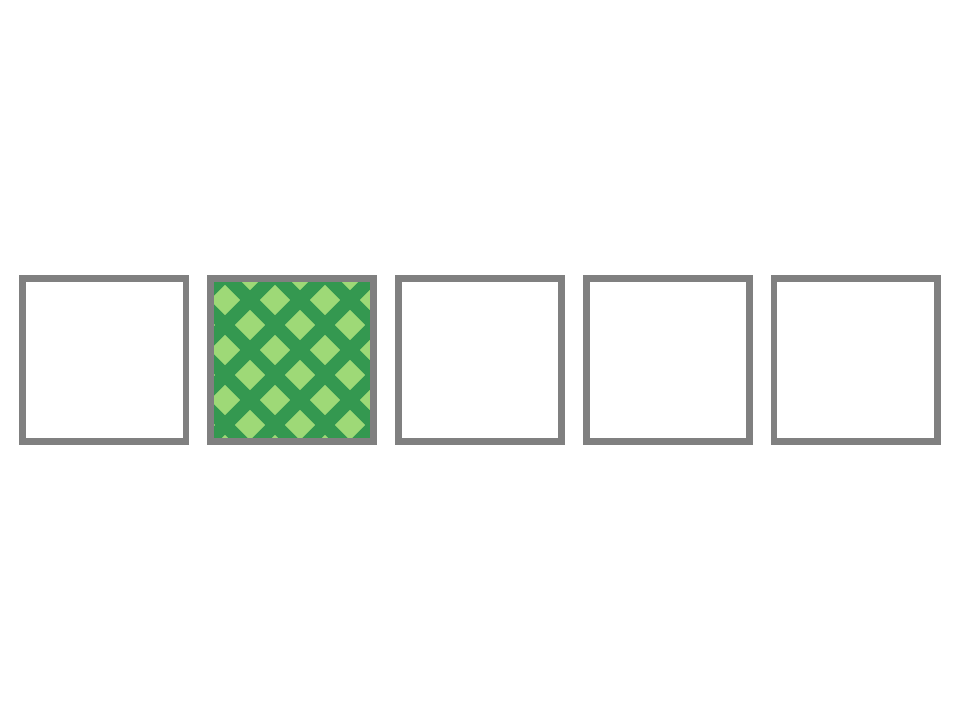} & \provideranks{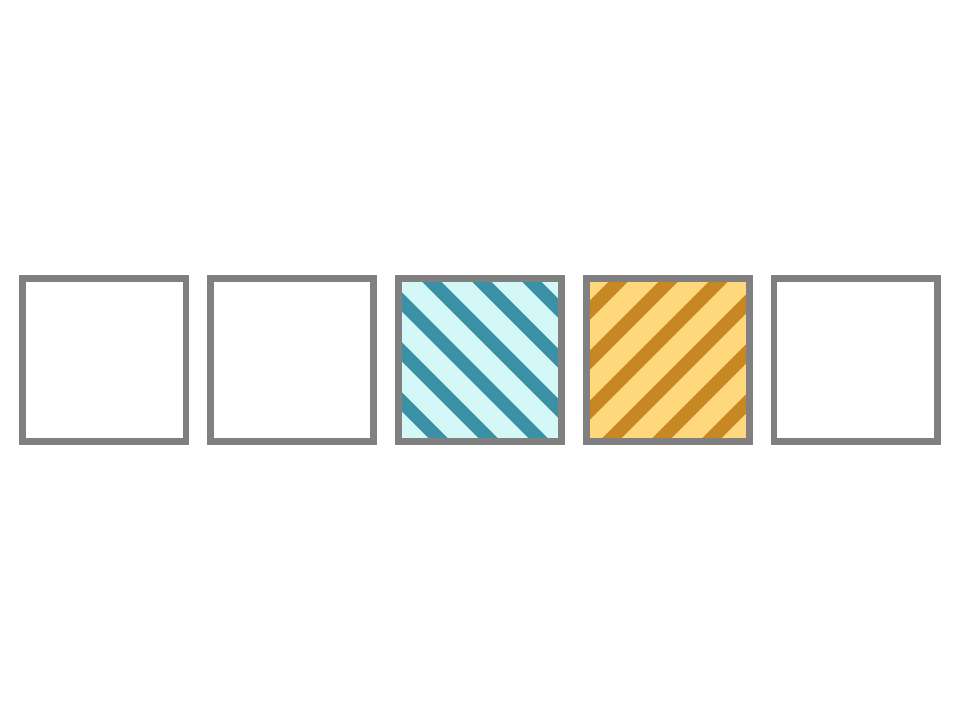} & \provideranks{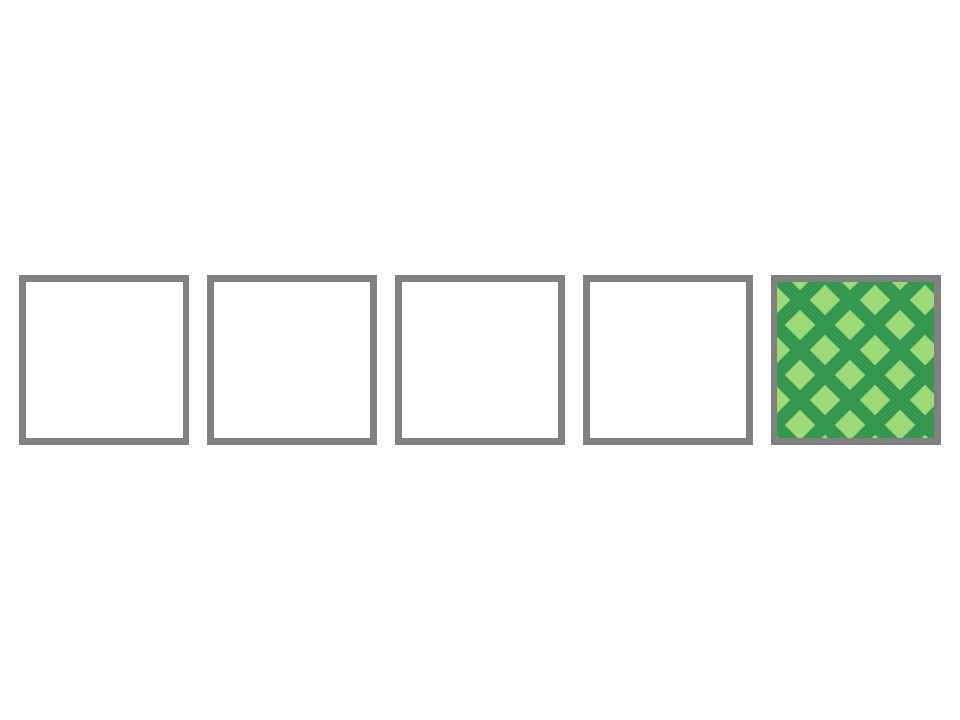} & \provideranks{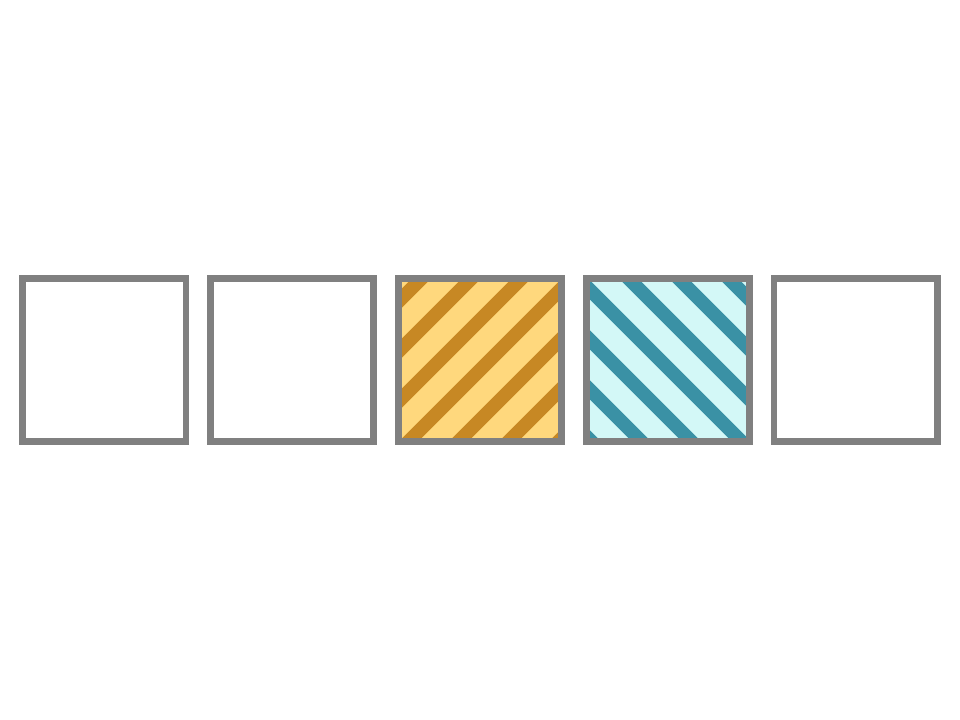} & \phantom{-}0.80 \\
114 & \provideranks{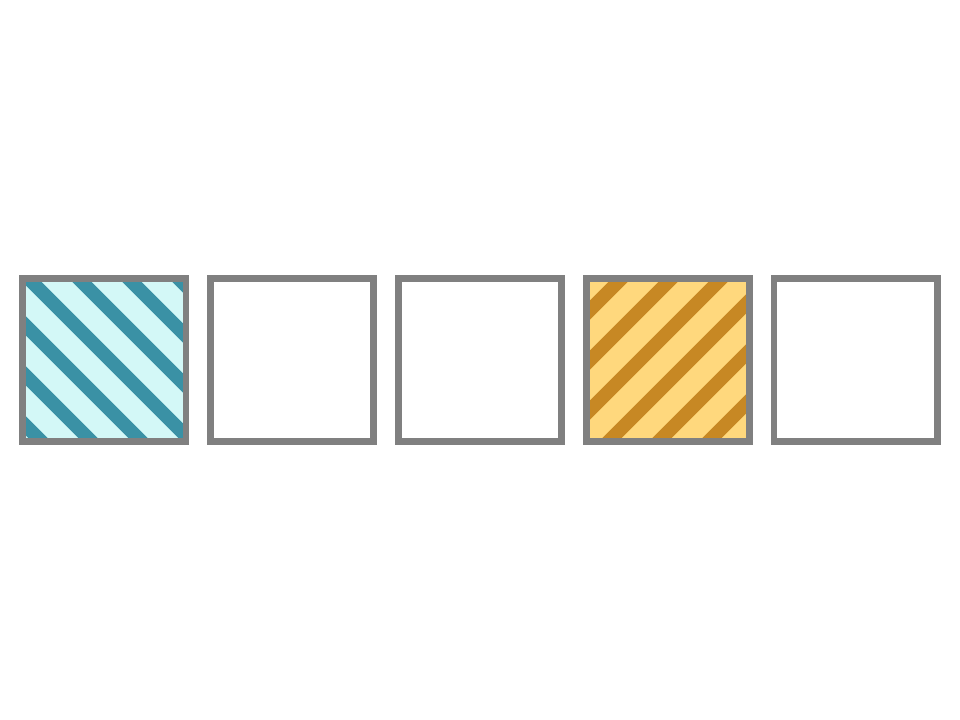} & \provideranks{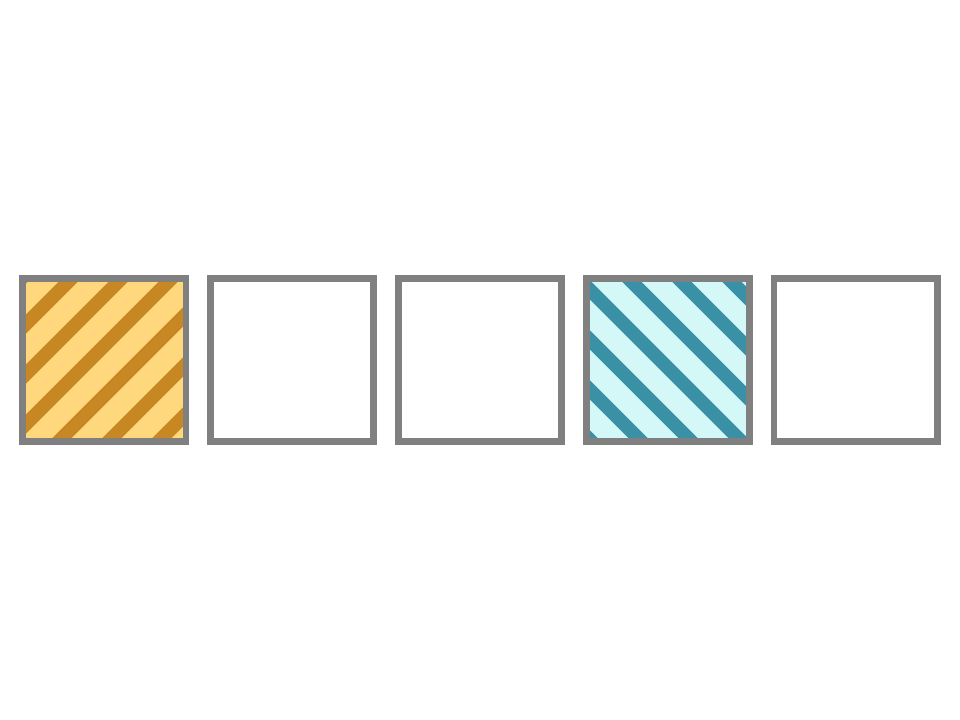} & \provideranks{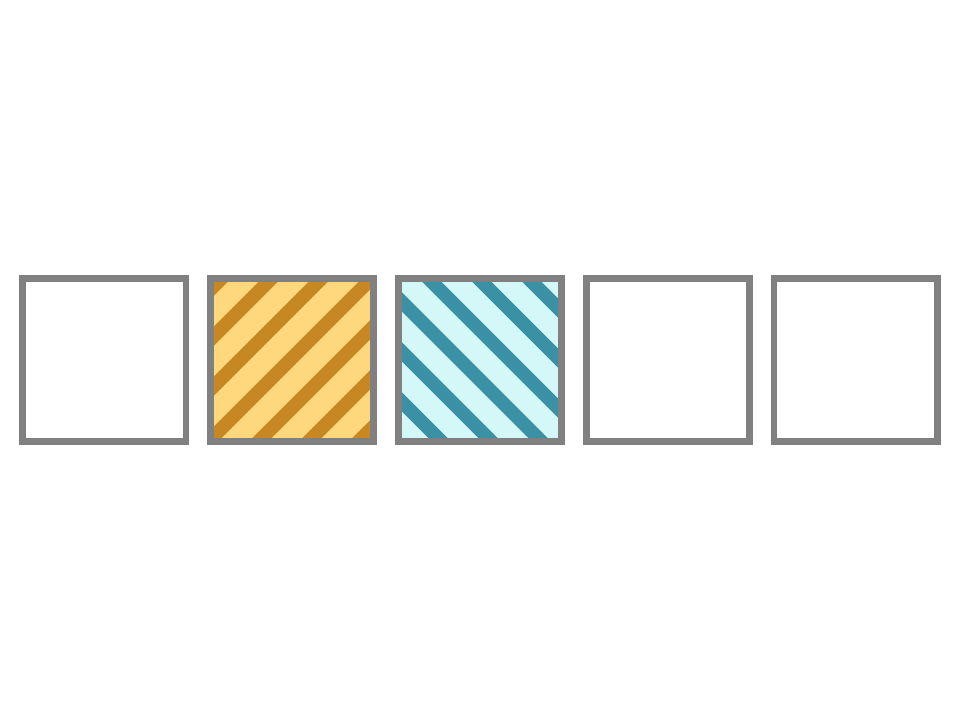} & \provideranks{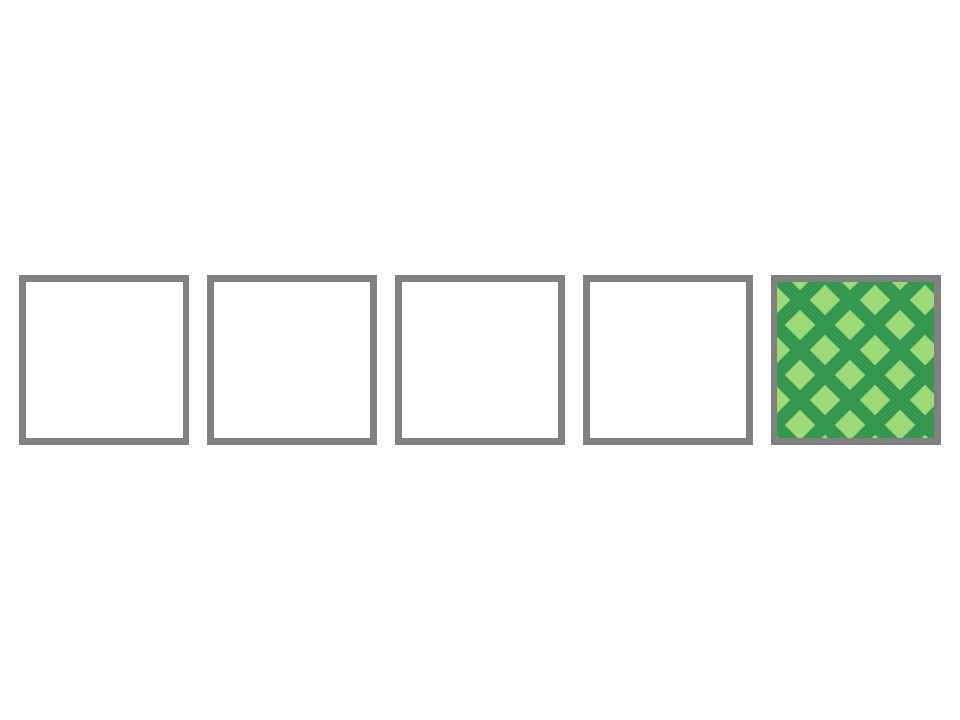} & \provideranks{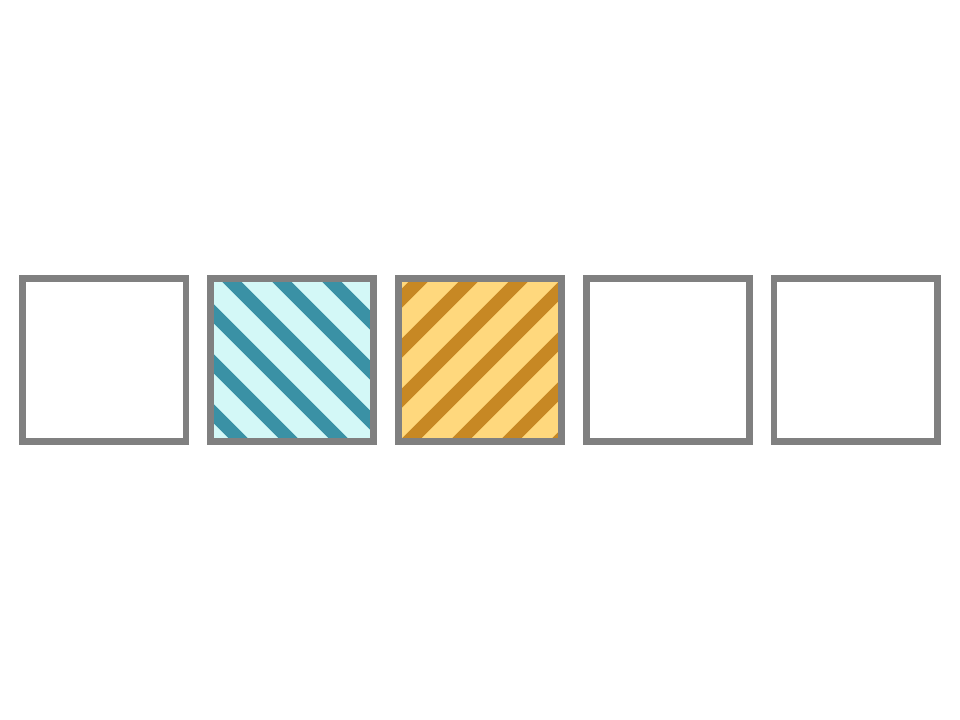} & -0.20 \\
116 & \provideranks{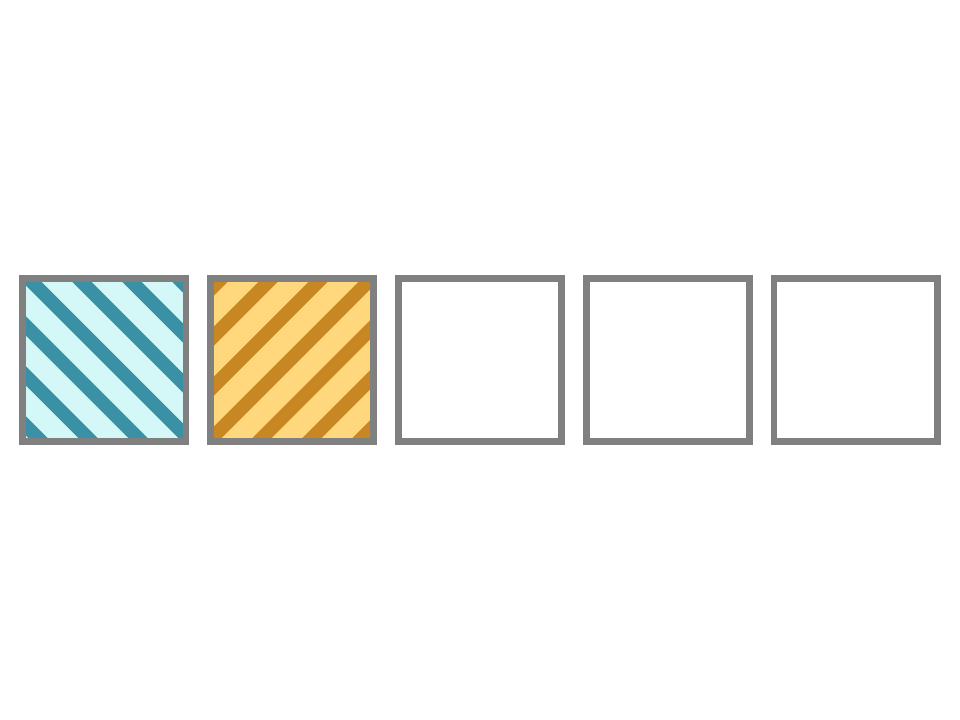} & \provideranks{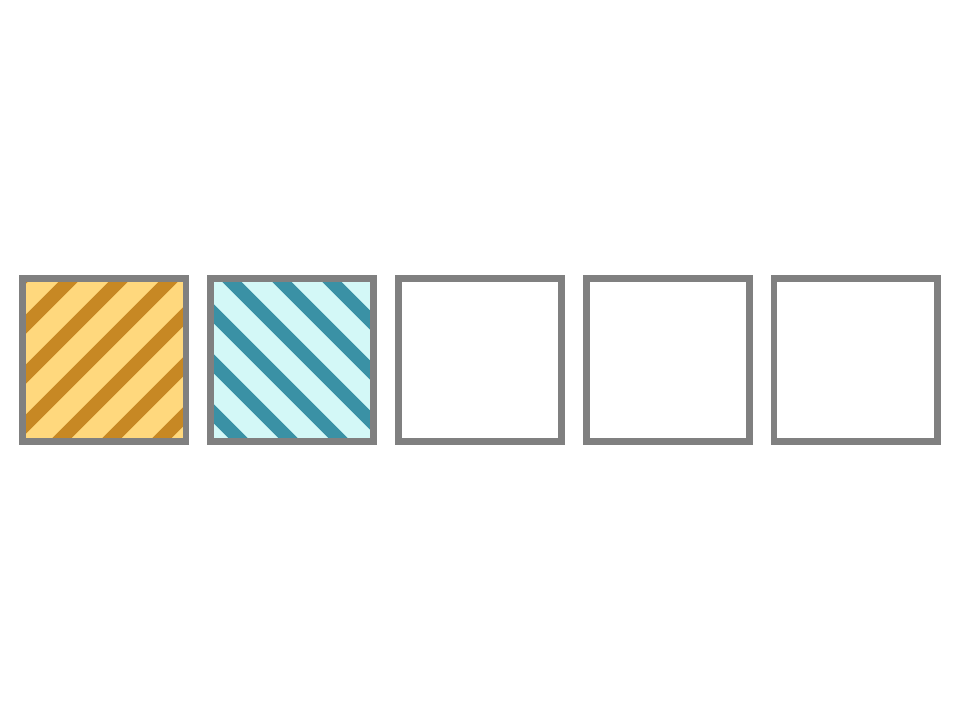} & \provideranks{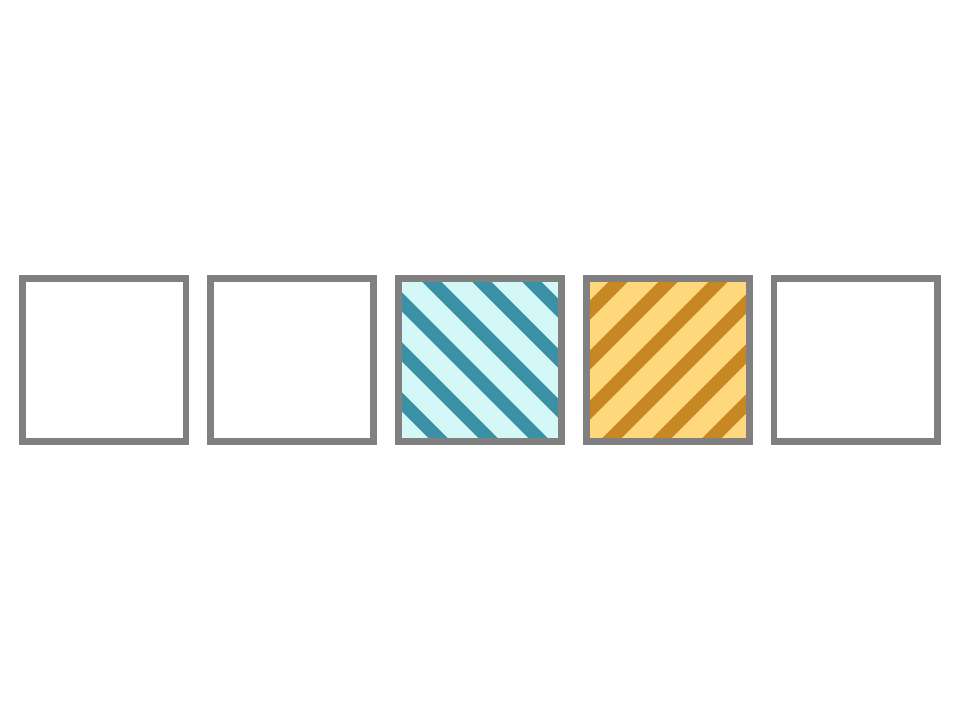} & \provideranks{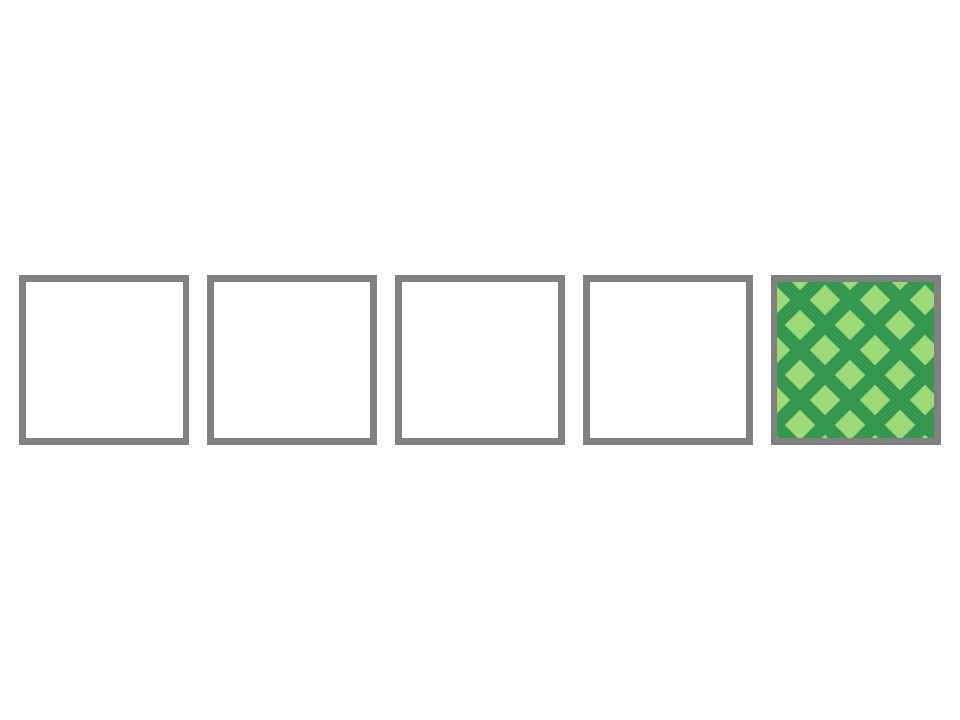} & \provideranks{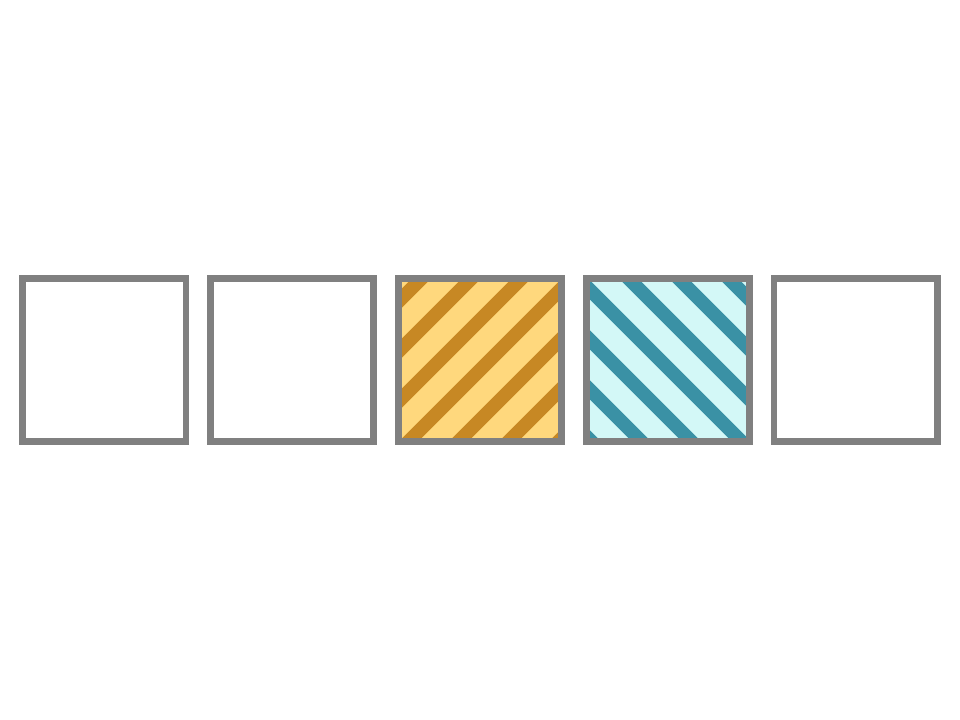} & \phantom{-}0.60 \\
117 & \provideranks{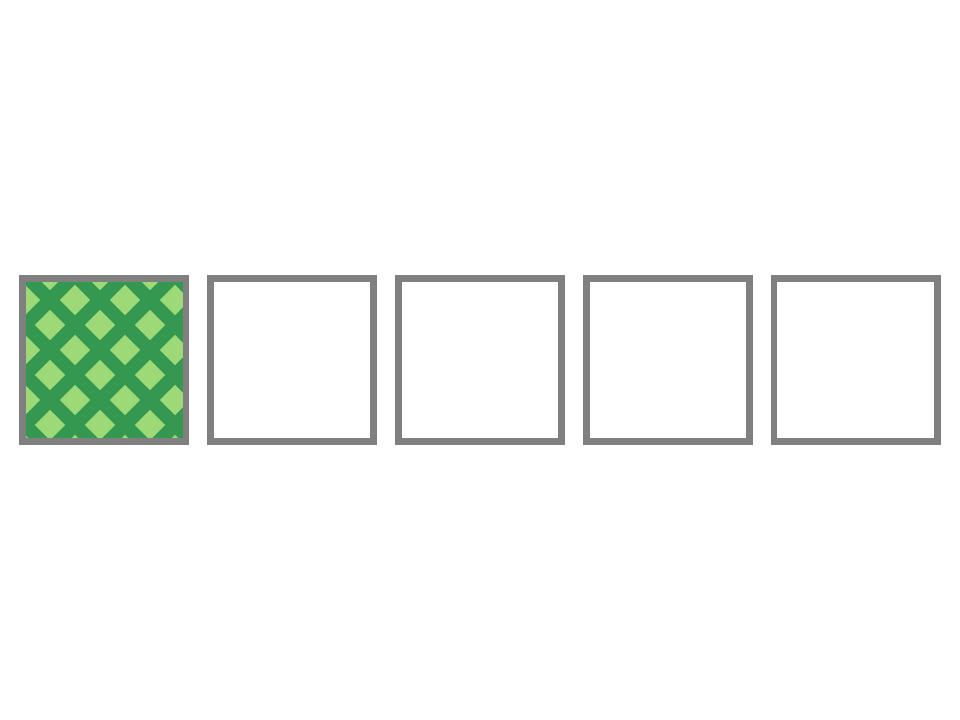} & \provideranks{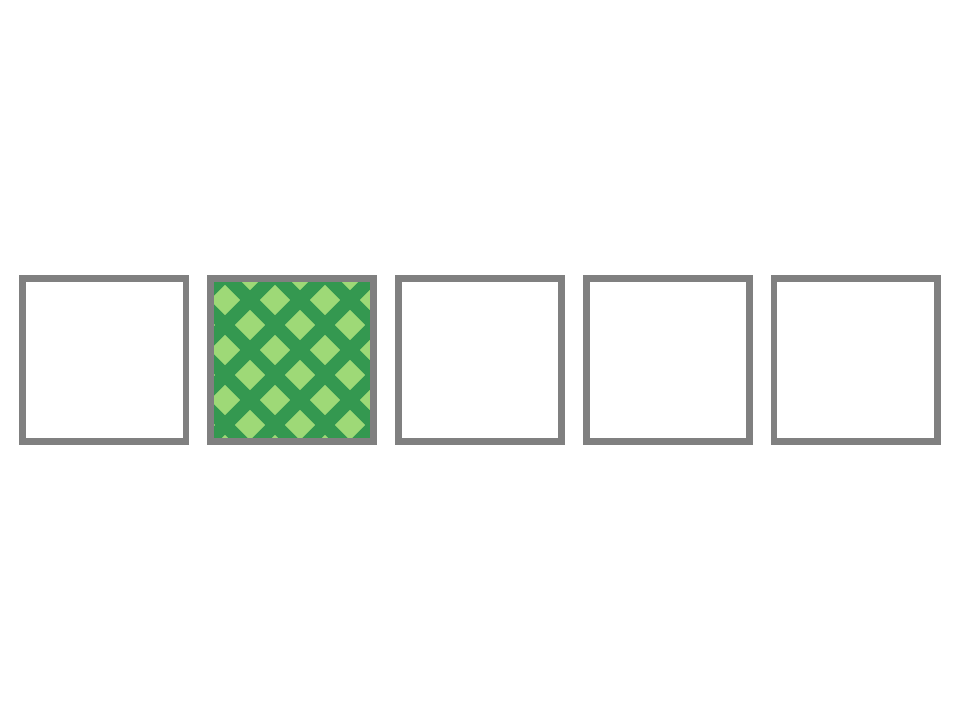} & \provideranks{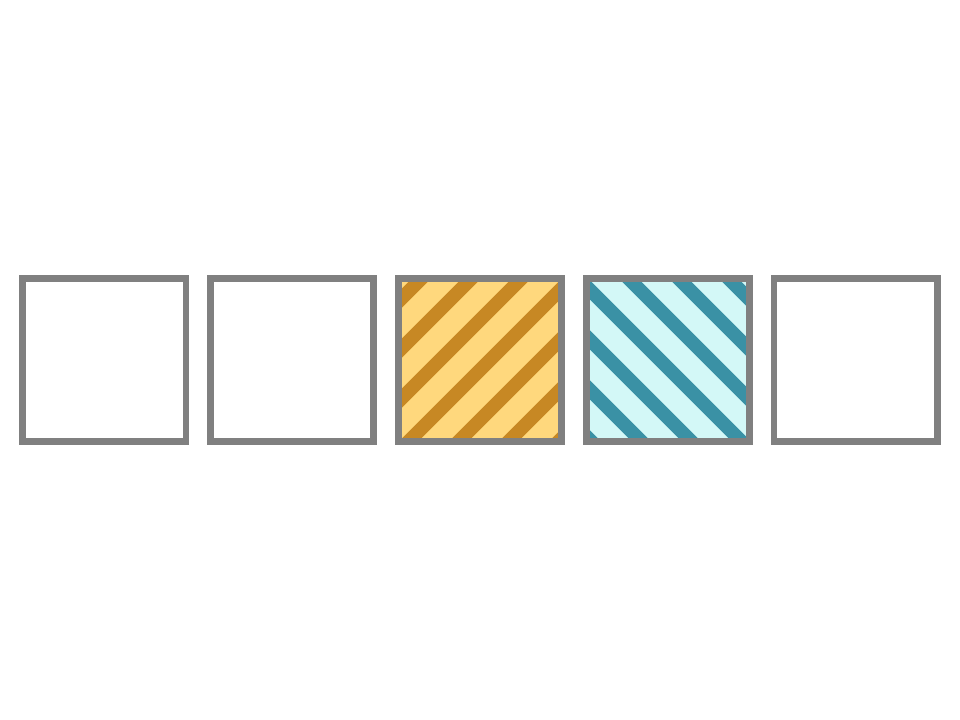} & \provideranks{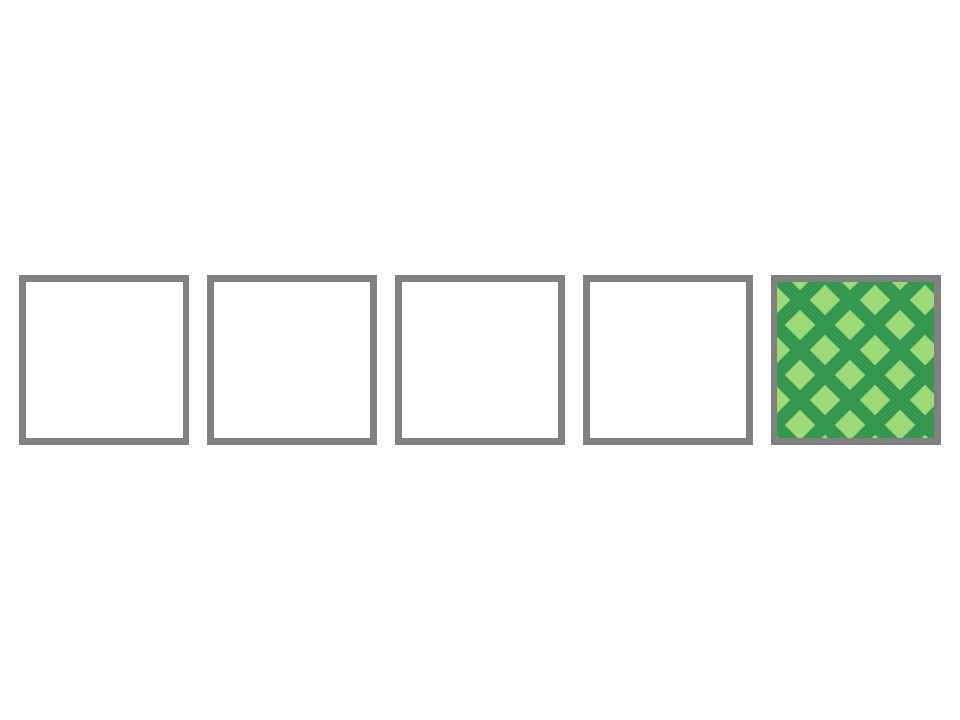} & \provideranks{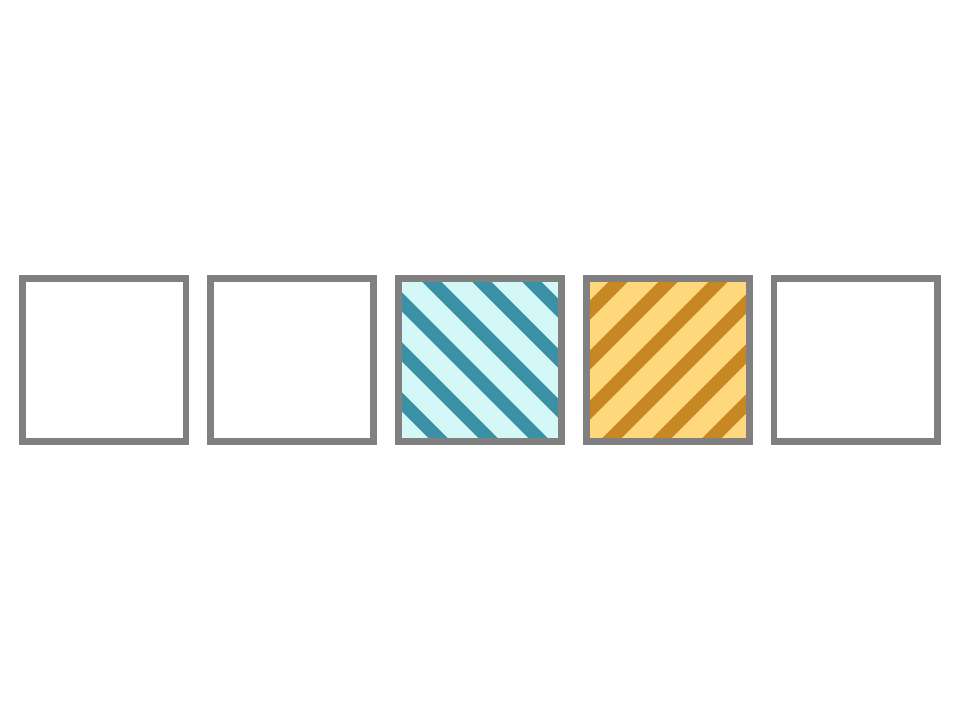} & \phantom{-}0.80 \\
\bottomrule
\end{tabular}
\vspace*{0.5em}
\caption{Correlation between the expert (\provideexample{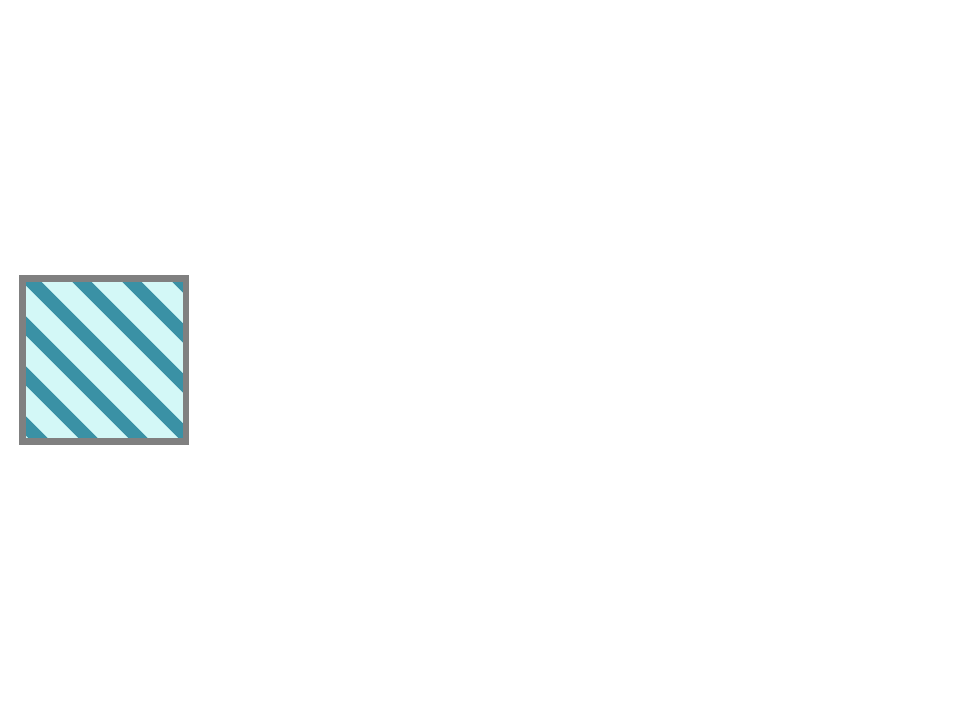}) and \mbox{\texttt{monoT5}} (\provideexample{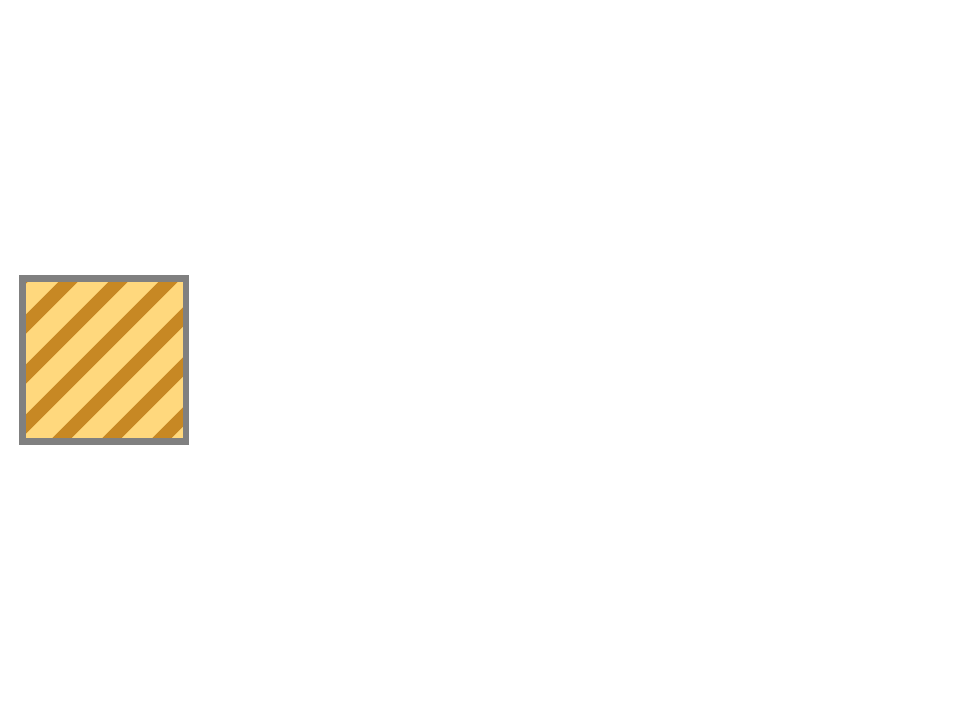}). Each cell shows the ranks 1-5 from left to right.
Equal ranks by the expert and \mbox{\texttt{monoT5}} are illustrated with an overlap in color and pattern (\provideexample{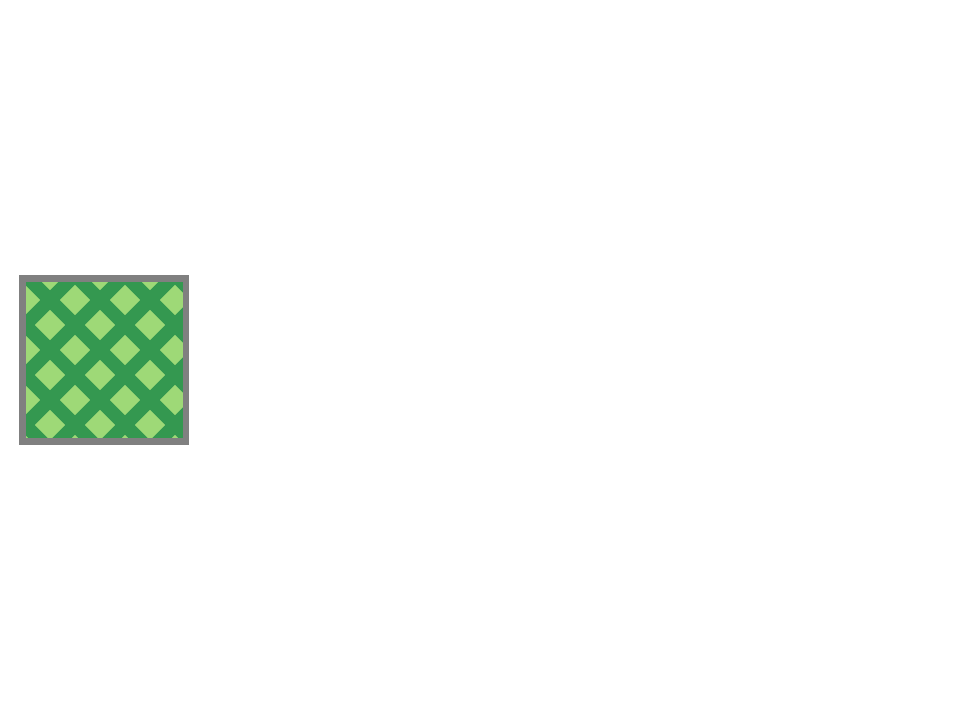}).
Each query is reported by its ID (qid).%
}
\label{table:rank-correlation}
\end{table*}

Finally, we compared the preferences of our chosen ranking model with that of an expert annotator. Table~\ref{table:rank-correlation} summarizes our results. For each of the 20~queries, the table indicates the rank assigned to LLM answers and document by  
\Ni
the expert and   
\Nii
\mbox{\texttt{monoT5}}.
Based on these ranks, the table also shows the rank correlation for each query as measured by Kendall's $\tau$. A consistent observation across almost all queries is that the expert assigned the best rank to \mbox{\texttt{ChatGPT}}, the second best to \mbox{\texttt{Lamma-2~13B}}, and the worst rank to \mbox{\texttt{GPT-2}}. For many queries, this is consistent with the ranks based on \mbox{\texttt{monoT5}} score. Most of the discrepancies between the expert and \mbox{\texttt{monoT5}} occurred with \mbox{\texttt{GPT-2 XL}} and the web document, as they often swapped places between the rankings.
Across all queries, we observe a mean correlation of $\tau = 0.64$ between expert and monoT5 rank with a 95~\% confidence interval of $[0.50, 0.78]$. 
\section{Discussion \& Limitations}
By studying the effect of prompting strategy and model size on NRP, we find that
\Ni
effectiveness increases with model size and
\Nii
instruction-tuned models benefit more from elaborate prompting.
However, we also find that NRP has difficulty distinguishing between highly effective models. Using the MultiMedQA prompt, larger models consistently generated answers that were ranked above all documents $D_q$.  
Whether this is an artifact of the dataset we used or whether this is also the case in other applications, needs to be investigated in future research.

In addition to the quality of the dataset, the efficacy of NRP is influenced by the underlying ranking model for capturing a quality dimension. As a consequence, the biases of a specific ranking model need to be considered when using it to evaluate LLMs with NRP for that dimension. For lexical models like TF-IDF and DPH, the biases are apparent: they use term-frequency to measure relevance and are biased towards LLM answers that contain many terms present in the query. For more sophisticated ranking models, the biases may be less apparent. In our experiments, we did not observe a consistent bias for \mbox{\texttt{monoT5}}.      

In our user study, we correlated the rankings produced by \mbox{\texttt{monoT5}} with those made by a healthcare professional. The high correlation between the expert and \mbox{\texttt{monoT5}} suggest that rankers trained on human-written documents retain their effectiveness for LLM answers, despite the different structure and length of these answers. Naturally, this finding is limited to the specific dataset we used in our experiments and by the fact that we collected annotations from only one expert. 

While our proposed method offers a new way to automatically score LLMs on open-ended questions, it depends on the availability of high-quality human annotations and ranking models that can capture them. This limits NRP to applications where these are available, and therefore makes it challenging to apply to low-resource languages and tasks where annotations are costly to obtain.  
\section{Conclusions}

We developed a method to scalably and automatically evaluate generated answers to open-ended questions. Our NRP~score evaluates model effectiveness by ranking answers in comparison with human-written documents. NRP~does not rely on `gold standard' answers, but uses existing texts. We investigated the factors that led to higher effectiveness when using the measure, and whether human evaluations of generated answers agree with the rankings from an effective ranking model.

We demonstrate that state-of-the-art ranking functions, once validated using a test collection with multi-dimensional relevance assessments, can be used to effectively discriminate high-quality from low-quality answers. The ranking functions can even be used to discriminate between answers from similar~LLMs with different prompting strategies. Furthermore, once the test collection and ranking function are finalized, other LLMs can be evaluated using the setup in an entirely offline and scaleable manner.

The experiments in this paper focused on evaluating generated answers for consumer health search questions on the CLEF 2021 eHealth dataset. For future work, we will develop and adapt additional test collections for other domains. Another avenue for future research is the extension of NRP to assess retrieval-augmented generation~(RAG) systems that ground their answers in documents by referring to them. This paper contributes a new way to automatically assess LLM capabilities by evaluating them in the context of human-annotated documents.

{\fontsize{9pt}{11pt}\selectfont 
\subsubsection*{Acknowledgements}
This publication has received funding from the European Union's Horizon Europe research and innovation programme under grant agreement No 101070014 (OpenWebSearch.EU, \url{https://doi.org/10.3030/101070014}).
}
\newpage
\renewcommand\bibsection{}
\bibliographystyle{splncs04nat}
\section*{References}
\bibliography{ecir25-evaluating-chs-answers-lit}
\end{document}